\documentclass[a4paper,fleqn,usenatbib,useAMS]{mnras}
\pdfoutput=1
\usepackage{newtxtext,newtxmath}
\usepackage[T1]{fontenc}
\usepackage{ae,aecompl}

\usepackage{graphicx}	
\usepackage{amsmath}	
\usepackage{amssymb}	
\usepackage{multirow}
\usepackage{xspace}
\usepackage[utf8]{inputenc}
\usepackage{pdflscape} 
\usepackage{longtable} 
\usepackage[para]{footmisc} 
\usepackage{xcolor}
\usepackage{subfiles}

\newcommand{\teff}{\ensuremath{T_{\textup{eff}}}\xspace}

\newcommand{\logg}{\ensuremath{\log g}\xspace}

\title[A first view of $\delta$~Sct and $\gamma$~Dor stars with TESS]{The first view of $\delta$~Scuti and $\gamma$~Doradus stars with the TESS mission}

\author[V. Antoci et al.]{
V. Antoci,$^{1}$\thanks{E-mail: antoci@phys.au.dk}
M. S. Cunha$^{2}$,	
D. M. Bowman$^{3}$,
S. J. Murphy$^{4,1}$,
D. W. Kurtz$^{5}$,
\newauthor
T. R. Bedding$^{4,1}$,
C. C. Borre$^{1}$,
S. Christophe$^{6}$,
J.~Daszy\'nska-Daszkiewicz$^{7}$,
\newauthor
L. Fox-Machado$^{8}$,
A. Garc\'{\i}a~Hern\'andez$^{9, 10}$,
H. Ghasemi$^{11}$,
R. Handberg$^{1}$,
\newauthor
H. Hansen$^{1}$,
A. Hasanzadeh$^{12}$,
G. Houdek$^{1}$,
C. Johnston$^{3}$,
A. B. Justesen$^{1}$,
\newauthor
F. Kahraman Alicavus$^{13, 14}$,
K. Kotysz$^{7}$,
D. Latham$^{15}$,
J.M. Matthews$^{16}$,
J. M{\o}nster$^{1}$,
\newauthor
E. Niemczura$^{7}$,
E. Paunzen$^{17}$,
J.~P.~S\'anchez Arias$^{18}$,
A. Pigulski$^{7}$,
J. Pepper$^{19}$,
\newauthor
T. Richey-Yowell$^{19, 20}$,
H. Safari$^{12}$,
S. Seager$^{21, 22, 23}$
B. Smalley$^{24}$,
T. Shutt$^{25}$,
A. S\'odor$^{26, 27}$,
\newauthor
J.-C. Su\'arez$^{28, 29}$,
A. Tkachenko$^{3}$,
T. Wu$^{30, 31, 32}$,
K. Zwintz$^{33}$,
S. Barcel\'o Forteza,$^{34}$,
\newauthor
E. Brunsden$^{25}$,
Z. Bogn\'ar$^{26, 27}$,
D. L. Buzasi$^{35}$,
S. Chowdhury$^{13}$,
P. De Cat$^{36}$,
\newauthor
J. A. Evans$^{24}$,
Z. Guo$^{37, 13}$,
J. A.  Guzik$^{38}$,
N. Jevtic$^{39}$,
P. Lampens$^{36}$,
M. Lares Martiz$^{10}$,
\newauthor
C. Lovekin$^{40}$,
G. Li$^{4}$,
G. M. Mirouh$^{41}$,
D. Mkrtichian$^{42}$,
M. J. P. F. G.  Monteiro$^{2, 43}$,
\newauthor
J. M. Nemec$^{44}$,
R. Ouazzani$^{6}$,
J. Pascual-Granado$^{10}$,
D. R. Reese$^{6}$,
M. Rieutord$^{45}$,
\newauthor
J. R. Rodon$^{10}$,
M. Skarka$^{46, 47}$,
P. Sowicka$^{13}$,
I. Stateva$^{48}$,
R. Szab\'o$^{26, 27}$,
\newauthor
and W.W.  Weiss$^{49}$
\\
$^{ }$The authors' affiliations are shown in Appendix B.
}

\date{Accepted XXX. Received YYY; in original form ZZZ}

\pubyear{2019}

\begin{document}
\label{firstpage}
\pagerange{\pageref{firstpage}--\pageref{lastpage}}
\maketitle

\begin{abstract}
{We present the first asteroseismic results for $\delta$~Scuti and $\gamma$~Doradus stars observed in Sectors 1 and 2 of the TESS mission. We utilise the 2-min cadence TESS data for a sample of 117 stars to classify their behaviour regarding variability and place them in the Hertzsprung-Russell diagram using Gaia DR2 data. Included within our sample are the eponymous members of two pulsator classes, $\gamma$~Doradus and SX~Phoenicis. Our sample of pulsating intermediate-mass stars observed by TESS also allows us to confront theoretical models of pulsation driving in the classical instability strip for the first time and show that mixing processes in the outer envelope play an important role. We derive an empirical estimate of 74\% for the relative amplitude suppression factor as a result of the redder TESS passband compared to the {\it Kepler} mission using a pulsating eclipsing binary  system. Furthermore, our sample contains many high-frequency pulsators, allowing us to probe the frequency variability of hot young $\delta$~Scuti stars, which were lacking in the {\it Kepler} mission data set, and identify promising targets for future asteroseismic modelling. The TESS data also allow us to refine the stellar parameters of SX~Phoenicis, which is believed to be a blue straggler. }

\end{abstract}




\section{Introduction}
\label{section: intro}

One of the important and long-standing goals within astronomy is to constrain the largely unknown interior physics of stars across the Hertzsprung--Russell (HR) diagram, and ultimately improve our understanding of stellar evolution. In particular, the interior rotation, mixing and angular momentum profiles represent significant uncertainties in theoretical models of stellar structure. All these uncertainties propagate from the main-sequence into later stages as these strongly influence the evolution of a star \citep{2009pfer.book.....M, 2013LNP...865....3M, 2018arXiv180907779A}. The only observational method for determining these unknowns in models is asteroseismology, which uses stellar oscillations to probe the sub-surface physics of stars \citep{2010aste.book.....A}. Asteroseismology allows us to test internal stellar physics across the entire HR diagram, since different classes of pulsating stars exist at various stages of stellar evolution.

Stellar pulsations in intermediate-mass stars are principally of two main types, characterised by their respective restoring mechanisms: pressure (p) modes, where pressure forces restore perturbations, and gravity (g) modes, where buoyancy does \citep{2010aste.book.....A}. However, for moderately and rapidly rotating stars, the Coriolis force is also important as a restoring force, leading to gravito-inertial modes, or Rossby (r) modes \citep{2018MNRAS.474.2774S}. Typically p~modes are most sensitive to the stellar envelope whereas g~modes probe the near-core region, with pulsations described by spherical harmonics such that each p or g~mode has a radial order, $n$, angular degree, $\ell$, and azimuthal order,~$m$. 

 In the last two decades, space telescopes such as WIRE \citep{2005ApJ...619.1072B}, MOST \citep{2007CoAst.150..333M}, CoRoT \citep{2009A&A...506..411A},  $Kepler$ \citep{2010ApJ...713L..79K} and BRITE \citep{2014PASP..126..573W} have provided the necessary continuous, long-term and high-precision time series photometry for the identification of large numbers of individual pulsation modes in single stars. Hence asteroseismology has been successfully applied to hundreds of solar-type and tens of thousands of red giant stars \citep{2013ARA&A..51..353C}, as well as dozens of pulsating white dwarfs and intermediate-mass main-sequence stars (e.g.,  \citealt{2017ApJ...847L...7A, 2018arXiv180907779A, 2019MNRAS.482.1757L}). 

Among the intermediate-mass main-sequence A and F stars, there are different pulsator classes, such as the $\delta$~Sct, $\gamma$~Dor and rapidly oscillating Ap (roAp) stars. These classes are located at the intersection of the main-sequence and the classical instability strip in the HR diagram, and exhibit p~modes and/or g~modes pulsations \citep{2000ASPC..210....3B, 2010aste.book.....A}. The high-frequency ($\nu \gtrsim 4$~d$^{-1}$), low-radial-order p~modes in $\delta$~Sct stars\footnote{The exact boundary between p and g~modes depends on the stellar parameters, and especially the rotation of the star.} are excited by the opacity mechanism operating in the He~{\sc ii} ionisation zone, as well as by turbulent pressure, which has also been shown to be responsible for excitation of moderate radial order p~modes within the classical instability strip \citep{2014ApJ...796..118A, 2016MNRAS.457.3163X}. On the other hand, low-frequency ($\nu \lesssim 4$~d$^{-1}$) g~modes in $\gamma$~Dor stars are excited by a convective flux modulation mechanism \citep{2000ApJ...542L..57G, 2010ApJ...713L.192G, 2005A&A...435..927D}. The instability regions of the $\delta$~Sct and $\gamma$~Dor stars partially overlap in the HR diagram \citep{2005A&A...435..927D}, and include stars on the pre-main-sequence, the main-sequence and post-main-sequence with typically masses in the range of $1.2 \leq M \leq 2.5$~M$_{\odot}$\footnote{In the case of low metallicity the mass can be lower than 1.2~M$_{\odot}$, see, e,g, Sec. \ref{subsection: SX Phe}}.

Space telescopes have provided continuous, long-term light curves of tens of thousands of pulsating stars. In particular, {\it Kepler} heralded a large and important  step forward for asteroseismology of $\delta$~Sct and $\gamma$~Dor stars, since its 4-yr light curves provided unprecedented frequency resolution and duty cycle, both of which are necessary for accurate mode identification. An unexpected discovery was that many pulsating A and F stars are hybrid pulsators exhibiting both p and g~modes \citep{2010ApJ...713L.192G, 2011A&A...534A.125U, 2011MNRAS.417..591B}, although pure pulsators of both types do exist (see e.g. \citealt{2015ApJS..218...27V, 2017ampm.book.....B, Barcelo2018a, 2019MNRAS.482.1757L}). The high incidence of hybrid stars remains unexplained, particularly if a single excitation mechanism should responsible for both the p and g~modes in hybrid stars \citep{2015MNRAS.452.3073B, 2016MNRAS.457.3163X}. This incomplete understanding of pulsational excitation in A and F stars is further complicated by the observation that a significant fraction of hybrids, $\gamma$ Dor and $\delta$~Sct stars are hotter than their theoretically predicted instability regions, and that constant stars also exist within the $\delta$~Sct instability domain \citep{2018MNRAS.476.3169B, 2019MNRAS.tmp..585M}. 

A major advance in asteroseismology has been the characterisation of internal rotation profiles of dozens of A and F stars using their g~mode pulsations \citep{2016A&A...593A.120V, 2017ApJ...847L...7A, 2019MNRAS.482.1757L}. From these studies, it has been inferred that a significant fraction of intermediate-mass main-sequence stars have, at first sight, near-uniform radial rotation profiles (e.g., \citealt{2014MNRAS.444..102K, 2015MNRAS.447.3264S, 2016A&A...593A.120V, 2018A&A...618A..24V, 2017MNRAS.465.2294O, 2018A&A...618A..47C, 2019MNRAS.482.1757L, 2019MNRAS.487..782L}), however in order to draw a final conclusion detailed studies in the manner of, e.g., \citet{2012A&A...547A..75O} and \citet{2019ApJ...871..135H} are required. Furthermore, the large number of excited g~modes in these stars allow the physics of convective core overshooting and chemical mixing to be probed, as well as accurate determinations of stellar masses and ages using forward seismic modelling \citep{2016A&A...592A.116S, 2018ApJS..237...15A, 2019MNRAS.tmp..495M}. These studies clearly demonstrate the asteroseismic potential of high quality space-based observations of intermediate-mass stars.

Although the pulsation mode density is typically high in $\delta$~Sct stars hampering mode identification, some stars have been shown to show regularities in their spectra consistent with the so-called large frequency separation which may aid in mode identification and modelling of p modes (see, e.g.,  \citealt{GH2015, GH2017a, Paparo2016b, Paparo2016a, 2017A&A...601A..57B}). The large frequency separation is the separation between consecutive radial order modes of the same degree and is a well-known quantity in the context of Sun and Sun-like stars \citep{2010aste.book.....A}. When using the same quantity for $\delta$ Sct stars, special care must be taken as these stars typically pulsate in lower radial order modes, where the large frequency separation deviates from the one measured at higher radial orders. Nevertheless, it was shown that these patterns are also proportional to the mean density of the stars \citep{2014A&A...563A...7S} which was observationally confirmed  \citep{GH2015, GH2017a}. 

More recently, the successful launch of TESS and the delivery of the first TESS mission data \citep{2014SPIE.9143E..20R, 2015JATIS...1a4003R}, which are in the southern ecliptic hemisphere, provide new insight into pulsation and rotation in intermediate-mass stars (e.g., \citealt{2019MNRAS.487.3523C, 2019arXiv190508835S}). The primary mission of TESS follows the groundwork of the {\it Kepler} mission, and aims to detect planetary transits in the brightest stars across the sky. TESS data cover almost the entire sky ($|b| > 6$~deg) at a cadence of 30~min and are publicly available to the community in the form of full frame images. However, a sub-sample of stars were selected to be observed at a 2-min cadence in each TESS observing sector and light curves of these stars are being delivered to the community a few months after observations have been obtained. Therefore, TESS provides a unique homogenous data set to test and constrain much of the currently unknown physics inside main-sequence A and F stars.

In this paper, we demonstrate the asteroseismic potential of TESS for the study of intermediate-mass stars, specifically for $\delta$~Sct and $\gamma$~Dor stars. We show a selection of different types of pulsator classes, including some of the class prototypes, and the significant advantages and inference that can be achieved using TESS mission data. In addition, we show how theoretical models regarding the excitation of pulsations in A and F stars can be tested. In the following subsections, we briefly outline the different pulsator classes within the A and F stars. 


\section{The pulsation zoo of A and F stars}

It is estimated that approximately half of the stars within the classical instability strip are pulsating at the $\umu$mag photometric precision provided by space telescopes \citep{2011MNRAS.417..591B, 2019MNRAS.tmp..585M}. It remains a mystery why the other half do not show pulsations, especially at that precision. Among the intermediate-mass stars there is a veritable zoo of different pulsator classes. The heterogeneity of variability in these stars allows us to test different physical phenomena, such as rotation, diffusion, and convection. Obviously, this is because the differences in observed pulsation characteristics are caused by differences in stellar structure, evolutionary stage and/or metallicity. Here, we briefly discuss the zoo of pulsating A and F stars.

\subsection{Ap and Am stars}

Stars within the $\delta$~Sct instability strip are predicted to pulsate in p~modes excited by the opacity mechanism operating in the He~{\sc II} ionisation zone \citep{1963ApJ...138..487C, 2010aste.book.....A} as well as by turbulent pressure (e.g., \citealt{2014ApJ...796..118A}). However, whether a star is unstable to pulsation strongly depends on mixing processes, such as rotation or the presence of a strong magnetic field, which can suppress the excitation of heat-driven (opacity mechanism) pulsation modes.

Slow rotation in intermediate-mass stars is typically caused by one of two mechanisms: (i) magnetic braking by a large-scale magnetic field; or (ii) tidal braking due to tidal forces caused by a close companion. Approximately 10~per~cent of A-type stars are observed to have a strong, large-scale magnetic field, which is likely of fossil origin and causes magnetic braking during the pre-main-sequence stage \citep{2000A&A...353..227S}. The slow rotation and strong magnetic field  allow atomic diffusion -- radiative levitation and gravitational settling of different ions -- to give rise to surface spectral peculiarities \citep{2000ApJ...544..933A}, which define the   sub-classes of the magnetic Ap stars, and suppress the excitation of low radial-order p~modes by the opacity mechanism \citep{2005MNRAS.360.1022S}.  However, a small fraction of these magnetic Ap stars pulsate in high-radial order modes; these are the rapidly oscillating Ap (roAp) stars \citep{1982MNRAS.200..807K}. The excitation of low-degree, high-radial order magneto-acoustic modes in roAp stars provides the opportunity to study the effects of pulsation and rotation in the presence of strong magnetic fields \citep{2000MNRAS.319.1020C, 2001MNRAS.325..373C, 2002AA...391..235B, 2006MNRAS.365..153C, 2004MNRAS.350..485S, 2005MNRAS.360.1022S}. We refer the reader to \citet{2019MNRAS.487.3523C} for the first results on roAp stars based on TESS data.

The second main sub-class of slowly rotating A and F stars are the non-magnetic Am stars \citep{1970PASP...82..781C, 1970ApJ...162..597B, 1978ApJ...221..869K, 2010A&A...523A..40A}. The presence of a close companion causes tidal braking such that atomic diffusion and gravitational settling separate chemical elements in the stellar envelope, deplete helium from the excitation region and inhibit the efficiency of pulsational excitation \citep{1970ApJ...160..641M, 1973A&A....23..221B,  2000A&A...360..603T}. However, recent studies focussed on the incidence of pulsations in Am stars have found a non-negligible fraction pulsating in p~modes \citep{2011A&A...535A...3S, 2017MNRAS.465.2662S} with a significant amount of excitation connected to the turbulent pressure in the hydrogen rather than the He II ionisation layer \citep{2014ApJ...796..118A,  2017MNRAS.465.2662S}. We present our analysis of TESS photometry of pulsating Am stars in Section~\ref{subsection: Am stars}.

\subsection{Metal-poor stars}

Among the A-type pulsators are two classes of metal-poor stars: the $\lambda$~Boo stars and SX~Phe stars. The $\lambda$~Boo stars are a spectroscopically defined class in which refractory chemical elements -- those with high condensation temperatures in pre-stellar disks -- are underabundant by 0.5 to 2~dex, but volatile elements such as C and O show solar abundance \citep{2002A&A...381..959H, 2002A&A...396..641A}. The peculiarity of these stars is believed to originate from selective accretion of gas with little dust, but the specifics remain uncertain \citep{1990ApJ...363..234V, 1994MNRAS.269..209K, 2015A&A...582L..10K}, and it is not clear whether the peculiarity is restricted to surface layers only.  Approximately 2~per~cent of A stars are $\lambda$~Boo stars \citep{1998AJ....116.2530G}, for which a catalogue has been compiled by \citet{2015PASA...32...36M}. At present, there is no  known difference between the pulsational characteristics of $\lambda$~Boo $\delta$~Sct stars and `normal' $\delta$~Sct stars. 

SX~Phe stars, on the other hand, are not a spectroscopically defined class but comprise Population~{\sc II} $\delta$~Sct stars, though not all of them are metal poor \citep{2017MNRAS.466.1290N}. Many SX~Phe stars are identified as blue stragglers in clusters, while those in the field are often identified kinematically. The dominant pulsations in SX~Phe stars are typically high-amplitude radial modes \citep{1990ASPC...11...64N, 2002ApJ...576..963T, 2013MNRAS.432.2284M} but can also have low amplitudes \citep{2014MNRAS.444..102K}. We analyse TESS data for SX~Phe itself in section~\ref{subsection: SX Phe}.

\subsection{High-amplitude $\delta$~Sct stars}

High-amplitude $\delta$~Scuti  stars (HADS) are usually recognized by their large amplitude, non-sinusoidal light curves and small rotational velocities, but their most essential property is the dominant presence of the fundamental and/or first overtone radial mode(s) \citep{2000ASPC..210..373M, 2000ASPC..210....3B}. In the HR diagram, ground-based observations indicated that HADS  occupy a narrow strip in the centre of $\delta$~Sct instability region, perhaps because the efficiency of mode excitation is maximised  in this region \citep{1996A&A...312..463P}. However, space photometry revealed that some HADS can be found throughout the entire instability strip \citep{2016MNRAS.459.1097B}.

Some studies have investigated whether the properties of HADS could be explained by them being in a post-main-sequence stage of stellar evolution (e.g. \citealt{1996A&A...312..463P, 2017ampm.book.....B}), since they obey a period-luminosity relation which allows an independent distance determination for some stellar systems (e.g., \citealt{1994AJ....108..222N, 2007AJ....133.2752M, 2012MNRAS.419..342C,2015AcA....65...81K, 2019MNRAS.486.4348Z}), such as globular clusters in which SX~Phe stars are abundant. However, there remains no clear consensus on a physical difference between HADS and their lower amplitude $\delta$~Sct counterparts. Important remaining questions concerning HADS include: the excitation (e.g., \citealt{2011A&A...528A.147P}); the relation between the single and double-mode HADS and the stellar parameters; and whether the pulsational properties of SX~Phe stars differ from those of Population~{\sc I} HADS. The all-sky TESS sample of HADS will be particularly useful for addressing these questions and will extend the study of a handful of {\it Kepler} stars, particularly the seismic modelling of HADS such as those by \citet{2011MNRAS.414.1721B} and \citet{2012MNRAS.419.3028B}. 

The TESS sample of HADS and SX~Phe stars in Sectors 1 and 2 includes at least 19 stars proposed for 2-min cadence observations based on a review of the ASAS-3 \citep{1997AcA....47..467P, 2002AcA....52..397P} database (Pigulski \& Kotysz, in preparation). The sample includes several known HADS, e.g.,  ZZ~Mic, RS~Gru, BS~Aqr and SX~Phe itself (see Sect.~\ref{subsection: SX Phe}). An example of a known HADS star observed by TESS, HD~224852 (TIC~355687188 ), is presented in detail in Sect.~\ref{subsection: HADS}.

\subsection{Pre-main-sequence stars}

Intermediate-mass pre-main-sequence stars within the instability regions can become unstable to p and g-modes during their evolution from the birthline to the zero-age main-sequence (ZAMS). Pre-main-sequence stars differ from their main-sequence and post main-sequence analogues of similar mass because of their interior structure \citep{1998ApJ...507L.141M}. Since pulsation modes carry information about the inner parts of stars and show a different pattern for stars on the pre- or the post-main-sequence phases \citep{2001A&A...372..233S}, it is possible to use asteroseismology to constrain the evolutionary stage of a star \citep{2007ApJ...671..581G}.  This is important because the evolutionary tracks for stars before and after the ZAMS intersect in this part of the HR diagram making it impossible to constrain a star's evolutionary stage from only its effective temperature and luminosity. 

In general, without asteroseismology, the pre-main-sequence stage of a star in this region of the HR diagram can only be assessed using certain observational features as indicators. Such indicators for young stars are emission lines in the spectra, infrared or ultraviolet excess, X-ray fluxes, and membership of a young open cluster or a star-forming region, i.e., younger than approximately 10 million years \citep{2004ApJ...614L..77S}. However, some of these features can also be misleading, since the comparatively old, low-mass asymptotic giant branch (AGB) and post-AGB stars show similar observational properties and populate the same region in the HR diagram as young stellar objects \citep{2014MNRAS.439.2211K}. Therefore, special care has to be taken when investigating the potential early evolutionary stage of a given star.

Currently, about 80 pre-main-sequence stars are known to be p- and g-mode pulsators of three different types: Slowly Pulsating B (SPB) stars \citep{2012MNRAS.420..291G}, $\delta$~Sct stars (e.g., \citealt{2014Sci...345..550Z}) and $\gamma$~Dor  stars \citep{2013A&A...550A.121Z}. Based on their basic properties, such as spectral types and effective temperature, TIC~150394126 and TIC~382551468 are two candidates for $\delta$~Sct and $\gamma$~Dor pulsations. The former has an infrared excesse \citep{2004ApJ...614L..77S} and the latter is potentially an Herbig Ae/Be type star, identifying them as likely being in their pre-main-sequence evolutionary stages. We analyse these stars in Section~\ref{subsection: pre-ms}.

\section{Frequency analyses of TESS data}
\label{section: data}

The TESS data analysed in this work are the 2-min  `Simple Aperture Photometry' (SAP) light curves provided by the TESS Science Team, which are publicly available from the Mikulski Archive for Space Telescopes (MAST)\footnote{\url{https://archive.stsci.edu}}. As part of the framework of the asteroseismic working group on pulsating stars of spectral type A and F within the TESS Asteroseismic Science Consortium (TASC)\footnote{\url{https://tasoc.dk}}, 117 known pulsating stars were allocated observing slots at a 2-min cadence by the TESS Mission in Sectors 1 and 2 of its  1-yr observing campaign in the southern ecliptic hemisphere. There were more $\delta$~Sct stars observed in the first two sectors, but for this study we decided to concentrate on those that were allocated to the TASC working group on A and F stars (TASC WG 4). Because g~modes have long periods, we did not specifically include known $\gamma$~Dor stars in our 2-min target list, since these stars will be observed with a cadence of 30 min, which is sufficient to determine their frequencies. We discuss one star that was $not$ observed during Sectors 1 and 2 later on. This target is the eponymous star $\gamma$~Doradus and is described in Section~\ref{subsection: gdor}.

\begin{figure}
\begin{center}
\includegraphics[width=0.48\textwidth]{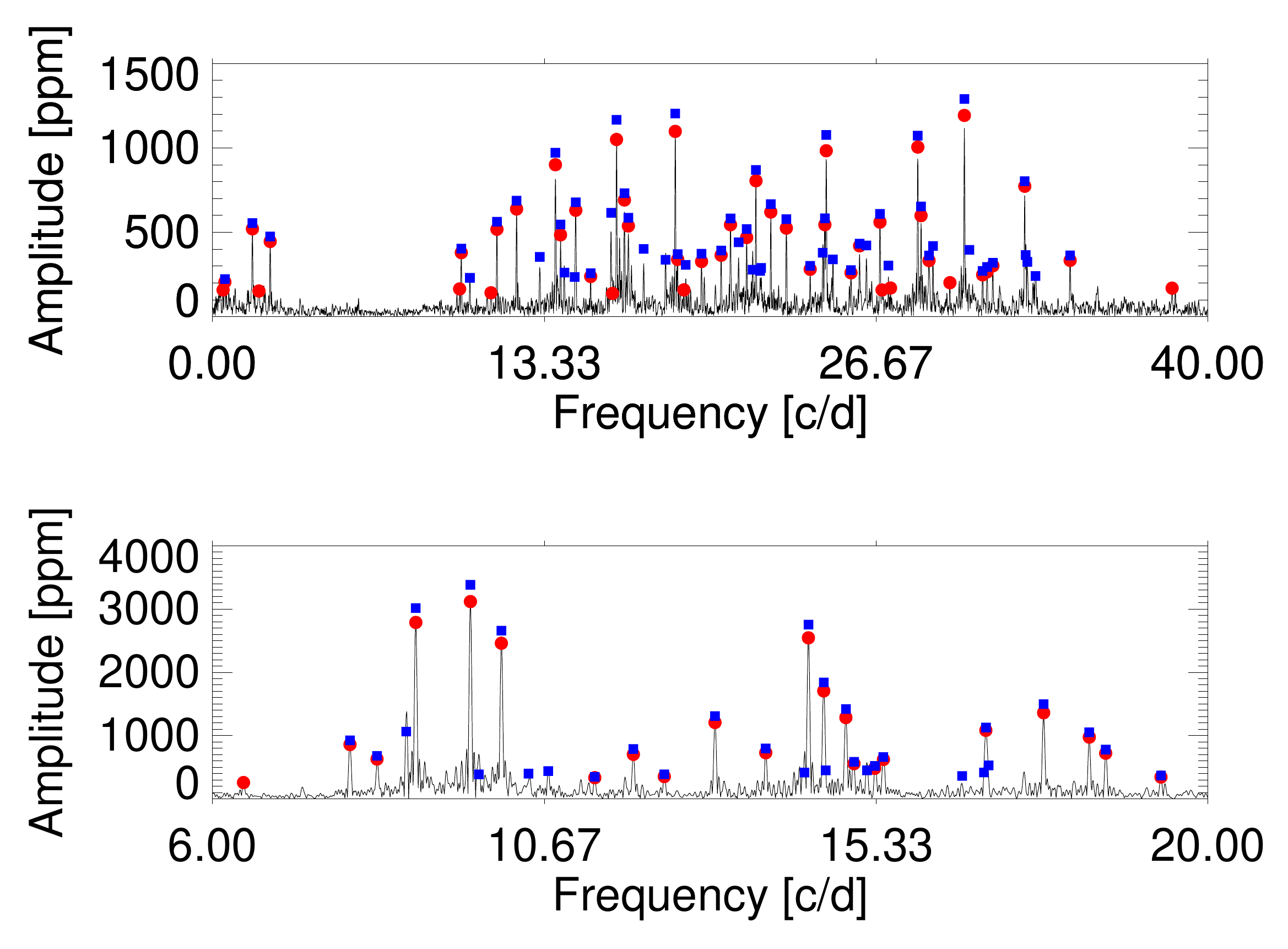}
\caption{Comparison between the frequency extraction pipelines from Leuven (blue squares) and Aarhus (red circles) showing good agreement between the two different pipelines. 
The amplitude values are offset for clarity. Upper panel: HD~210111 (TIC~229059574). Lower Panel: TX Ret (TIC~38587180). }
\label{fig:pipelines}
\end{center}
\end{figure}

We analysed the TESS light curves of our sample in different independent groups, using various frequency extraction software routines. Commonly used programs to perform frequency analysis of light curves are {\sc PERIOD04} \citep{2004IAUS..224..786L} and SigSpec \citep{2007A&A...467.1353R}, while the KU~Leuven and Aarhus University ECHO (Extraction of CoHerent Oscillations) pipelines were used also. We discuss these pipelines in Sections~\ref{subsection:Leuvenpipeline} and \ref{subsection: echo}, and conclude that they yield compatible results for peaks in the amplitude spectrum with signal-to-noise (S/N) $\geq 4$. A comparison of the significant frequencies identified in two stars by the Leuven and Aarhus pipelines is shown in Fig.~\ref{fig:pipelines}. We plan to compile a catalogue of pulsating A and F stars observed by TESS at the 2-min cadence in the entire southern ecliptic hemisphere, which will include all significant pulsation frequencies and identification of their combination frequencies.

\subsection{The KU Leuven pipeline}
\label{subsection:Leuvenpipeline}

The KU Leuven iterative pre-whitening pipeline was used to extract significant periodic variability in time series data. It uses Lomb-Scargle (\citealt{1982ApJ...263..835S}) periodogram and multi-frequency non-linear least-squares optimisation to extract significant frequencies and their associated amplitudes and phases. Uncertainties on frequencies, amplitudes and phases are calculated according to \citet{1999DSSN...13...28M}, and corrected for the correlated nature of oversampled time series data following \citet{2003ASPC..292..383S}. The pipeline was developed by \citet{2009A&A...506..111D}, and improved by \citet{2012A&A...542A..55P} and \citet{2015A&A...574A..17V}; detailed discussions and applications to pulsating stars observed by CoRoT and {\it Kepler} were provided by these studies.

With this pipeline, peaks in the amplitude spectrum can be extracted in order of decreasing amplitude or decreasing S/N, as specified by the user. Normally, a lower limit of S/N $\geq 4$ \citep{1993A&A...271..482B} is chosen, where the noise is calculated as the average amplitude in a symmetric 1-d$^{-1}$ window in the residual periodogram after pre-whitening. From extensive testing, it has been found that using a window around the extracted peak for calculating a local noise estimate (and hence the S/N) is more pragmatic and reliable for iterative pre-whitening than using a $p$-value \citep{2012A&A...542A..55P}. Otherwise, hundreds or even thousands of individual frequencies can be found to be statistically significant, yet with an increasing probability of an artificial frequency having been injected during the prewhitening process into a light curve when the number of iterations becomes large. Typically, this limit is a few hundred frequency extraction iterations for a 4-yr {\it Kepler} light curve of a $\delta$~Sct or $\gamma$~Dor star, but this depends on the quality and lengths of the data, and assumes that peaks are coherent and so varies from star to star. 
 
 \subsection{ECHO}
\label{subsection: echo}

The ECHO program (Extraction of CoHerent Oscillations) was developed at  Aarhus University. It extracts coherent signals from photometric light curves using an optimized, iterative pre-whitening technique. Coherent signals are iteratively identified using a Lomb-Scargle periodogram that is calculated using the \citet{1989ApJ...338..277P} algorithm, which allows a fast computation for unevenly sampled data. The frequency, amplitude and phase of an oscillation is computed using the algorithm described by \citet{1995A&A...301..123F}. The statistical uncertainties on the amplitudes, frequencies and phases are calculated according to \citet{1999DSSN...13...28M}. The statistical significances of the extracted signals are estimated from the periodogram using Scargle's significance criteron \citep{1982ApJ...263..835S}. The extracted signals are continuously optimized at each iteration, taking into account the influence of one signal on another. This is done by reinserting previously extracted signals back into the pre-whitened light curve and redetermining the maximum amplitude of the peak. This may be slightly different from the previously found value, due to the influence of neighbouring peaks \citep{handbergphd}.

Using simulated light curves, \citet{2014MNRAS.439.3453B} showed that iterative pre-whitening can lead to a very large number of false detections. This is because imperfect subtraction of signals will inject new, statistically significant signals into the light curve. For this reason, the reliability of each extracted oscillation is determined by comparing the amplitude of an extracted signal $A_{\text{f}}^{\text{ex}}$ to the corresponding amplitude $A_{\text{f}}^{\text{og}}$ in the unaltered, original light curve. If the amplitudes do not agree within a certain limit $\alpha$, such that

\begin{eqnarray}
\alpha < \frac{ A_{\text{f}}^{\text{ex}}}{A_{\text{f}}^{\text{og}}} < \frac{1}{\alpha},
\end{eqnarray}

\noindent the extracted signal is rejected. From simulated light curves, we found that $\alpha = 0.75$ is a suitable value, keeping the number of false detections to a minimum while still extracting most of the real signals present in the light curve. These simulations were based on 4~years of $Kepler$ data. For shorter observations, our tests showed that a more conservative factor of  $\alpha = 0.8$ is required. This method is similar to that used by \citet{2015A&A...574A..17V}, although they started with $\alpha = 0.5$ and allowed it to vary during pre-whitening. In our use of the Leuven and ECHO pipelines, we enforced a maximum of number of 100 frequencies that could be extracted, although it was not typically needed since peaks became insignificant before this criterion was met. This is because the pre-whitening of so many peaks can introduce additional signals, and make the results unreliable. 

In addition, we must be aware that observations of `only' 27-56 days may not sufficient to resolve closely-spaced frequencies clearly known to be present in many $\delta$ Sct and $\gamma$ Dor stars. \citet{2017A&A...603A..13K} have gone beyond the formal frequency resolution of 1/$\Delta$T in their frequency analysis\footnote{where $\Delta$T is the lengths of the data set.}, using a probabilistic approach. We, however, prefer to use the traditional Rayleigh criterion because extracting unresolved frequencies, can lead to over-interpreting the data and wrong input for asteroseismic modelling. We recommend using formally unresolved peaks with extreme caution, and only based on careful testing and the use of additional independent information (e.g., for roAp stars, where an assumption of the oblique pulsator model requires exact splitting in the rotational multiplets). 
The statistical significance of extracted peaks is an important topic that goes beyond the scope of this paper as we are only showing a first view into TESS $\delta$ Sct and $\gamma$ Dor stars. At this stage we are expanding ECHO to also include the 'traditional' SNR test.

\section{Target stars}
\label{section: targets}

In this section, we present the details of all TASC WG4 A and F stars that are not bona fide or suspected Ap or roAp stars observed in a 2-min cadence in Sectors 1 and 2. The variability type for each star was identified, based on TESS data, by visual inspection of the light curves and their Fourier spectra. Spectroscopic and photometric parameters from the literature for all the stars, where available, are assembled in Table~\ref{table: star_parameters1} in the appendices, including claimed detections of binarity. Note that here we do not evaluate the robustness of these literature values and leave it to the reader to investigate these in detail.

In Table~\ref{tab:stellar_parameters_new}, we list the newly derived effective temperatures using the spectral energy distribution method (SED) (see Section \ref{SED}), new spectroscopically derived parameters (see Section \ref{spec}), the dereddened luminosities from Gaia and Hipparcos  data (Section \ref{Gaia}), and the TESS stellar variability type. The latter is one of the following categories: (1) $\delta$~Sct star: $\nu \ge 4$~d$^{-1}$; (2) $\gamma$~Dor star: $\nu \le 4$~d$^{-1}$; (3) hybrid star: when frequencies were detected in the $\gamma$~Dor and the $\delta$~Sct  frequency ranges. Note that in many cases the peaks in the $\gamma$~Dor range were not resolved and more data are required to fully characterise the frequencies. (4) binarity: eclipses and/or ellipsoidal modulation were detected; (5) rotational variable (rot): variability is consistent with rotational modulation of spotted stellar surfaces. (6) no detection: no significant variability was detected. We also note that $\nu$ Pic (TIC~260416268) was observed during sectors 1 and 2 but the amplitude differs significantly between the sectors suggesting a problem in the data extraction.



\onecolumn

\begin{landscape}

\setlength\LTcapwidth{\textwidth} 
\setlength\LTleft{0pt}            
\setlength\LTright{0pt}           

\begin{longtable}{|r|c|l|c|c|c|c|c|c|c|c|c|c|c|}


\caption{Stellar parameters for the TESS $\delta$~Sct and $\gamma$~Dor pulsators observed at 2-min cadence during the first two pointings (Sectors 1 and 2) in the ecliptic Southern hemisphere. The columns indicate: the TESS Input Catalogue (TIC) number, the TESS magnitude, an alternative identifier of the star, spectroscopic (Spec.) and photometric (Phot) effective temperatures and log g values and their associated uncertainties from literature respectively. The reference is indicated next to each value. In addition, we list the projected rotational velocity and binarity status as available in literature. The full table is available in the online version in the appendix. }
\label{table: star_parameters}

\\ \hline

\multicolumn{1}{|c|}{TIC} &  TESS  & \multicolumn{1}{c|}{alternative}    &  $\teff [K]$  &  $\sigma_{\teff}$  &   $\teff [K]$  &  $\sigma_{\teff}$  &   $\logg$  &  $\sigma_{\logg}$  &   $\logg$  &  $\sigma_{\logg}$  &  $v \sin i$  &  $\sigma_{\text{vsini}}$ &   binarity \\
 & mag &\multicolumn{1}{c|}{identifier}  & Spec. & Spec. & Phot. & Phot. & Spec. & Spec. & Phot. & Phot. &[km~s$^{-1}$]&[km~s$^{-1}$]& \\\hline

\endhead
\hline
\endfoot

9632550 & 9.06 & BS Aqr & 7245\footnotemark[1] & 75\footnotemark[1] & - & - & 3.82\footnotemark[1] & 0.12\footnotemark[1] & - & - & 23\footnotemark[1] & - &  - \\
12470841 & 8.44 & HD 923 & - & - & 7832 \footnotemark[2] & - & - & - & - & - & - & - & - \\
12784216 & 8.92 & HD 213125 & - & - & 6519\footnotemark[3] & 178\footnotemark[3] & - & - & 4.03\footnotemark[3] & 0.26\footnotemark[3] & - & - & -\\
12974182 & 6.78 & HD 218003 & 7471\footnotemark[4] & 95\footnotemark[4] & 7584\footnotemark[5] & 208\footnotemark[5] & 3.86\footnotemark[4] & 0.18\footnotemark[4] & 4.05\footnotemark[5] & - & - & - & - \\
... & ... & ... & ... & ...& ... & ... & ... &... & ... &... & ...& ... & ... \\

\footnotetext[1]{\cite{Kunder2017}} 
\footnotetext[2]{\cite{McDonald2012}}
\footnotetext[3]{\cite{huber2016}} 
\footnotetext[4]{\cite{Stevens2017}} 
\footnotetext[5]{\cite{McDonald2017}} 

\end{longtable}




\setlength\LTcapwidth{\textwidth} 
\setlength\LTleft{0pt}            
\setlength\LTright{0pt}           

\begin{longtable}{|r|l|c|c|c|c|c|c|l|c|} 

\caption{Stellar parameters and variability type of the 2-min cadence TESS $\delta$ Scuti and $\gamma$~Dor pulsators determined in this work. In this table we list the TESS Input Catalogue (TIC) name, an alternative name for each star, the effective temperatures derived using the SED method (Section \ref{SED}), the luminosities derived based on the Gaia and when available Hipparcos parallaxes and the respective uncertainties (Section \ref{Gaia}). In column 9 we describe the variability type determined from the TESS data. The question mark means that the variability type is uncertain (e.g. due to unresolved peaks).  HAGD denotes a High Amplitude $\gamma$ Dor star. The last column describes the chemical peculiarity as described in \citet{Renson2009}. The full table is available in the online version in the appendix.}   

\label{tab:stellar_parameters_new}

\\ \hline

	\multicolumn{1}{|c|}{TIC} 	&	\multicolumn{1}{c|}{alternative}  &	$T_{\rm eff}$ [K] &	 $\sigma T_{\rm eff}$ & $\log (\rm L/{\rm L}_{\odot}$) & $\sigma \log (\rm L/{\rm L}_{\odot}$) & $\log (\rm L/{\rm L}_{\odot}$) & $\sigma \log (\rm L/{\rm L}_{\odot}$) 	&	variability 	& chem.	\\
	
			&\multicolumn{1}{c|}{identifier} &	SED	 	& 		&	Gaia		&			&	HIPP.	&		&	type	&	pec.	\\ \hline			
\endhead
\hline
\endfoot			

9632550 &  BS Aqr  & 6760 & 140 & 1.62 & 0.03 & 1.96 & 0.94 &  HADS &\\
12470841 &  HD 923   & 8090 & 230 & 1.64 & 0.02 & 2.01 & 0.43 &  one large peak only &\\
12784216 &  HD 213125  & 6390 & 120 & 1.14 & 0.01 &  &  &  $\delta$~Sct/hybrid?&\\
12974182 &  HD 218003  & 7670 & 170 & 1.103 & 0.008 & 1.08 & 0.04 &  rot?/binary?& \\
... & ... & ... & ... & ...& ... & ... & ... &... & ...  \\

\hline

\end{longtable}

\end{landscape}
\clearpage
\twocolumn

\subsection{Effective temperatures derived from Spectral Energy Distributions}\label{SED}
Effective temperature can be determined from the stellar spectral energy
distribution (SED). For our target stars these were constructed from literature
photometry, using, $B$ and $V$ magnitudes from either Tycho \citep{1997A&A...323L..57H} or APASS9 \citep{2015AAS...22533616H}, USNO-B1 $R$ magnitudes \citep{2003AJ....125..984M}, $J$, $H$ and $K$ from 2MASS \citep{2006AJ....131.1163S}, supplemented with CMC14 $r'$ \citep{2002A&A...395..347E} and APASS9 $g'$, $r'$, $i'$ photometry.

Stellar energy distributions can be significantly affected by interstellar reddening, which was calculated as described in Section \ref{Gaia} The SEDs were de-reddened using the analytical extinction fit of \citep{1983MNRAS.203..301H}.
The stellar $T_{\rm eff}$ values were determined by fitting solar-composition
\cite{1993KurCD..13.....K} model fluxes to the de-reddening SEDs. The model fluxes were convolved with photometric filter response functions. A weighted Levenberg-Marquardt non-linear least-squares fitting procedure was used to find the solution that minimized the difference between the observed and model fluxes. Since $\log g$ is poorly constrained by our SEDs, we fixed $\log g =4.0$ for all the fits. The uncertainties in $T_{\rm eff}$ includes the formal least-squares error and adopted uncertainties in $E(B-V)$ of $\pm$0.02 and $[M/H]$ of $\pm$0.5 added in quadrature.

\subsection{New spectroscopic determinations of $T_{\rm eff}$ based on archival data}\label{spec}

Spectroscopic data were taken from the ESO archive\footnote{\url{http://archive.eso.org/cms.html}}. 
We used medium and high signal-to-noise ratio ($\geq$\,65) spectra from FEROS (Fiber-fed Extended Range Optical Spectrograph; R$\sim$48000), HARPS (High Accuracy Radial velocity Planet Searcher, R$\sim$115000), and UVES (Ultraviolet and Visual Echelle Spectrograph, R$\sim$65000). Observational details are given in Table\,\ref{infodata}. 

We applied two methods to derive spectroscopic parameters for the 14 stars that had available archival data. 
For the first method we used, the hydrostatic, plane-parallel line-blanketed, local thermodynamic equilibrium ATLAS9 model atmospheres \citep{1993KurCD..13.....K}. The synthetic spectra were generated by using the SYNTHE code \citep{1981SAOSR.391.....K}. The projected rotational velocities, $v \sin i$, were calculated by fitting the profiles of non-blended metal lines \citep{2008oasp.book.....G}. The H$\beta$ lines were used to derive $T_{\rm eff}$ by considering the minimum difference between the synthetic and observed spectra. In this analysis, a solar metallicity and the obtained $v \sin i$ values were fixed. In addition, we assumed $\log g = 4.0$ (dex). This was done because the hydrogen lines are not sensitive to $\log g$ for stars with  $T_{\rm eff} \le 8000$~K \citep{2002A&A...392..619H}. Our results are listed in Table\,\ref{hydrogenteff}, with 1$\sigma$ uncertainties.

In addition, the Grid Search in Stellar Parameters (GSSP; \citet{2015A&A...581A.129T}) software package was used as an alternative method for determining the atmospheric parameters of our stars. GSSP is a multi-purpose spectrum analysis tool that is designed for precision spectrum analysis of single and binary stars. As the name suggests, the method is based on a grid-search in 6-D parameter space (metallicity [M/H], $T_{\rm eff}$, $\log g$, micro- (v$_{\rm micro}$) and macro-turbulent (v$_{\rm macro}$) velocities, and $v \sin i$) utilising atmospheric models and synthetic spectra. Each synthetic spectrum in a grid is compared to the normalised observed spectrum of a star with the chi-square ($\chi^2$) merit function being used to judge the goodness of fit. Parameter errors are computed from $\chi^2$ statistics by projecting all grid points onto each individual parameter so that correlations are taken into account. GSSP employs the SynthV radiative transfer code \citep{1996ASPC..108..198T} for calculation of synthetic spectra and {\sc LLmodels} software package \citep{2004A&A...428..993S} for computation of atmosphere models. Both codes operate under the assumption of local thermodynamical equilibrium and allow for extra functionality such as vertical stratification of elemental abundances in the atmosphere of a star. 
The $\log g$ value is not well constrained in the temperature range in question, which is the reason for fixing  it to 4.0 dex (as was also done in the previous method). This is because the hydrogen Balmer lines are largely insensitive to changes in the surface gravity in this particular temperature regime, while significant broadening of metal lines and correlations of $\log g$ with [M/H] and v$_{\rm micro}$ results in typical uncertainties as large as 0.3--0.5 dex.

The photometric and spectroscopic effective temperatures determined here typically agree within the uncertainties. There are few exceptions all of which are either (spectroscopic) binaries or HADS. For the latter the observed  T$_{\rm eff}$ depends on the pulsation cycle (see Sec. \ref{subsection: SX Phe} for a discussion). 

\begin{table}
\centering
  \caption{Spectroscopic data.} 
  \begin{tabular}{rlccrc}

\\ \hline

\multicolumn{1}{c}{TIC}  & \multicolumn{1}{c}{alternative}     & Instrument  & Observation   & S/N & \multicolumn{1}{c}{Nr. of} \\
         &  \multicolumn{1}{c}{name}    &             &         &   & \multicolumn{1}{c}{spectra}  \\
 \hline
 9632550     &  BS Aqr 	   & FEROS      &  2007 July       & 200 & 4\\
 32197339    &  HD 210139  & HARPS      &  2005 July       & 105 & 11\\
38587180    &  TX Ret  	   & HARPS      &  2011 Dec.      & 270 & 1\\
52258534    &  BG Hyi 	   & HARPS      &  2011 July       & 125 & 2\\
139825582  &  CF Ind  	   & HARPS      &  2011 July       & 190 & 1\\
183595451  &  AI Scl 	   & UVES       &  2001 Dec.       & 280 & 1\\
224285325  &  SX Phe 	   & FEROS      &  2007 June & 330 & 6\\
229154157 & HD 11667& HARPS & 2006 Nov. & 65 & 2\\
253917376  &  UV PsA   	  & HARPS      &  2011 July        & 150 & 1\\
265566844  &  BX Ind  	  & HARPS      &  2011 July        & 125 & 1\\
303584611  &  BS Scl 	  & HARPS      &  2011 July        & 205 & 2\\
355687188  &  HD 224852 & FEROS      &  2007 June       & 100 & 8\\
382551468  &  CD-53 251  & HARPS      &  2008 Nov.       & 65 & 28\\
394015973  &  BE Ind         & HARPS      &  2011 July       & 150 & 1\\\
\\ \hline

\end{tabular}
\label{infodata}
\end{table}

\begin{table*}
\centering
\label{hydrogenteff}
  \caption{ Spectroscopic parameters determined using two different methods as described in detail in Section \ref{spec}. In columns three and four we list the $T_{\rm eff}$ determined from the hydrogen lines and GSSP methods, respectively. The parameters agree within their uncertainties. The $v\sin i$ values were also calculated by two different teams, and agree well within the error bars (here we chose to display the values determined with the GSSP method). The metallicity and v$_{\rm micro}$ were determined using the GSSP code. In the last column LPV stands for line profile variations. The asymmetries observed in the H$\beta$ line in  HD 224852 are likely due to the data extraction pipeline. }
  \begin{tabular}{@{}rlllrrrl@{}}

\\ \hline
\multicolumn{1}{c}{TIC}& \multicolumn{1}{c}{alternative}& \multicolumn{1}{c}{$T_{\rm eff}$ [K]}    &  \multicolumn{1}{c}{$T_{\rm eff}$ [K]} & \multicolumn{1}{c}{$v\sin i$}   & \multicolumn{1}{c} {[M/H] }& \multicolumn{1}{c}{v$_{\rm micro}$}  & comments\\

 & \multicolumn{1}{c}{name} & \multicolumn{1}{c}{H lines} &  \multicolumn{1}{c}{GSSP} & \multicolumn{1}{c}{[km~s$^{-1}$]}  & &  \multicolumn{1}{c}{[km~s$^{-1}$]}  &   \\
							
 \hline
9632550     &  BS Aqr         &  7100\,$\pm$\,150  &  7240\,$\pm$\,290 & 21 \,$\pm$\, 6 &$ -0.2 \,\pm$\, 0.3 & 3.3 \,$\pm$\, 2.5 & SB2\\
32197339   &  HD 210139  &  7600\,$\pm$\,200  &  7960\,$\pm$\,95 & 8 \,$\pm$\, 1 &$ -0.05 \,\pm$\, 0.08 & 2.9 \,$\pm$\, 0.6 & pronounced LPVs \\
38587180   & TX Ret          &  7200\,$\pm$\,150  &  7310\,$\pm$\,40 & 70 \,$\pm$\, 3 & $-0.01 \,\pm$\, 0.04 & 3.8 \,$\pm$\, 0.5 & \\
52258534   & BG Hyi          &  6900\,$\pm$\,120  &  6980\,$\pm$\,50 & 25 \,$\pm$\, 1 & $-0.35 \,\pm$\, 0.07 & 4.3 \,$\pm$\, 0.5 & \\
139825582 &  CF Ind          &  7100\,$\pm$\,150  & 7090\,$\pm$\,40 & 227 \,$\pm$\, 13 & $-0.22 \,\pm$\, 0.06 & 3.6 \,$\pm$\, 0.5 & pronounced LPVs\\
183595451 & AI Scl            &  7300\,$\pm$\,120  & 7315\,$\pm$\,40 & 102 \,$\pm$\, 5 & $-0.05 \,\pm$\, 0.07 & 4.0 \,$\pm$\, 0.6 & \\
224285325 & SX Phe          & 7500\,$\pm$\,150  & 7500\,$\pm$\,115 & 18 \,$\pm$\, 2 & $-0.1 \,\pm$\, 0.2 & 1.9 \,$\pm$\, 0.4 & \\
229154157 & HD 11667     &  7300\,$\pm$\,200  & 7185\,$\pm$\,75 & 16 \,$\pm$\, 1 & $-0.06 \,\pm$\, 0.07 & 3.1 \,$\pm$\, 0.5 & SB2, with LPVs\\
253917376 & UV PsA         &  7000\,$\pm$\,200  & 7025\,$\pm$\,50 & 76 \,$\pm$\, 4 & $-0.16 \,\pm$\, 0.06 & 4.0 \,$\pm$\,0.5 & \\
265566844 &  BX Ind          &  6800\,$\pm$\,220  & 6790\,$\pm$\,60 & 5 \,$\pm$\, 1 & 0.04 \,$\pm$\, 0.06 & 1.3 \,$\pm$\, 0.3 & \\
303584611 & BS Scl           &  7900\,$\pm$\,150  & 7955\,$\pm$\,55 & 32 \,$\pm$\, 1 & $-0.08 \,\pm$\, 0.05 & 3.2 \,$\pm$\, 0.4 & \\
355687188 & HD 224852   &  7000\,$\pm$\,150  & 7335\,$\pm$\,90 & 14 \,$\pm$\, 1 & $-0.04 \,\pm$\, 0.08 & 3.0 \,$\pm$\, 0.5 &  asymmetric H$\beta$ \\
382551468 & CD-53 251    &  6500\,$\pm$\,200  &  6800\,$\pm$\,80 & 13 \,$\pm$\, 1 & $-0.08 \,\pm$\, 0.07 & 2.0 \,$\pm$\, 0.4 & pronounced LPVs\\
394015973 &  BE Ind          &  7500\,$\pm$\,200  &  7500\,$\pm$\,50 & 45 \,$\pm$\, 2 & 0.06 \,$\pm$\ 0.06 & 3.3 \,$\pm$\, 0.5 & \\\hline
\end{tabular}
\label{hydrogenteff}
\end{table*}

\subsection{Luminosities derived from the Gaia parallaxes}\label{Gaia}

We calculated the absolute magnitude and luminosity for each target (see Table \ref{tab:stellar_parameters_new}) using  the parallaxes from both the re-reduced Hipparcos data \citep{Leeuwen07} and Gaia DR2 \citep{Gaia2}. This was done to check for biases in our parameters, especially for nearby and bright stars. All but two stars from our sample (HD~4125 and $\alpha$ Pic) have Gaia DR2 parallaxes, but only 65 stars have Hipparcos parallaxes including HD~4125 and $\alpha$ Pic. 

To estimate the total interstellar extinction, we used several published reddening maps  \citep{2018A&A...616A..17A,chen98,schlegel98,drimmel03,green15,green18}. The distances and their uncertainties were calculated directly  from the parallaxes. Almost all stars are within 1\,kpc from the Sun, which means that the extinction is small but not negligible (total absorption $A_{\mathrm{V}}$\,$<$\,0.25\,mag, with a mean and median of 0.05 and 0.04\,mag, respectively). We calculated a mean value from the above-listed references, which are all consistent within 0.01\,mag. This value of 0.01\,mag was adopted as the extinction uncertainty for all targets. We chose to do so because there are no estimates of the individual errors from the different maps and methods.

To calculate luminosities for our targets, we used the bolometric corrections for the $V$ magnitudes from \citet{Flower96}. The bolometric correction is at a minimum for A stars and does not influence the luminosity calculation significantly. The effective temperatures were taken from Sect. \ref{SED}. However, there is no homogeneous source of $V$ magnitudes that includes all our targets. We therefore calculated the averages from the magnitudes published by \citet{kharchenko01} and \citet{Henden16}, and transformed the Gaia DR2 $G$ magnitudes according to the calibration by \citet{2018A&A...616A..17A}. Within these three data sets, we found no outliers larger than $0.015$\,mag.  For the error calculation, we took a full error propagation of the individual errors into account. 

\citet{2018A&A...616A...2L} found that the Gaia parallaxes require a zero-point offset of 0.03 mas, although, the offset is dependent on the colour, magnitude and position on the sky \citep{2018A&A...616A...2L, 2019ApJ...878..136Z}. \citet{2019MNRAS.tmp..585M} have experimented with applying the suggested offset of 0.03 mas, but found that this correction results in unrealistically low luminosities for A and F stars. They deduce that the zero-point offset may be smaller for bluer stars. Since there is no consensus on the exact correction, we follow \citet{2019MNRAS.tmp..585M} and do not apply any zero-point offset in our analyses. We also note that \citet{2018A&A...616A..17A} discourage users of applying the offset to individual parallaxes.

The stellar parameters with their respective uncertainties are listed in Table \ref{tab:stellar_parameters_new}. If we compare the absolute magnitudes derived from the Hipparcos and Gaia DR2 data sets using the identical apparent magnitude, reddening and Bolometric correction, we find only three targets for which the deviation is larger than 3$\sigma$, specifically $\nu$ Pic, BX Ind, and $\theta$ Gru. This proves that the Gaia DR2 data have no significant discrepancies (in a statistical sense) when compared to the Hipparcos ones. We also note that these stars are bright which may explain the deviation. In addition, $\theta$ Gru was identified as a binary (see \ref{table: star_parameters1}), which may apply for the other two as well.

\begin{figure}
\begin{center}
\includegraphics[width=1\columnwidth]{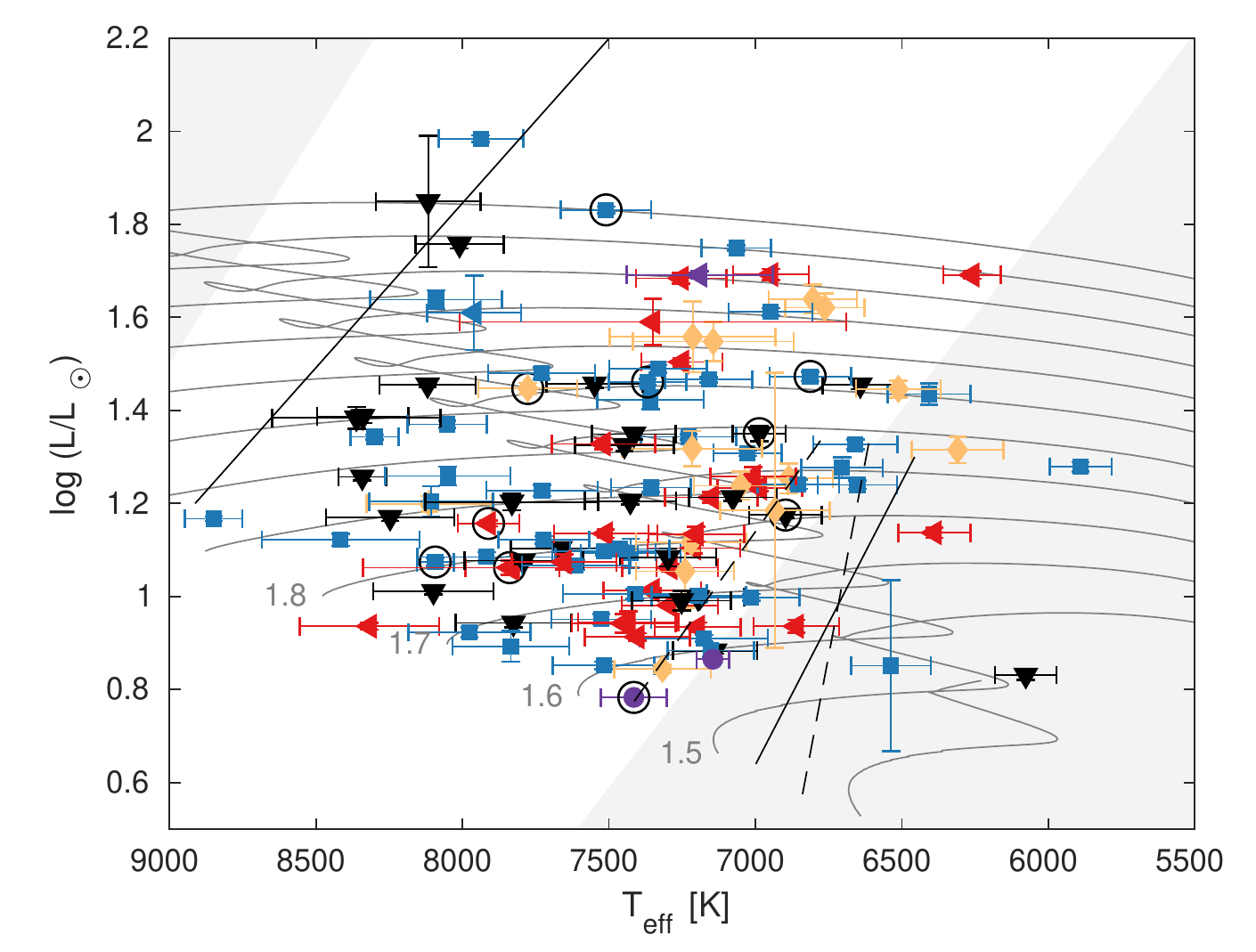}
\caption{Intermediate-mass A and F stars observed by TESS in sectors 1 and 2 in the HR~diagram. Non-variable stars are shown as black downward facing triangles, $\delta$~Sct stars as blue  squares, $\gamma$~Dor stars as  violet circles, hybrids or suspected hybrids in yellow diamonds, and HADS stars as left-facing red triangles. Stars marked with a black circle are known binary stars. The blue and red edges of the $\delta$~Sct instability domain (marked by the straight, continuous black lines) are taken from \citet{2000ASPC..210..215P} and \citet{2005A&A...435..927D}, respectively. The blue edges from \citet{2000ASPC..210..215P} and \citet{2005A&A...435..927D} more or less overlap. The $\gamma$~Dor instability strip marked by the dashed lines is from \citet{2005A&A...435..927D}. In addition, we highlight in white the $\delta$ Sct instability strip found empirically by \citet{2019MNRAS.tmp..585M} within which >20\% of the population pulsates (peaking at 70\%). The evolutionary tracks were computed with the Warsaw-New Jersey models \citep{1977AcA....27...95D, 1998A&A...333..141P,1999AcA....49..119P} using no rotation and solar metallicity. The evolutionary tracks, labeled with masses expressed in $M_\odot$, are only displayed to guide the eye. }
\label{fig:stars}
\end{center}
\end{figure}

In Fig.~\ref{fig:stars} we plot our stars in the HR~diagram using Gaia luminosities and $T_{\rm eff}$ determined in this work using the SED method. Due to binarity, large error bars in their Gaia measurements or missing parallaxes the following stars have been omitted: $\theta$ Ret (V = 6.05), HD 223991 (V = 6.35), HD 216941 (V = 9.49), and HD 210767 (V = 7.78).

\section{Understanding the excitation mechanism of $\delta$~Sct stars using Am stars}
\label{subsection: Am stars}

In the absence of mixing, which is usually related to rotation and efficient convection in the outer envelopes, atomic diffusion,  which includes radiative levitation and gravitational settling, plays an important role in A stars. As a result, Am stars display underabundances of  Sc and Ca and overabundances of Ba, Sr and Y, among other elements, in their photospheres \citep{1974ARA&A..12..257P}.

Traditionally, the excitation mechanism associated with $\delta$~Sct pulsations is the $\kappa$-mechanism acting in the He~{\sc II} ionisation layer \citep{1963ApJ...138..487C}. However, observations \citep{2011Natur.477..570A, 2019MNRAS.tmp..585M} clearly indicate that this mechanism alone is insufficient. This is especially the case for the chemically peculiar Am stars where He is expected to be depleted from the He~{\sc II} ionisation layer where the $\kappa$-mechanism operates. Even when including atomic diffusion \citep{2000A&A...360..603T}, models in only a very narrow region in the instability strip were found to have unstable modes, which is clearly contradicted by observations (e.g., \citealt{2017MNRAS.465.2662S}). 

Here we investigate to what extent the pulsational stability changes when helium is depleted from the outer layers. More specifically we have used the same approach as described in \cite{2001MNRAS.323..362B} to model the helium abundance in the outer layers, to mimic atomic diffusion. For our models with a chemically homogenous envelope, the helium abundance by mass is Y= 0.28, whereas for our depleted models the helium abundance was decreased to Y = 0.15 in the outer layers, extending down beneath the He~{\sc II} ionisation layer. 

\begin{figure}
\begin{center}
\includegraphics[width=0.48\textwidth]{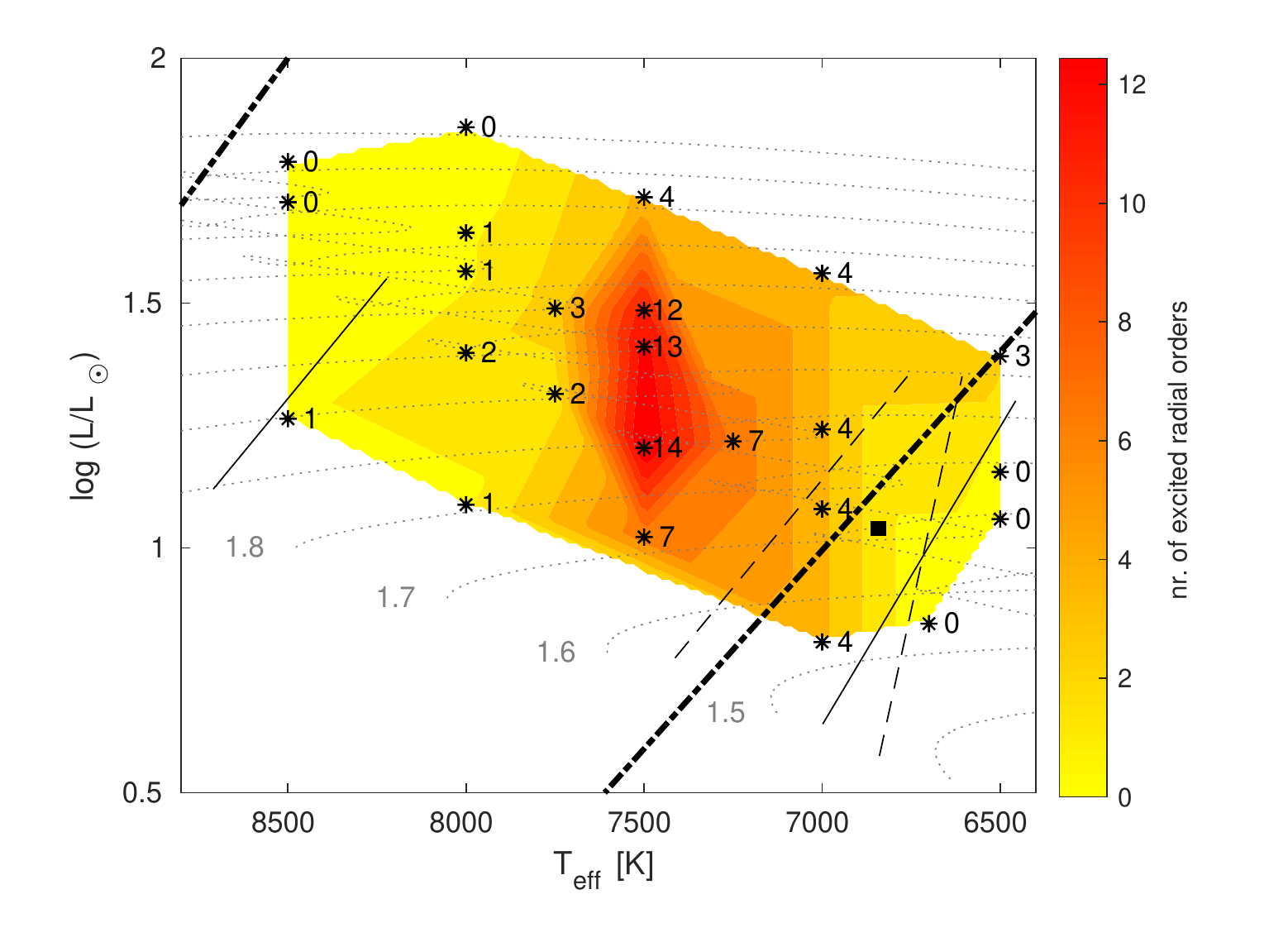}

\caption{Theoretical instability strip from models including time-dependent non-local convection treatment (Section \ref{subsection: Am stars}). The asterisks indicate the location of our theoretical models and the adjacent number indicates  the number of excited radial orders. 
In straight, continuous black lines we mark the theoretical blue and red edges of the $\delta$~Sct instability domain from \protect\citet{2000ASPC..210..215P} and \protect\citet{2005A&A...435..927D}, respectively. The blue edges from \protect\citet{2000ASPC..210..215P} and \protect\citet{2005A&A...435..927D} are more or less overlapping. The black square is the red edge computed by \protect\cite{2000ASPC..210..454H}. For convenience we also indicate the $\gamma$~Dor instability strip in dashed lines \protect\citep{2005A&A...435..927D}. In addition, we also depict the empirical $\delta$ Sct instability strip boundaries from \protect\citet{2019MNRAS.tmp..585M} at which the pulsator fraction is 20\% (dashed-dotted line). The evolutionary tracks were computed with the Warsaw-New Jersey models \protect\citep{1977AcA....27...95D, 1998A&A...333..141P, 1999AcA....49..119P} using no rotation and solar metallicity. The evolutionary tracks are only displayed to guide the eye. }

\label{fig:dsct_IS}

\includegraphics[width=0.48\textwidth]{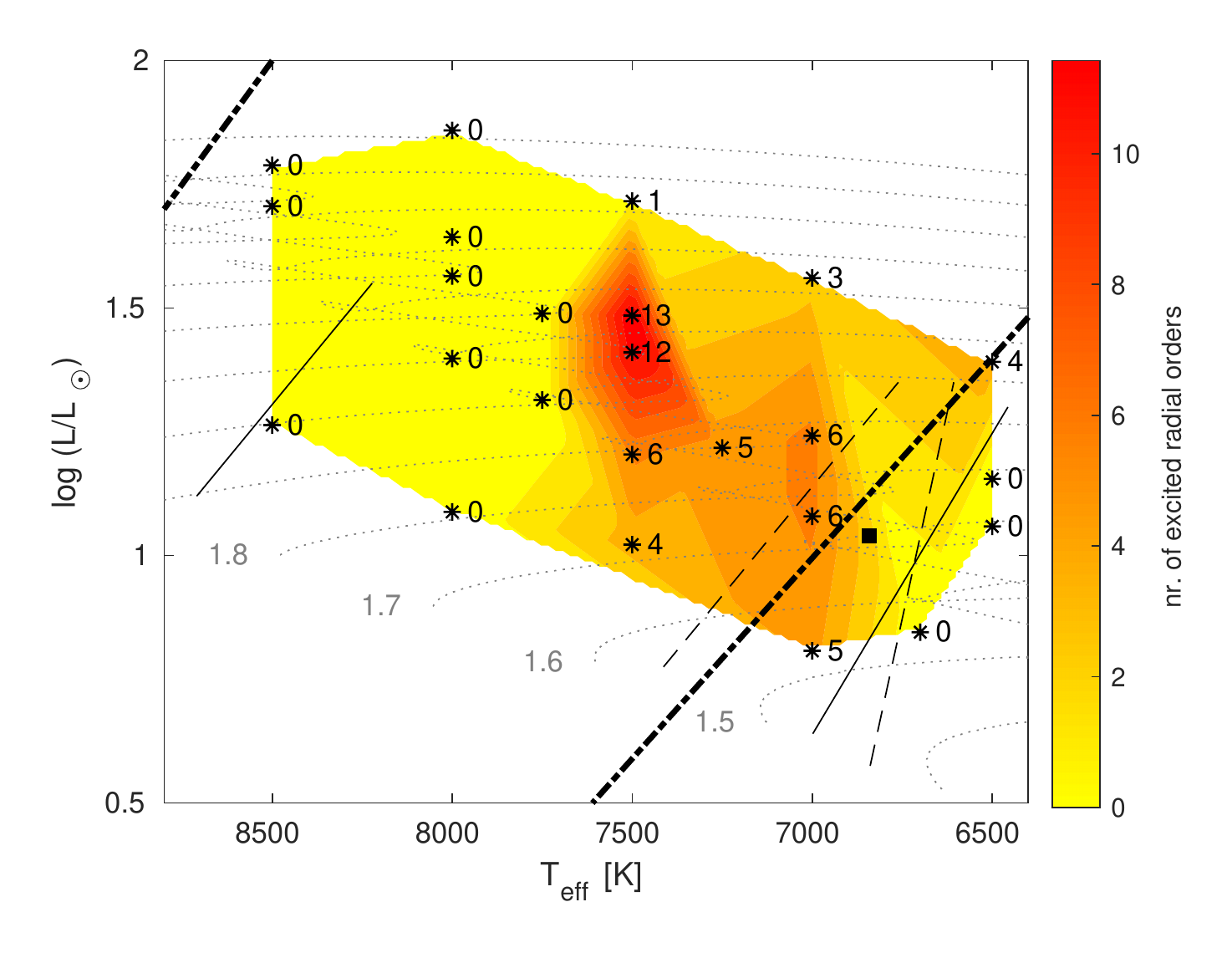}
\caption{Theoretical instability strip with depleted helium in the outer layers mimicking helium settling. Models include time-dependent non-local convection treatment (Section \ref{subsection: Am stars}). The asterisks indicate the location of our theoretical models and the adjacent number indicates  the number of excited radial orders. The instability domains and evolutionary tracks are the same as in Fig. \ref{fig:dsct_IS}. }
\label{fig:dsct_IS_He15}
\end{center}
\end{figure}

In Fig. \ref{fig:dsct_IS} and Fig. \ref{fig:dsct_IS_He15} we show results of our stability computations, in which we include the pulsationally perturbed radiative ($\kappa$-mechanism) and convective heat fluxes as well as the perturbed momentum flux (turbulent pressure). These computations were obtained from models following the procedures described in detail in \cite{Houdek1999} and \cite{2001MNRAS.323..362B}. These models adopt the time-dependent non-local convection formulation by \cite{1977LNP....71..349G, 1977ApJ...214..196G}. In addition to the mixing length parameter, $\alpha_{\rm MLT}$, this formulation uses three additional non-local convection parameters, $a, b$ and $ c$. Here we adopt the values $a^2=b^2=c^2=950$, which allowed \cite{2014ApJ...796..118A} to reproduce roughly 85~per~cent of the observed frequency range of HD~187547 (a pulsating Am star) and \cite{2001MNRAS.323..362B} to reproduce oscillations in roAp stars.

In Fig. \ref{fig:dsct_IS} we plot the theoretical instability region of our `mixed models' (i.e., He non-depleted) and  Fig. \ref{fig:dsct_IS_He15} illustrates the instability domain of our depleted models. The asterisks show the locations of the computed models in the HR diagram and the number of excited radial orders found for that particular model is indicated. The contour levels in this plot delineate the number of consecutive excited radial-order modes. The redder the area, the larger number of modes to be excited. Yellow indicates no excitation at all. We note that in both cases, i.e. in the homogenous and the depleted envelope models, there seems to be a maximum in the number of excited radial orders at an effective temperature of 7500~K. Interestingly, the largest number of unstable modes roughly corresponds with the temperature where \citet{2019MNRAS.tmp..585M} found the largest fraction of pulsating stars, although slightly shifted in luminosity.

\begin{figure}
\begin{center}
\includegraphics[width=0.48\textwidth]{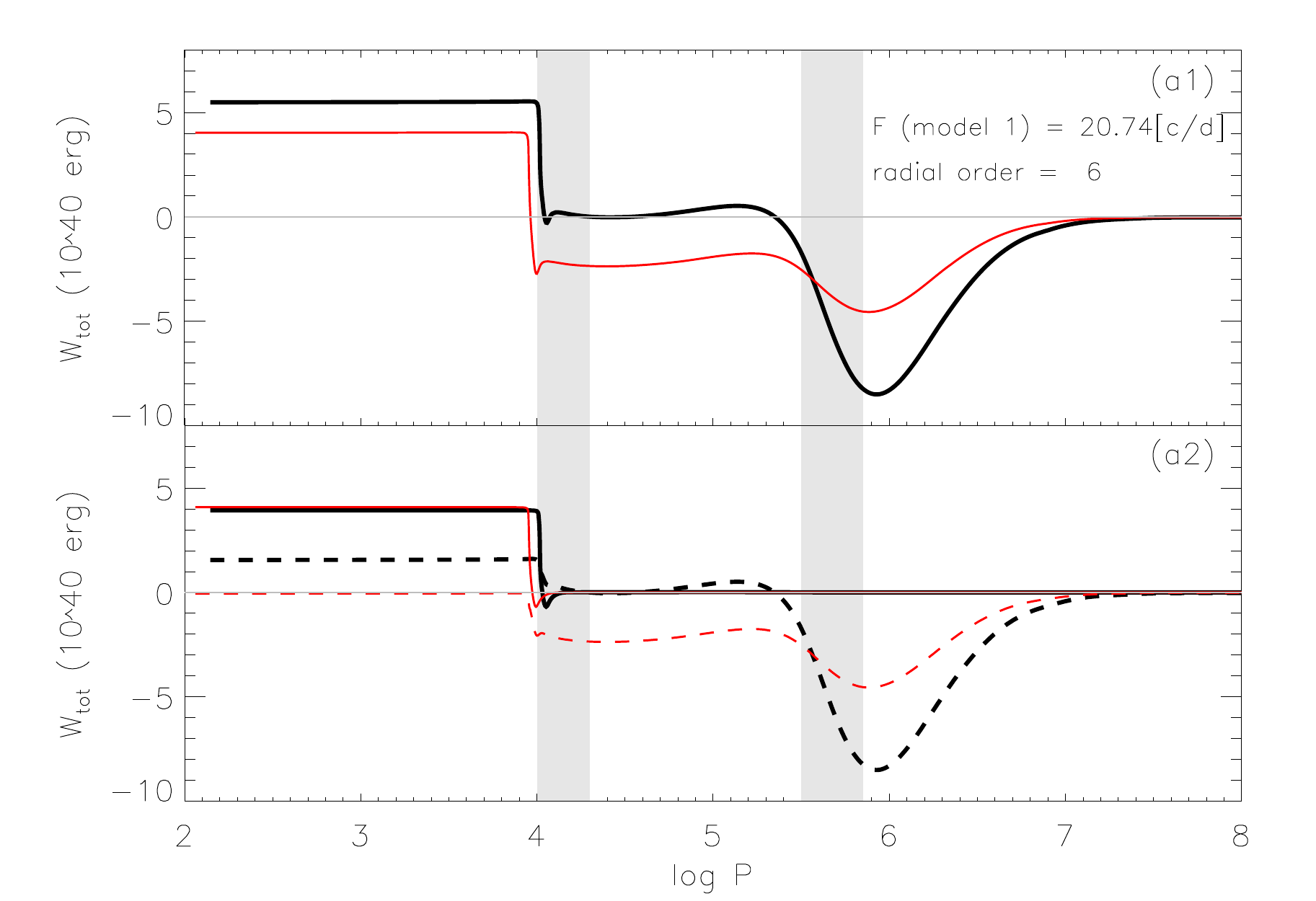}
\includegraphics[width=0.48\textwidth]{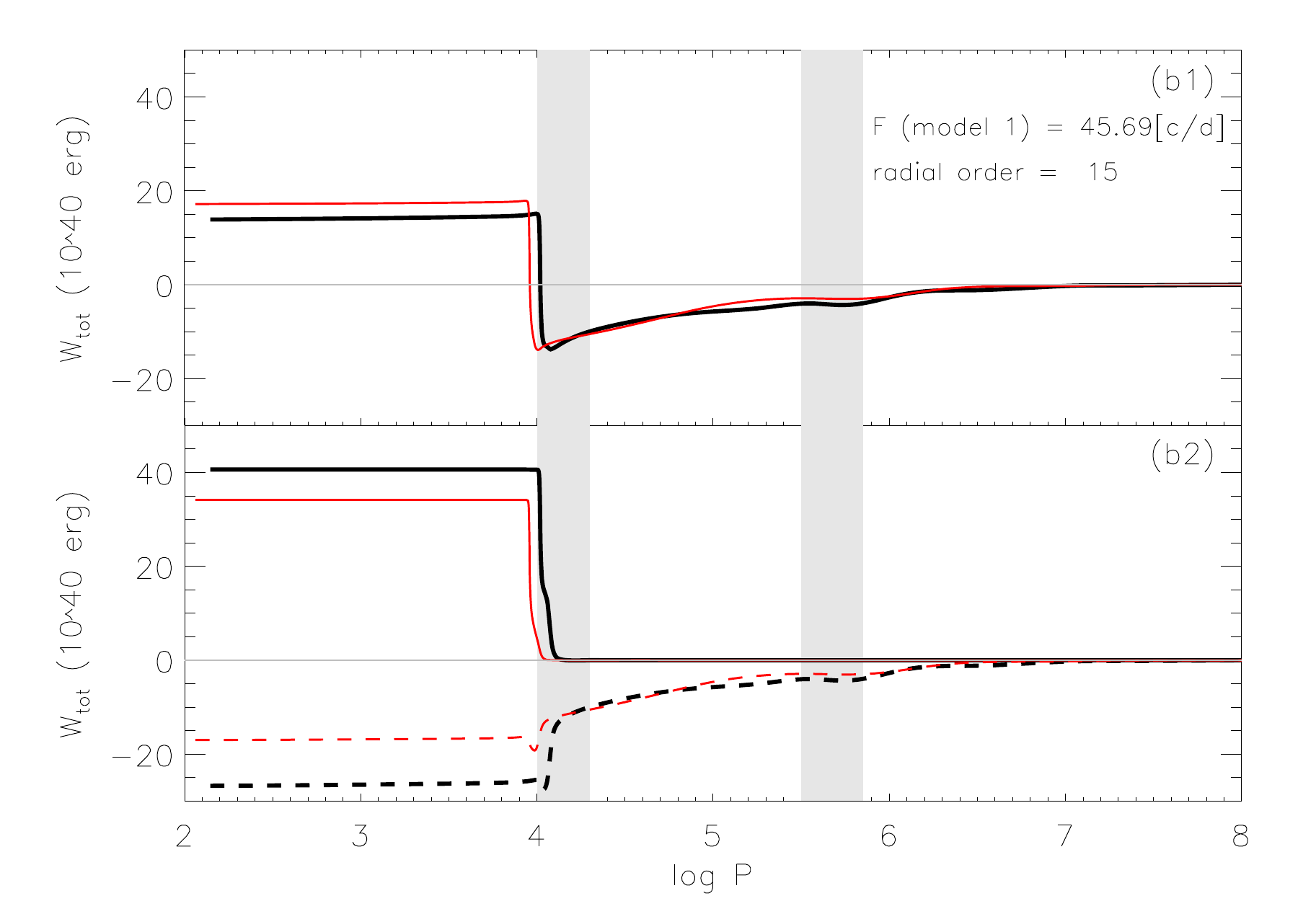}

\caption{ Accumulated work integrals (solid curves in panels (a1) and (b1)) as a function of the total pressure for two different radial orders, $n=6$ and $n=15$, for the homogeneous model (black) and the He-depleted model (red in the online and grey in the printed version). The modes are unstable whenever the accumulated work integrals are positive towards the surface of the star, as displayed in (a1) and (b1). In panels (a2) and (b2) we show the contributions of the gas pressure (dashed line) and the turbulent pressure (solid line). The grey region at $\log P=6$ indicates the{\sc He~II}, and at $\log P=4$ the {\sc H~I} and {\sc He~I} ionization zones. From panels (a2) and (b2) it is evident that turbulent pressure in the {\sc H~I} and {\sc He~I} ionisation zones plays a significant role in the excitation of both modes. Specifically in the case of $n=15$, it is the only driving agent, while the gas pressure (responsible for the $\kappa$-mechanism) is damping pulsations in the {\sc He II} ionisation zone. The model parameters are in both cases the following: M$=2.3$~M$_{\sun}$, $T_{\rm eff} = 7500$ K, $\log L = 1.48$~L$_{\sun}$ and correspond to the model with 13 unstable radial orders in Fig. \ref{fig:dsct_IS} and 12 in  Fig. \ref{fig:dsct_IS_He15}, respectively. }
\label{fig:modes}
\end{center}
\end{figure}

In Fig. \ref{fig:modes}, panels (a1) and (b1), we show the accumulated work integrals for a radial mode with $n=6$ and a mode with $n=15$, respectively, in our chemically homogeneous and He-depleted models. Regions where the accumulated work integral increases towards the surface of the star contribute to the driving of the mode while regions where it decreases contribute to the damping. Panels (a2) and (b2) further show the individual contributions to the total accumulated work integral of the gas (dashed line) and of the turbulent pressure (solid line). For both the homogenous (Fig.~\ref{fig:dsct_IS}) and the He-depleted envelope models (Fig.~\ref{fig:dsct_IS_He15}), the turbulent pressure contributes crucially to mode instability.

\begin{figure}
\begin{center}
\includegraphics[width=0.48\textwidth]{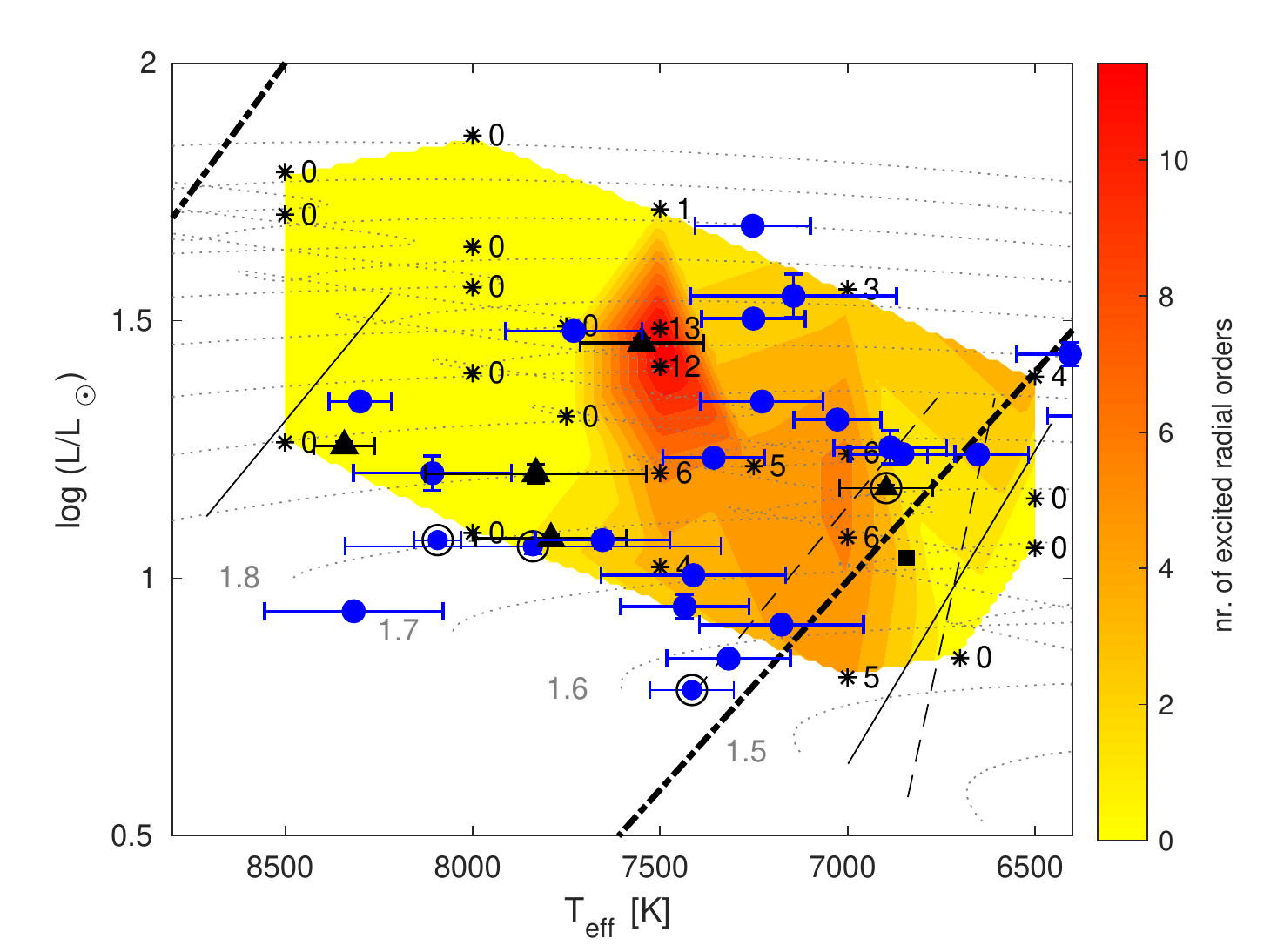}
\caption{Comparison between the observed TESS Am stars and the theoretical instability strip including He depletion in the outer envelope. All stars are identified to be Am stars based on the Renson catalogue \citep{Renson2009}. The blue circles depict the pulsating Am stars; the black triangles identify constant Am stars. Known binaries (see Table \ref{table: star_parameters}) are marked by black rings. The square, asterisks and all lines have the same meaning as in Fig. \ref{fig:dsct_IS}.}
\label{fig:AMstars}
\end{center}
\end{figure}

In this paper, we do not aim to fully reproduce the instability domain of $\delta$~Sct stars, but to show that diffusion, especially of He, plays an important role in driving pulsations in $\delta$~Sct stars, and that turbulent pressure, acting in the H ionisation zone is a significant driving agent. Therefore, excitation by turbulent pressure offers a solution to the long-standing question of the driving mechanism in pulsating Am stars. In Fig. \ref{fig:AMstars} we plot all Am stars from \citet{Renson2009} that were observed by TESS in sectors 1 and 2. We distinguish between constant (black triangles) and pulsating Am (blue) stars, and mark known binaries with black rings. We find good agreement between the observed Am stars and the predicted instability region when turbulent pressure is included. Our findings also agree well with earlier ground-based observations, showing that pulsating Am stars tend to be found in the cooler part of the instability region \citep{2017MNRAS.465.2662S}. 

To conclude this section, we show that even in the case where He is depleted from the outer layers, excitation of pulsations is still possible and matches the observed distribution of known pulsating Am stars observed by TESS. In addition, we see an increase in the number of unstable radial orders and magnitude of the linear growth rates in stars with $T_{\rm eff} \simeq 7500$~K. Here we demonstrate the potential of how TESS observations of many Am stars will improve our understanding of pulsation driving in intermediate-mass stars. Before drawing any final conclusions, however, we need to perform in-depth analyses and to vary parameters in our computations, e.g., metallicity, the mixing length and the non-local convection parameters $ a, b,$ and $c$. In addition, we will also explore the excitation at different helium depletion values. This will be done in a future paper.

\section{Examples of variability}
\label{section: case studies}

As TESS is observing almost the entire sky, we have the unprecedented opportunity to obtain continuous, high-precision light curves of a variety of pulsating stars as described in Section~\ref{section: intro}. The 2-min TESS sample in sectors 1 and 2 presented in this work includes some prototypes, i.e., SX~Phe and $\gamma$~Dor. In this section, we demonstrate the power of TESS data for a variety of different pulsator classes amongst A and F stars and the potential of future asteroseismic modelling. The stars shown here were chosen either because they are eponymous for their classes (SX~Phe and $\gamma$~Dor), very bright ($\alpha$ Pic) or to cover the entire variety of the stars studied here: $\delta$ Sct, $\gamma$ Dor, $\lambda$ Boo, HADS, pre main-sequence, eclipsing binary stars.


\subsection{TIC~229059574 = HD~210111}
\label{subsection: lambda boo}

TIC~229059574 (HD~210111), is a $\lambda$~Boo star in a double-lined spectroscopic binary (SB2) system \citep{2012MNRAS.419.3604P}. SB2 line profiles were seen in all 24 high-resolution spectra taken by \citet{2012MNRAS.419.3604P} with UVES, from which the authors concluded that this is a pair of equal-mass stars with both stars having $T_{\rm eff} \approx$ 7400~K, $\log g$ = 3.8, \mbox{[Fe/H] = $-$1.0~dex} and $v\sin i = $ 30~km\,s$^{-1}$. By computing composite spectra with various atmospheric parameters, they also concluded that neither component has a solar metallicity, from which we infer that they are both $\lambda$~Boo stars.

\citet{1994IBVS.4094....1P} first discovered the $\delta$~Sct variability of HD~210111, and \citet{2006A&A...455..673B} carried out the most extensive analysis of its variability with a multi-site ground-based campaign over a 30-d period. The TESS light curve is of similar duration, but almost uninterrupted and of space-telescope quality.

\begin{figure}
\begin{center}
\includegraphics[width=0.48\textwidth]{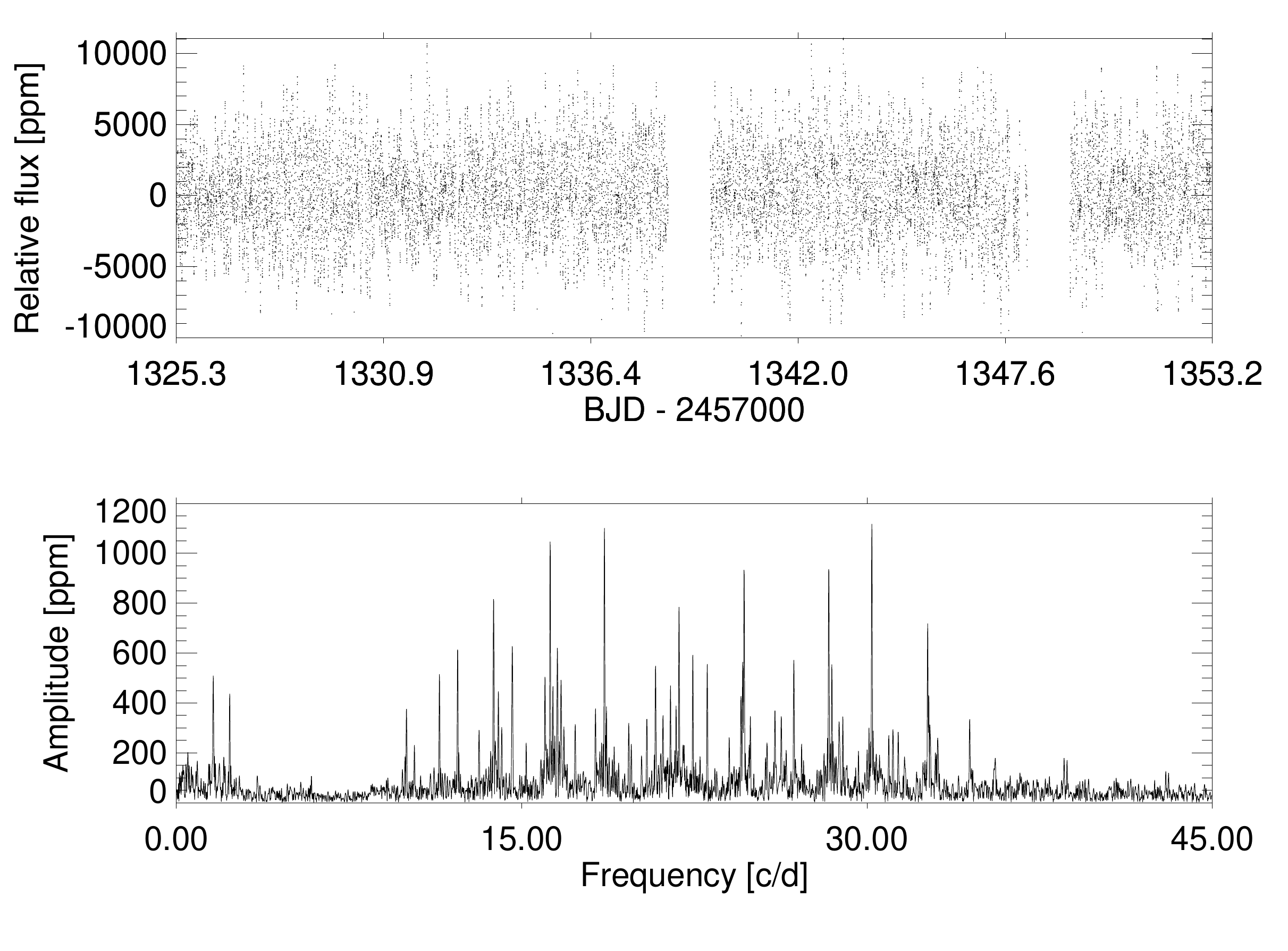}
\caption{TESS light curve (top) and amplitude  transform (bottom) of the $\lambda$~Boo star HD~210111.}
\label{fig:LB_FT}
\end{center}
\end{figure}

The TESS light curve from sector~1 shows HD~210111 to be highly multi-periodic with frequencies in the range $10 - 40$~d$^{-1}$ and amplitudes of order $100 - 1000$~ppm (Fig.~\ref{fig:LB_FT}). The highest peak lies at 30.2~d$^{-1}$, consistent with the non-radial pulsation with a 49-min period identified by \citet{1999A&A...350..553B}. Many of the 18 oscillation frequencies reported by \citet{2006A&A...455..673B} are present in the TESS data, but the amplitudes are significantly different. This may be due to the different passbands, although amplitude variability is not uncommon in $\delta$~Sct stars \citep{2008A&A...478..855L, 2015A&A...579A.133B, 2016MNRAS.460.1970B}. In the TESS passband, HD~210111 has 59 peaks with amplitudes exceeding 200~ppm, where our analysis is terminated because of the high density of peaks. 

Visual inspection of the amplitude spectrum of the light curve (Fig.~\ref{fig:LB_FT}) suggests there may be a common frequency spacing between $\sim2$ and 2.5~d$^{-1}$. When using all 59 peaks above 200~ppm, a histogram of frequency differences showed no outstanding features compared to a randomly generated set of frequencies covering the same frequency range. However, a subset of the strongest peaks (with amplitudes above 500 ppm) showed an excess of peaks spaced by $\sim$2.2~d$^{-1}$. Common spacings can sometimes be attributed to rotational splittings, or to the large frequency separation $\Delta \nu$, which is the frequency spacing between consecutive radial orders\footnote{The exact spacing depends on the radial orders and can only be directly related to the Sun or Sun-like stars in case high radial order modes are observed.}. The $v\sin i$ of these stars is low, at about 30~km\,s$^{-1}$, so it is unlikely that the rotational splitting of one or both stars is being observed, unless the inclination $i$ happens to be low. We therefore find it more plausible that the identified spacing corresponds to the large frequency separation or half this value. A more accurate assessment using a Gaia luminosity is prevented by the SB2 nature of the target. It is also not clear whether solar-metallicity tracks should be used. The reported [Fe/H] is -1~dex, but in $\lambda$~Boo stars the low metallicity might be a surface phenomenon only. Asteroseismology is one of the tools that can be used to probe the global stellar metallicity, but HD~210111 is perhaps too complex a target to be a good starting point because of it being a binary of two similar stars. At this stage it is unclear whether only one or both components are pulsating. A more detailed analysis is beyond the scope of this paper.

\begin{figure}
\begin{center}
\includegraphics[width=0.48\textwidth]{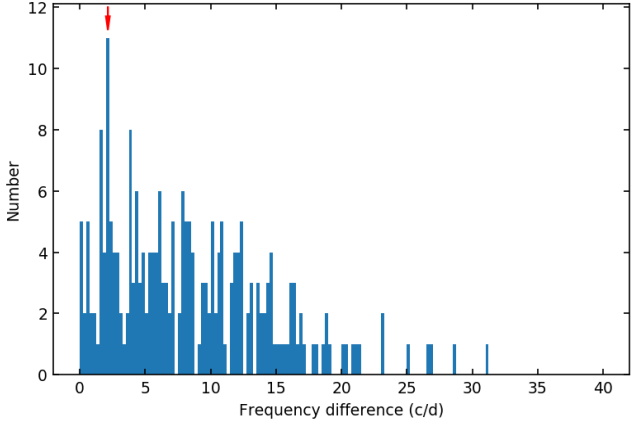}
\caption{A histogram of the frequency differences from the peaks extracted from HD~210111 that have amplitudes exceeding 500~ppm. The peak at 2.2\,d$^{-1}$ is marked with a red arrow. Integer multiples of this peak (4.4, 6.6, $\dots$\,d$^{-1}$) are also expected. }
\label{fig:LB_freq_diff}
\end{center}
\end{figure}

\subsection{TIC 167602316 = $\alpha$ Pictoris}
\label{subsection: alphapic}

$\alpha$~Pictoris is the brightest star in the sample presented here ($V=3.3$) and was not known to be a pulsating star prior to TESS observations. This makes it one of the brightest known $\delta$~Sct stars. This star has $T_{\rm eff} = 7550 \pm 35$~K \citep{2012A&A...537A.120Z}, a Hipparcos luminosity of $\log\,({\rm L/\rm L}_\odot) = 1.47 \pm 0.14$ and is rapidly rotating with $v \sin i = 206$~km~s$^{-1}$ \citep{2012A&A...537A.120Z}.  

\begin{figure}
\begin{center}
\includegraphics[width=0.48\textwidth]{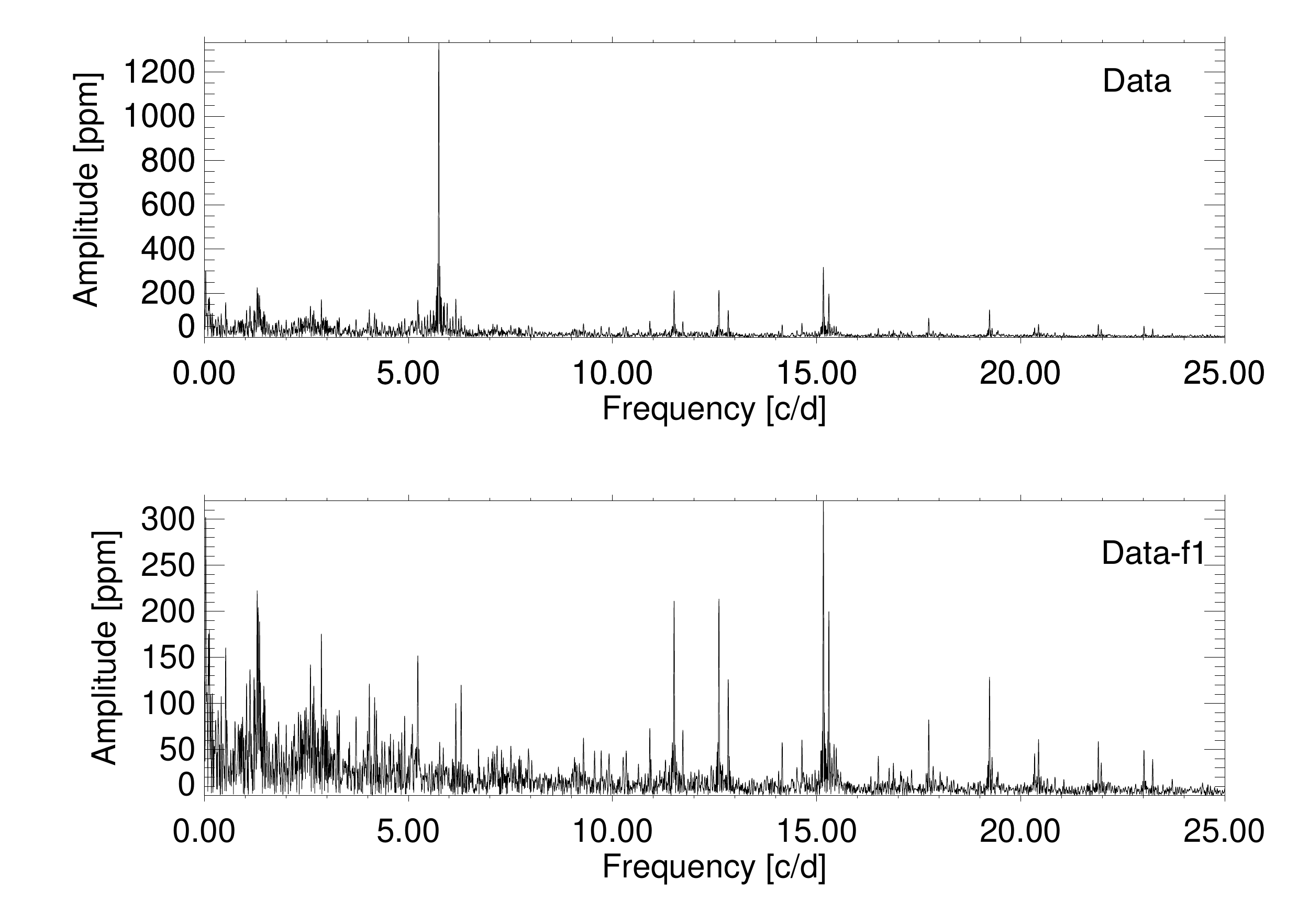}
\caption{Amplitude spectra of the original data of $\alpha$~Pic (top) and after pre-whitening the dominant mode (bottom). This star is in the TESS continuous viewing zone on the Southern hemisphere which makes it a promising target for further study. }
\label{fig:alphapic}
\end{center}
\end{figure}

$\alpha$~Pic has a rich pulsation spectrum (Fig.~\ref{fig:alphapic}), with 39 significant peaks after 2 sectors of observations. From the lower panel of Fig.~\ref{fig:alphapic} it can be clearly seen that many additional peaks are observed in the low-frequency region that are consistent either with g and/or Rossby modes. In the high frequency range we detect a spacing that is consistent with the large frequency separation of about 2.7~d$^{-1}$ and fits well the stellar parameters given Table~\ref{tab:stellar_parameters_new}.

Based on the known stellar parameters, the highest peak at $f_1=5.74$~d$^{-1}$ could be the fundamental radial mode. The frequency $f_3$ is close to $2f_1$, which may imply that this frequency is the second harmonic, however it is not an exact match (the difference is 0.017~d$^{-1}$), which is similar to the resolution. This issue will be resolved with one year of data, as this star is in the continuous TESS viewing zone.  No other obvious combination frequencies were detected.

\subsection{TIC 224285325 = SX~Phe}
\label{subsection: SX Phe}

\begin{figure}
\begin{center}
\includegraphics[width=0.48\textwidth]{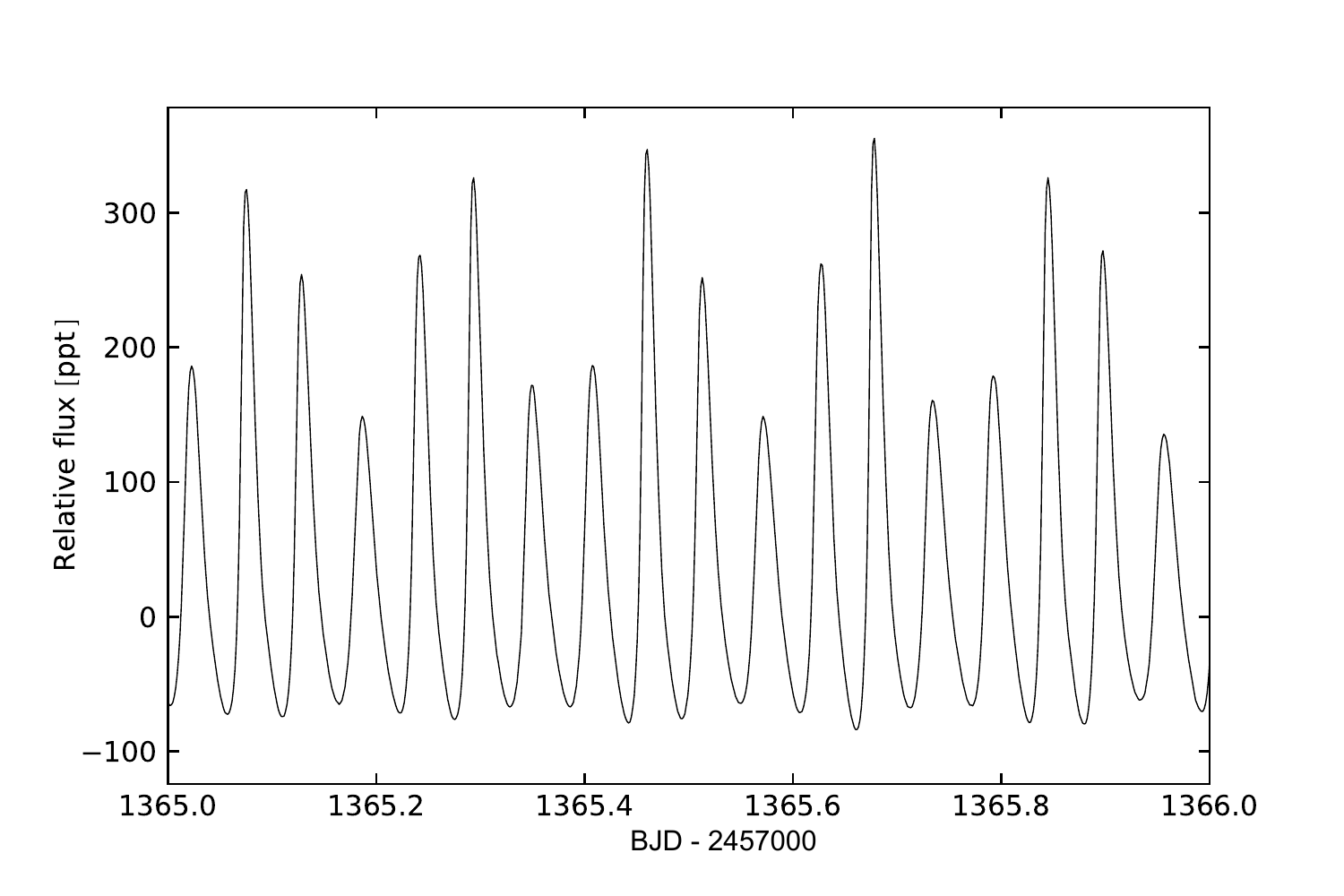}
\includegraphics[width=0.48\textwidth]{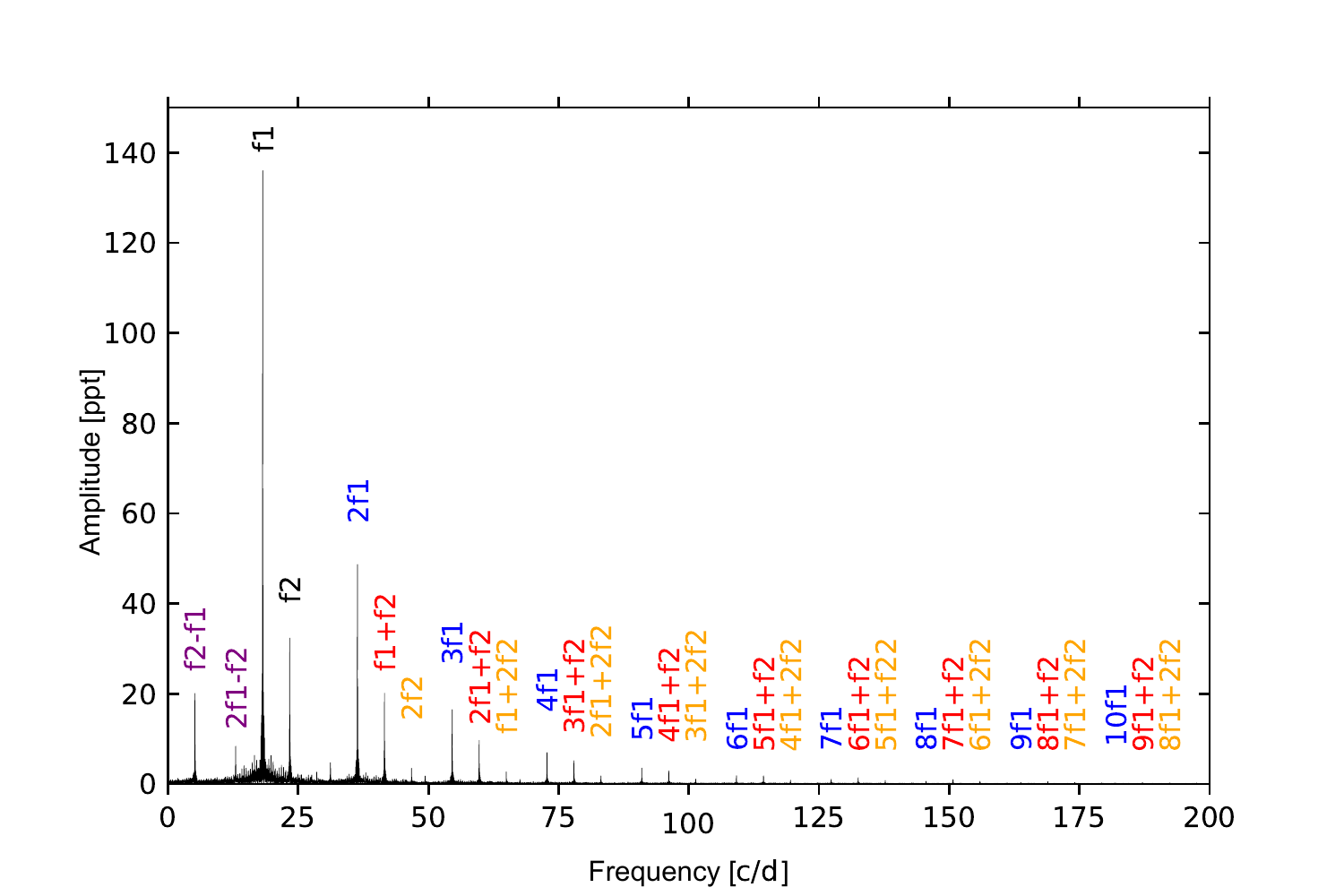}
\caption{A 1-d extract of the TESS light curve of SX~Phe (top). Amplitude spectrum of the entire TESS light curve of SX~Phe (bottom). The two strong independent mode frequencies are labelled along with combination frequencies belonging to three distinct families, and two other combinations at lower frequencies.}
\label{fig: SX Phe}
\end{center}
\end{figure}

SX~Phe is a remarkable HADS star, whose strongest peak exceeds 100~ppt. The light curve is highly asymmetric, with minima of a consistent depth, but maxima spanning a wide range in brightness (Fig.~\ref{fig: SX Phe}). Such light curves are common among high-amplitude SX~Phe stars dominated by one or two mode frequencies and their combinations (e.g., the \textit{Kepler} SX~Phe star KIC~11754974; \citealt{2013MNRAS.432.2284M}). The shapes of the light curves arise from specific phase relationships between parent pulsation modes and their combination frequencies \citep{2015MNRAS.450.3015K}.

The amplitude spectrum of the light curve of SX~Phe itself is indeed dominated by two parent radial modes and their many combination frequencies (Fig.~\ref{fig: SX Phe}). These combinations and their harmonics continue to high frequencies: even at 7$f_1$ (127~d$^{-1}$), the pulsation amplitude is over 1~ppt. We extracted frequencies down to an amplitude of 1~ppt. Five other apparently independent peaks exist with amplitudes above 1~ppt. Although it is possible to assign combinations to some of them, the combinations are not simple (i.e. they do not have low coefficients), so we conclude that some of these peaks could be independent mode frequencies and speculate that resonance may play an important role. A frequency list is provided in Table \ref{SXphefre}.

\begin{table}
\caption{The list of extracted significant frequencies of SX~Phe from TESS Sector 1 and 2 data.}
\centering
\label{SXphefre}
\begin{tabular}{llcc}
\\ \hline
$f_n$ &Frequency& Amplitude & Combinations \\
     & [d$^{-1}$]& [ppt] &  \\
     &  & $\pm 0.040$ &  \\

\hline

$f_1$ & 18.193565(6) & 136.284 &    \\
$f_2$ & 23.37928(2) & 33.079 &  \\
$f_3$ & 36.38712(2) & 48.757 &  $2f_1$ \\
$f_4$ & 41.57284(4) & 20.286 & $f_1+f_2$ \\
$f_5$ & 5.185718(4) & 20.133 &   $f_2-f_1$ \\
$f_6$ & 54.58069(5) & 16.653 &  $3f_1$ \\
$f_7$ & 13.00784(8) & 10.237 & $2f_1-f_2$ \\
$f_8$ & 59.766413 & 9.801 &  $2f_1+f_2$ \\
$f_9$ & 72.7742(1) & 6.939 &   $4f_1$ \\
$f_{10}$ & 77.9599(2) & 5.047 &   $3f_1+f_2$ \\
$f_{11}$ & 31.2014(2) & 4.948 &  $3f_1-f_2$ \\
$f_{12}$ & 46.7585(2) & 3.572 &   $2f_2$ \\
$f_{13}$ & 90.9678(2) & 3.527 & $5f_1$ \\
$f_{14}$ & 96.1535(3) & 2.893 &   $4f_1+f_2$ \\
$f_{15}$ & 22.6303(3) & 2.774 &  \\
$f_{16}$ & 28.5650(3) & 2.692 &   $2f_2-f_1$ \\
$f_{17}$ & 64.9521(3) & 2.632 &  $f_1+2f_2$ \\
$f_{18}$ & 27.5864(4) & 2.074 &   \\
$f_{19}$ & 114.3471(4) & 1.931 &   $5f_1+f_2$ \\
$f_{20}$ & 109.1613(4) & 1.831 &   $6f_1$ \\
$f_{21}$ & 49.3950042 & 1.798 &    \\
$f_{22}$ & 83.1456(5) & 1.738 &  $2f_1+2f_2$ \\ 
$f_{23}$ & 132.5406(5) & 1.495 &   $6f_1+f_2$ \\
$f_{24}$ & 40.823(3) & 1.198 & $f_1+f_20$ \\
$f_{25}$ & 127.355(1) & 1.118 &  $7f_1$ \\
$f_{26}$ & 17.2645(7) & 1.105 &    \\
$f_{27}$ & 45.7806(7) & 1.072 &   \\

\hline
\end{tabular}
\end{table}

The strongest mode corresponds to the frequency $f_1 = 18.193565(6)$~d$^{-1}$, which is high for the fundamental radial mode but rather typical for SX Phe-type stars (e.g., \citealt{2011AJ....142..110M, 2019MNRAS.486.4348Z}). The frequency ratio of $f_1/f_2$ is 0.778192(6), is consistent with the identification of $f_1$ as the fundamental radial mode and $f_2$ as the first radial overtone. Normal $\delta$~Sct stars have values around 0.772, but SX~Phe itself has a rather high value of 0.778192(6), owing mostly to its low metallicity. \citet{1996A&A...312..463P} showed that higher values of this frequency ratio correlate well with lower metallicity, and SX~Phe itself has a low metallicity, with $Z \approx 0.001$. In addition, \citet{2014IAUS..301...31P} shows that other SX Phe stars can also have such a high ratio.

Using the Warsaw-New Jersey evolutionary code (e.g., \citealt{1998A&A...333..141P,1999AcA....49..119P})
and the nonadiabatic pulsational code of \citet{1977AcA....27...95D}, we computed models for SX~Phe with the aim of fitting $f_1$ and $f_2$ as the radial fundamental and first overtone modes, respectively, using the so-called Petersen diagram (e.g., \citealt{1996A&A...312..463P}). We searched a wide range of effective temperatures ranging from 7200 K to 8700 K. The OPAL opacity tables and the solar element mixture of \citet{2009ARA&A..47..481A} were adopted. We considered different metallicity values, $Z$, and initial hydrogen abundance, $X_0$, in the ranges (0.001, 0.002) and (0.66, 0.74), respectively. The value of the mixing length parameter, $\alpha_{\rm MLT}$ = 1.5 and overshooting from the convective core was not included. We note that, at the metallicities and effective temperatures specified above, the choice of the $\alpha_{\rm MLT}$ does not have a high impact on the exact values of the mode frequencies. Our pulsations models, confirm that the observed frequencies $f_1$ and $f_2$ can only correspond to the radial fundamental and first overtone modes, respectively. Thus, for further seismic modelling we will only consider these radial orders.  

As a start, we used the model parameters M = 1.0 M$_{\odot},$  $Z = 0.001$, $X_0 = 0.70$ and an age of roughly 3.9 Gy\footnote{Given that SX Phe is consistent with the scenario of a blue straggler, the age in this context represents the time since its formation through mass transfer or a merger of two stars.} as given by \citet{1996A&A...312..463P}. Our best seismic model for these input values has an effective temperature of $T_{\rm eff}=7660$ K, a luminosity 
$\log {\rm L/L}_{\odot}=0.811$ and a frequency ratio of the radial fundamental mode to the first overtone of $f_1/f_2=0.780014$. The observed counterparts are $\log {\rm L/L}_{\odot}=0.842\pm0.09$ and a frequency ratio of 0.778192(6). The theoretical values for the radial fundamental and the first radial overtone are 18.193565 d$^{-1}$ and  $f_2=23.324654$ d$^{-1}$, respectively.  Both radial modes are unstable (excited). However, as one can see, the difference between the theoretical and observed value of $f_1/f_2$ is 0.001822 and is above the numerical accuracy which is around the fifth decimal place, making it clearly significant. 
In the next step,  we varied the metallicity and found a model with $Z=0.0014$
and a mass $M=1.15 M_{\odot}$, which fits better the observed frequency ratio.
The model has the following parameters: $T_{\rm eff}=8270$, $\log {\rm L/L}_{\odot}=0.984$, an age of $\sim$2.5 Gy and a frequency ratio of $f_1/f_2=0.77841$. The individual radial modes frequencies are $f_1=18.19357$ d$^{-1}$ and  $f_2=23.37268$ d$^{-1}$.
This model gives a better fit but still there is a room for improvement, especially because the radial fundamental and the first overtone in this model are both stable. 

\begin{figure} 
\centering
\includegraphics[width=\columnwidth]{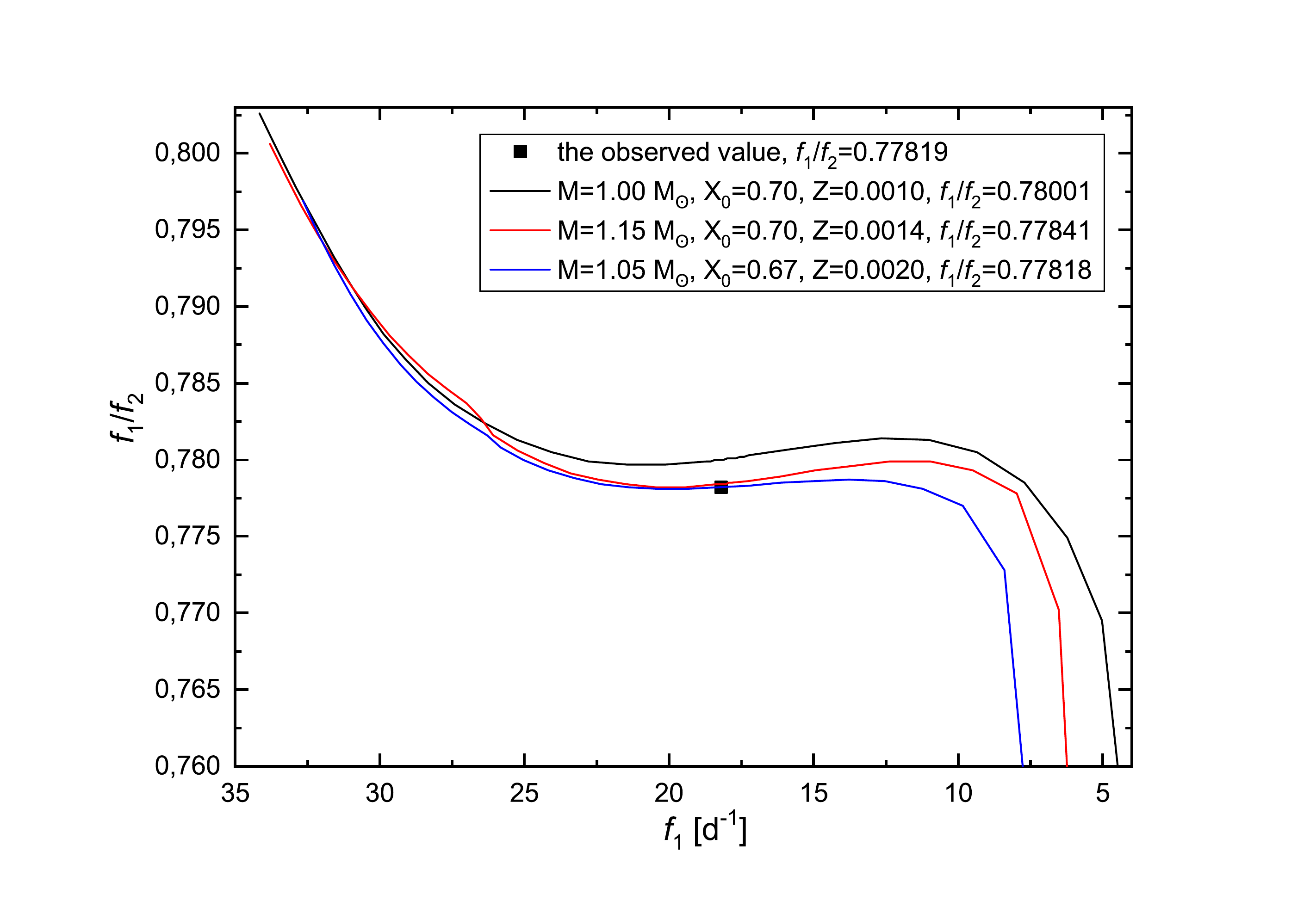} 
\caption{Petersen diagram illustrating the frequency ratio of the first radial overtone to the radial fundamental as a function of the radial fundamental mode for SX Phe. The model reproducing the observations best is depicted in blue. We refer the reader to the text for further details.}

\label{fig:petersen} 
\end{figure}

As a next step, we changed the initial abundance of hydrogen.  The model reproducing the observed frequency ratio up to the fifth decimal place has the following parameters: $M=1.05~M_{\odot}$, $X_0=0.67$, $Z=0.002$, $T_{\rm eff}=7760$ K, $\log {\rm L/L}_{\odot}=0.844$, an age of 2.8 Gy and it was interpolated to the dominant observed frequency $f_1=18.193565$ d$^{-1}$. The value of the second frequency is 23.37976 d$^{-1}$, and differs by only 0.00048 d$^{-1}$ from the observed value. Thus, the theoretical value of the frequency ratio is 0.77818, whereas the observed value is 0.77819. Taking into account the numerical accuracy, which is not better than five decimal places as specified above, we can conclude that this model reproduces the observed frequencies of the two radial modes of SX Phet well. In addition, this model also predicts instability (excitation) of both radial modes, the radial fundamental and first overtone. In Fig.~\ref{fig:petersen}, we show the evolution of the frequency ratio of the first radial overtone to the radial fundamental as a function of the radial fundamental mode (the Petersen diagram) for the three seismic models described above. The observed value is marked as an open square. 

The lower hydrogen abundance (higher helium abundance) is not unlikely  because, as other SX Phoenicis variables, SX Phe itself can be a blue straggler. Such objects are presumably formed by the merger of two stars or by interactions in a binary system (e.g., \citealt{2011AJ....142..110M, 2017MNRAS.466.1290N}). As a consequence, these stars may have enhanced helium abundance. This is also consistent with the models fitting KIC 11145123, which is also a suspected blue straggler \citep{2014MNRAS.444..102K}. In addition, the model luminosity also agrees with the Gaia luminosity and the model effective temperature is within $2\sigma$ of the spectroscopically derived value ($T_{\rm eff}=7500 \pm 150$ K, and $3.7\sigma$ within the photometrical ($T_{\rm eff}=7210 \pm 150$ K). Measured differences in effective temperature are to be expected, because as noted by \citet{1993AJ....106.2493K}, the effective temperature of a star displaying large amplitudes of pulsations depends on the pulsation phase at which the observations were made. In the case of SX Phe the spectroscopic $T_{\rm eff}$ was determined based on one spectrum taken at a single unknown pulsation phase.

\subsection{TIC~49677785 = HD~220687}
\label{subsection: EB}

HD~220687 (TIC~49677785) is a known eclipsing binary system hosting a pulsating primary \citep{2007AcA....57...61P}. Recently, HD~220687 was observed for approximately 80~d by the K2 mission \citep{2014PASP..126..398H}. From these K2 data and estimates of the primary's surface gravity and effective temperature from RAVE spectroscopy, \citet{2019AJ....157...17L} obtained a binary model with the Wilson Devinney Code, albeit without an estimate for the mass ratio from radial velocities. After subtracting their binary light curve solution, \citet{2019AJ....157...17L} found 35 significant frequencies with a S/N $\geq 4$ using a noise window of the size $5$~d$^{-1}$ centred around each peak; all of these frequencies were inferred to be $\delta$~Sct pulsations.

\begin{figure} 
\centering
\includegraphics[width=\columnwidth]{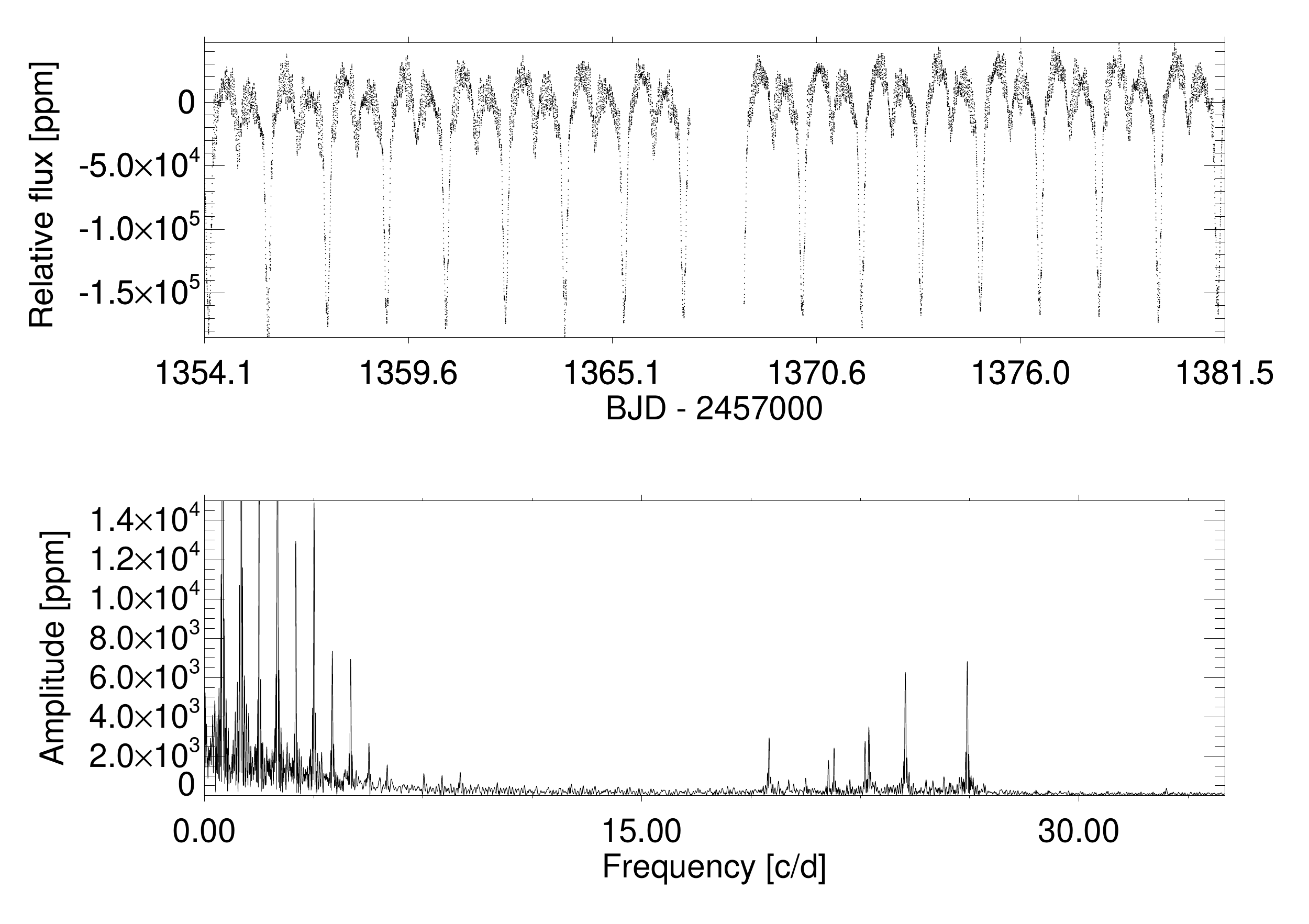} 
\caption{HD~220687. Upper panel: light curve. Lower panel: amplitude spectrum. The equally spaced peaks at low frequencies are due to the eclipses clearly visible in the light curve displayed in the upper panel.}
\label{TIC49677785} 
\end{figure}

Since the K2 observations, HD~220687 was observed by the TESS mission in Sector 2 for 27.4~d (Fig. \ref{TIC49677785}). Given the lack of an accurately constrained mass ratio and the presence of remaining harmonics after the binary model subtraction, we chose to not adopt the binary solution by Lee et al. (2019). Instead, we directly subtracted the orbital frequency ($f_{\rm orb} = 0.62725$~d$^{-1}$) and all its significant harmonics up to 80~d$^{-1}$ and then examined the remaining variance. From iterative pre-whitening pipeline we obtained 25 significant frequencies, 11 of which can be explained as harmonics or combination frequencies. We note that seven of the frequencies reported by Lee et al. (2019) are either harmonics of the orbital frequency, or can be explained as second-order combination frequencies with the orbital frequency. Finally, 12 of the 25 frequencies in the TESS light curve of HD~220687 are above the $Kepler$/K2 long-cadence Nyquist frequency of 24.49~d$^{-1}$ which emphasizes the usefulness of short-cadence observations for studying $\delta$~Sct stars.

The TESS and K2 light curves of HD~220687 also offer the only opportunity we are aware of to determine an empirical estimate for the amplitude suppression factor of pulsation modes observed in the {\it Kepler}/K2 and TESS passbands. Since the TESS passband (6000 -- 10\,000~\AA) is redder than that of {\it Kepler}/K2 (4300--8300~\AA), the photometric pulsation mode amplitudes in early-type stars appear smaller in TESS photometry compared to bluer passbands since the wavelength range mainly probes the Rayleigh-Jeans tail, as opposed to the peak of the blackbody function. The ratio of pulsation mode amplitudes in the TESS and {\it Kepler}/K2 passbands for A and F stars was estimated to be of the order 50~per~cent by \citet{2019A&A...621A.135B}, yet this value included many assumptions.

Here, we provide an empirical measurement of the ratio of pulsation mode amplitudes for the TESS and {\it Kepler}/K2 passbands using HD~220687 (TIC~49677785; EPIC~245932119). We downloaded the K2 target pixel files of HD~220687, created custom aperture masks and detrended light curve spanning approximately 80~d using the methodology described by \citet{2018A&A...616A..77B}. The calculation of an amplitude spectrum using a Discrete Fourier Transform (\citealt{1975Ap&SS..36..137D, 1985MNRAS.213..773K}) allowed the determination of the amplitudes of pulsation modes and harmonics associated with the eclipses present in the light curve. We performed independent frequency extraction using the K2 and TESS light curves via iterative pre-whitening and corrected the observed amplitudes of each extracted peak for the amplitude suppression caused by the sampling frequency of each instrument using

\begin{equation}
A = A_0~{\rm sinc}\left(\frac{\pi}{N}\right) = A_0~{\rm sinc}\left(\frac{\pi \nu}{\nu_{\rm samp}}\right) ~ ,
\label{equation: integration}
\end{equation}
	
\noindent where $A$ and $A_0$ are the observed and corrected pulsation mode amplitudes, respectively, $N$ is the number of data points per pulsation cycle, $\nu$ is the pulsation mode frequency and $\nu_{\rm samp}$ is the instrumental sampling frequency \citep{2014PhDT.......131M, 2015MNRAS.449.1004B}. 

Under the assumptions of negligible amplitude modulation of its pulsation modes (see e.g. \citealt{2016MNRAS.460.1970B}) and negligible contamination in HD~220687, which is reasonable given that HD~220687 is $V = 9.6$ and no other nearby sources are apparent, the ratio of the corrected pulsation mode amplitudes gave an empirical estimate of the TESS/{\it Kepler} amplitude suppression factor. For the dominant, independent pulsation mode amplitudes in HD~220687, we measured an average ratio of their amplitudes in the TESS to {\it Kepler}/K2 passbands to be $74 \pm 1$~per~cent. Thus, the reduction in pulsation mode amplitudes when observing A and F stars with TESS is non-negligible compared to bluer wavelength instruments such as {\it Kepler}/K2, but importantly this passband suppression does not prevent the detection of pulsation modes. On the other hand, the vast number of A and F stars observed in the 2-min TESS cadence is a major advantage for the study of high-frequency pulsations in $\delta$~Sct stars, given that the bias introduced by the 30-min {\it Kepler} cadence into the observed frequency distribution of these stars is much more significant \citep{2015MNRAS.453.2569M, 2018MNRAS.476.3169B}. We note that the suppression factor is not a constant but depends on the spectral type of the target star and to a lesser extent on the geometry of the pulsation modes (limb-darkening). 

It is worth noting that $\delta$~Sct/$\gamma$~Dor pulsators found in close binaries, are  not only in detached systems but also in those that are undergoing or have experienced mass transfer \citep{2016ApJ...826...69G, 2017ApJ...837..114G}.
Among the 303 pulsating EB identified by Gaulme \& Guzik (2019, in prep.) in about 3000 Kepler EB stars, most are $\delta$~Sct and $\gamma$~Dor pulsators (264 out of 303, ~87\%). We expect that the TESS 2-min cadence data can yield about 300 eclipsing binaries per sector, and more than 1/10 are expected to contain pulsating stars. 

\subsection{TIC 150394126 = HD 46190}
\label{subsection: pre-ms}

\begin{figure}
\begin{center}
\includegraphics[width=0.48\textwidth]{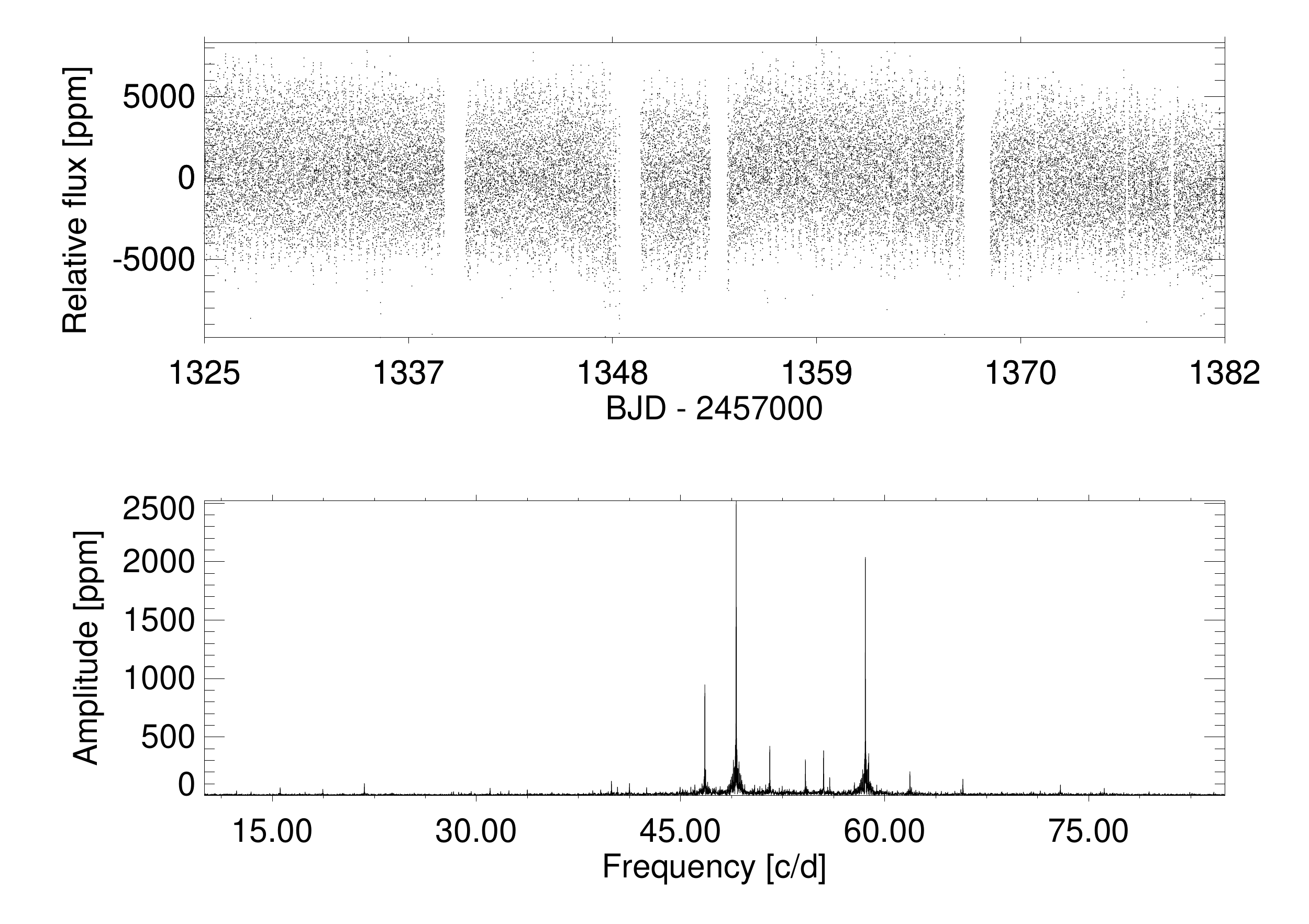}
\caption{TESS light curve (top) and amplitude spectrum (bottom) of the young star HD 46190.}
\label{fig:HD46190}
\end{center}
\end{figure}

HD~46190 was observed in sectors 1 and 2 in the TESS mission (Fig. \ref{fig:HD46190}); its light curve has a total time base of 56~d. Prior to TESS observations, this star was thought to be a young star with a spectral type of A0, but its pulsational variability was unknown. We include HD~46190 in this paper, as it is a potential pre-main-sequence $\delta$~Sct star.

 In total 65 frequencies between 15 and 76~d$^{-1}$ were  found to be statistically significant by both the KU Leuven and Aarhus ECHO pipelines. Many of these peaks can be explained by combination frequencies, depending on the number of parent modes assumed. However, some of these identified linear combinations are quite complex, and hence could also be accidental matches. 

There are no estimates of fundamental parameters or binarity from spectroscopy in the literature, and only an approximate effective temperature of $8450 \pm 600$~K in the catalogue of \citet{2018AJ....155...39O} and Gaia DR2. Our determined effective temperature based on the SED method is $8850 \pm 100$~K (Table~\ref{tab:stellar_parameters_new}).  \citet{2004ApJ...614L..77S} report a debris disk and an infrared excess around HD~49160 using Spitzer spectroscopy, and identify this star as a Vega-like main-sequence star.

The highest-amplitude p-mode frequency in HD~46190 is at 49.09 d$^{-1}$, which supports the interpretation that this star is a hot and young star near the ZAMS and the blue edge of the classical instability strip \citep{2014Sci...345..550Z}. Yet without fundamental parameters from spectroscopy, we cannot infer if HD~46190 is a pre-main-sequence or main-sequence star. Once these data become available, and should a pre-main-sequence status be established, we can determine its properties based on the range of modes excited in the star. For pre-main-sequence stars, \citet{2014Sci...345..550Z} showed that there is a significant relationship between the pulsational properties and its evolution. Such a relation could not be established for $\delta$ Sct stars on the main-sequence or the post main-sequence.

\subsection{TIC~355687188 = HD~224852}
\label{subsection: HADS}

The variability of the HADS star TIC~355687188 (HD~224852, BV~1007, NSV~14800, A8~V, $V = 10.2$) was discovered by \citet{1967IBVS..195....1S}, but the nature of this variability was only revealed from the ASAS-3 observations by \citet{2002AcA....52..397P}, who reported a period of 0.122072~d and classified the star as $\delta$~Sct. \citet{2005A&A...440.1097P} carried out a detailed analysis of the ASAS-3 data and found the star to be a double-mode pulsator with the frequency ratio of 0.77339 matching the ratio of the fundamental and first-overtone radial modes for HADS stars. In addition, the amplitude of the first overtone mode is higher than that of the fundamental mode. The full model describing the variability of the star in the ASAS data included 10 terms: the two radial mode frequencies, three harmonics of these frequencies and five combination frequencies.

\begin{figure} 
\centering
\includegraphics[width=\columnwidth]{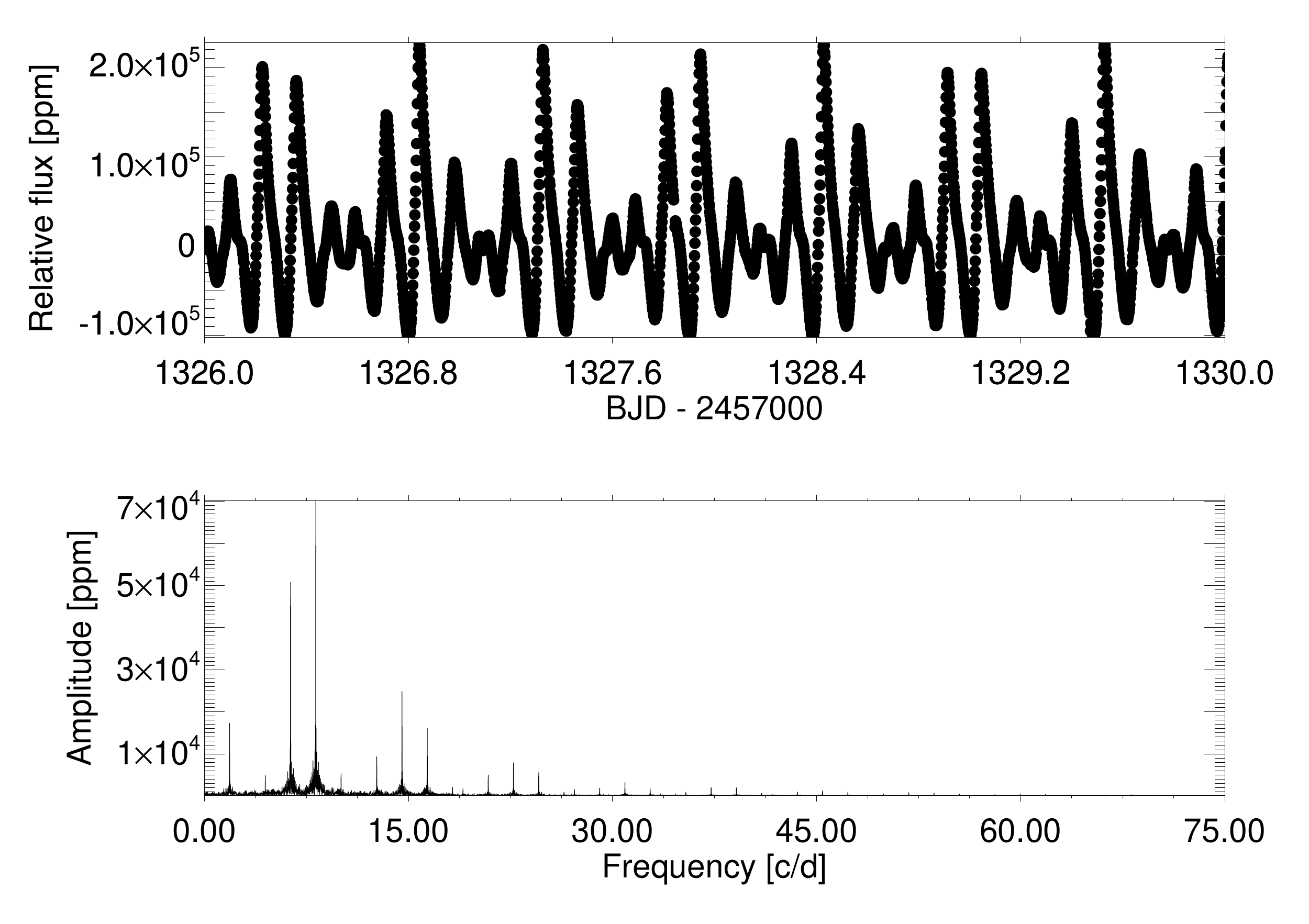} 
\caption{Upper panel: A 4-day section of the TESS light curve of HD~224852. The highly non-sinusoidal light curve and the beating between the two dominant modes are clearly visible. Lower panel: amplitude spectrum. The large majority of these peaks are combination frequencies and harmonics of the 4 independent modes. Although there are statistically significant peaks up to frequencies of 75 d$^{-1}$, for illustrative reasons we only plot up to 75 d$^{-1}$. }

\label{fig-lc} 
\end{figure}

HD~224852 was observed in Sectors 1 and 2 of the TESS mission with 2-min cadence; these data span 56.2~d with small ($\sim$1-d) gaps in the middle and between each observing sector. A 4-d section of the TESS light curve is shown in Fig.~\ref{fig-lc}, in which the strong beating of the two radial modes can be seen. Frequency analysis of the TESS data indicates that the observed variability in HD~224852 can be accounted for by only four independent modes, their harmonics and combination terms. In total, the model including all significant frequencies contains 96 sinusoidal terms yet only four independent pulsation modes (Table~\ref{tab-fs}), 12 harmonics (of $f_1$ and $f_2$ only), and 80 combination frequencies.

Both radial modes have non-sinusoidal light curves, which is in accordance with the findings of \citet{2005A&A...440.1097P}. The first overtone mode has a larger amplitude than the fundamental mode, which is more commonly found for HADS stars in the LMC \citep{2010AcA....60....1P} as opposed to the Milky Way. We infer that the independent frequencies $f_3$ and $f_4$ are non-radial modes. However, we cannot exclude that $f_4$ may be a higher-overtone radial mode. This can only be verified by detailed seismic modelling. We also note that no low-frequency modes were detected in HD~224852 above the detection threshold of 0.023~ppt corresponding to S/N $\approx$ 2.8. The two radial modes show slight amplitude modulation during the TESS observations. In particular, the amplitude of $f_1$ drops by about 0.7~per~cent and $f_2$ by about 0.4~per~cent during this time. This effect has also been observed for other high-amplitude radial pulsators, such the HADS star KIC~5950759 \citep{2017ampm.book.....B}.

\begin{table}
\caption{Characterization of the amplitude spectrum of  HD~224852.}
\label{tab-fs}
\begin{tabular}{cr@{.}lccr}
\hline
Mode & \multicolumn{2}{c}{Frequency} & Harmonics & Combinations & S/N \\
     & \multicolumn{2}{c}{[d$^{-1}$]}& detected & detected  &\\
\hline
$f_1$ & 8&1919470(07)  & 7 & 78 & 8855.39\\
$f_2$ & 6&3355745(10) & 5 & 79 & 6414.12\\
$f_3$ & 7&9646(05) & 0 & 5 & 14.47\\
$f_4$ & 13&1380(18) & 0 & 0 & 4.28\\
\hline
\end{tabular}
\end{table}

\subsection{TIC 219234987 = $\gamma$~Dor}
\label{subsection: gdor}

$\gamma$ Dor is a bright ($V=4.20$) F1V \citep{2006AJ....132..161G} and is eponymous for an entire class of gravity/Rossby modes pulsators, the $\gamma$~Dor stars. Although this paper concentrates only on sectors 1 and 2, we decided to make an exception and to also include this prototype in this first-light article.

TESS observed $\gamma$~Dor in sectors 3, 4 and 5. The light curve was extracted by performing aperture photometry on target pixel files downloaded from MAST using the {\sc LIGHTCURVE} python package \citep{geert_barentsen_2019_2565212}. Bright as it is, $\gamma$~Dor saturates TESS CCDs. We settled on a classic approach for the preliminary analysis presented here. The photometric data extraction will be refined in the future using the halo photometry technique \citep{2017MNRAS.471.2882W}. The frequency analysis was subsequently done following an iterative pre-whitening procedure with the {\sc PERIOD04} software. The light curve and amplitude spectrum are plotted in Fig. \ref{fig:gDor} along with the 17 extracted frequencies with S/N $\geq$ 4.

$\gamma$~Dor has been extensively studied in the literature. Ground-based photometric \citep{1992Obs...112...53C,1994MNRAS.267..103B,1994MNRAS.270..905B,2008A&A...492..167T} and spectroscopic observations \citep{1996MNRAS.281.1315B,2018MNRAS.475.3813B} dedicated to this star allowed the identification of a handful of frequencies. \citet{2018MNRAS.475.3813B} recently performed a comprehensive study of all available ground-based data. They found four consistent frequencies (1.3209, 1.3641, 1.4742, 1.8783 d$^{-1}$) that can all be confirmed with TESS data. The frequency at 0.31672 d$^{-1}$, previously identified as a daily alias, is also present here, suggesting that it is indeed intrinsic to the star. We also corroborate the 2.7428~d$^{-1}$ combination frequency. The amplitude spectrum shows amplitude excess at multiples of the main frequency group around $\sim$1.3 d$^{-1}$, which is most likely due to harmonic and combination frequencies \cite[e.g.][]{2015MNRAS.450.3015K, 2018MNRAS.477.2183S}, however, we cannot explore this further given the limited frequency resolution. Nevertheless, based on the observations and theory by \citet{2019MNRAS.487..782L} and \citet{2018MNRAS.477.2183S}, we can speculate that the power excess around 1.4 d$^{-1}$ and 2.8 d$^{-1}$ could correspond to prograde l=1 and l=2 g modes, respectively. Further observations are needed to resolve the modes and confirm this hypothesis. No significant p~modes were detected, ruling out a hybrid nature. TESS observations in synergy with available spectroscopic \citep[e.g.][]{2006AJ....132..161G,2012A&A...542A.116A} and astrometric \citep{2018A&A...616A...1G} measurements will be presented in a forthcoming paper. 

{
\begin{figure}
                \resizebox{\hsize}{!}{\includegraphics{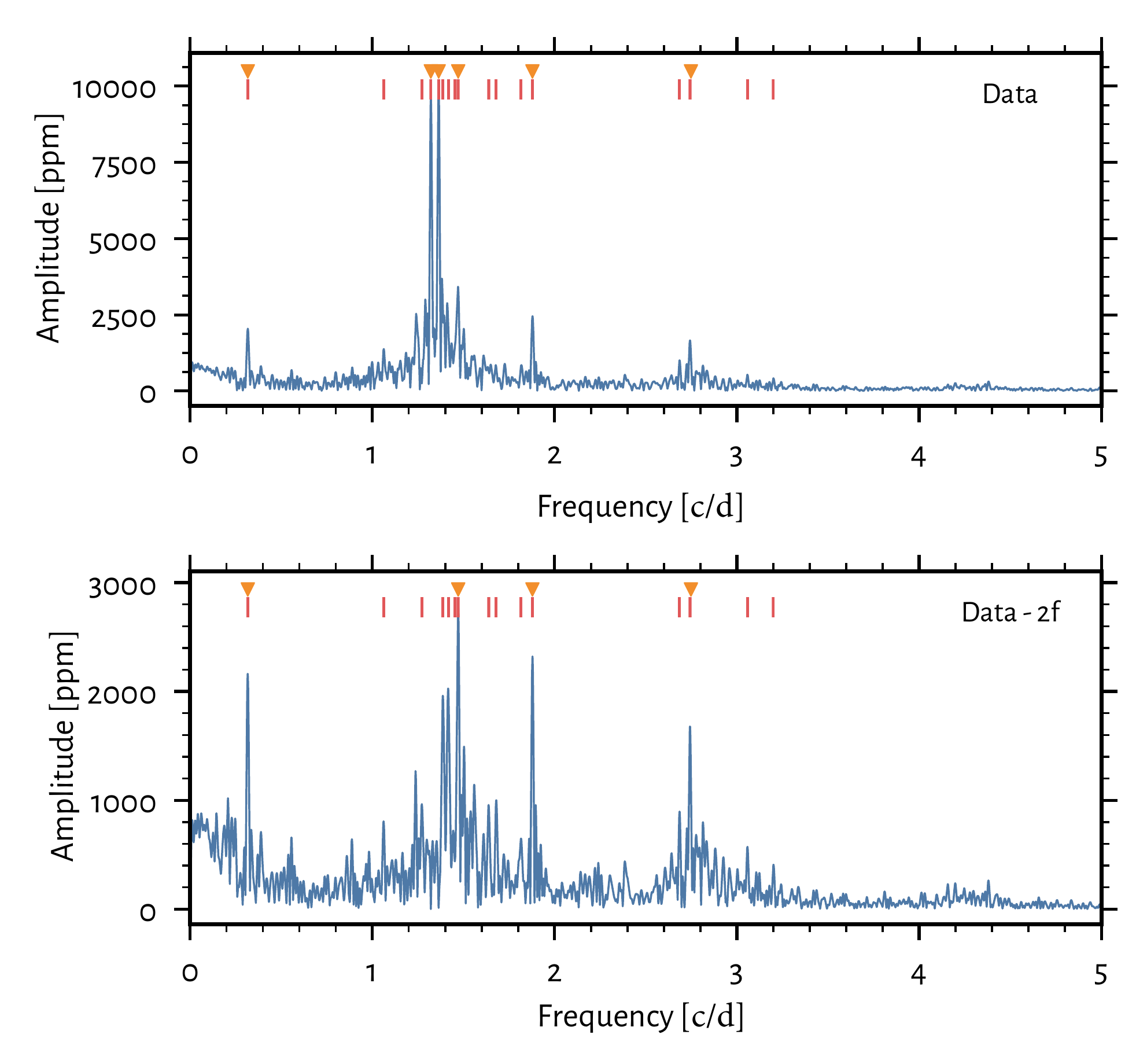}}
                \caption{Amplitude Spectra of $\gamma$~Doradus from TESS observations before ({\it top}) and after pre-whitening the first two frequencies ({\it bottom}). Red vertical bars mark the 17 frequencies with S/N $> 4$ extracted from TESS data. Orange triangles indicate frequencies found in \citet{2018MNRAS.475.3813B}.  }
                \label{fig:gDor}
\end{figure}

\subsection{TIC 154842794 = $\upi$~PsA}
\label{subsection: gamma Dor}

TIC 154842794 ($\upi$~PsA, HD~217792) is a F1V type star \citep{2006AJ....132..161G} and a newly classified $\gamma$ Dor pulsator. \citep{2015ApJ...804..146D} find the following stellar parameters: $T_{\rm eff} = 7400 \pm 250$~K, $\log\,g = 4.3$, and $M = 1.5$~M$_{\odot}$, with the effective temperature agreeing well with our determination of $T_{\rm eff} = 7440 \pm 110$~K. Additional values that may be relevant for the reader are $v_{\rm turb}$ = 2 km~s$^{-1}$ from \citet{2006AJ....132..161G} and [Fe/H] $= -0.25 \pm 0.06$ from \citet{2016ApJ...826..171G} and a radius of $R = 0.87$~R$_{\odot}$ from \citet{2001A&A...367..521P}. We note that the radius of this star may be underestimated, as it is unrealistically low for an early F type star on or near the main-sequence. 

\citet{1965IBVS...86....1S} classified the star as a Cepheid with a period of 7.975 d and amplitude of 0.3 mag. It is thus likely the complex beating of the low-amplitude pulsations resulted in the misclassification. $\pi$ PsA is a known spectroscopic binary \citep{1961MNRAS.123..183B} with a low mass companion and an orbital period of 178~d \citep{2012AJ....144...62A,2004A&A...424..727P} and eccentricity 0.53 \citep{2005ApJ...629..507A}. The velocity amplitude of the primary is 21.3~km\,s$^{-1}$ \citep{2004A&A...424..727P}. It has also been flagged as a possible astrometric binary \citep{2005A&A...442..365J}. None of the frequencies are multiples of the orbital frequency. The star has an IR excess \citep{2014ApJS..211...25C,2007ApJ...658.1289T}, indicating the presence of a debris disk around the star and has been identified as a variable with a frequency of 0.94305~d$^{-1}$ \citep{2002MNRAS.331...45K}. This frequency was confirmed using the original Hipparcos data. The sole Hipparcos frequency, calculated from 104 data-points, is inconsistent with any of the frequencies found in the TESS data. 

The TESS light curve indicates the presence of a flare at BJD = 2458370.162 (Fig. \ref{fig:flare}), which might originate either from a background star or a possible companion. Detailed analyses as described in \cite{2017MNRAS.466.3060P} are required to identify the flare origin, but are out of the scope of this paper.

\begin{figure}
                \resizebox{\hsize}{!}{\includegraphics{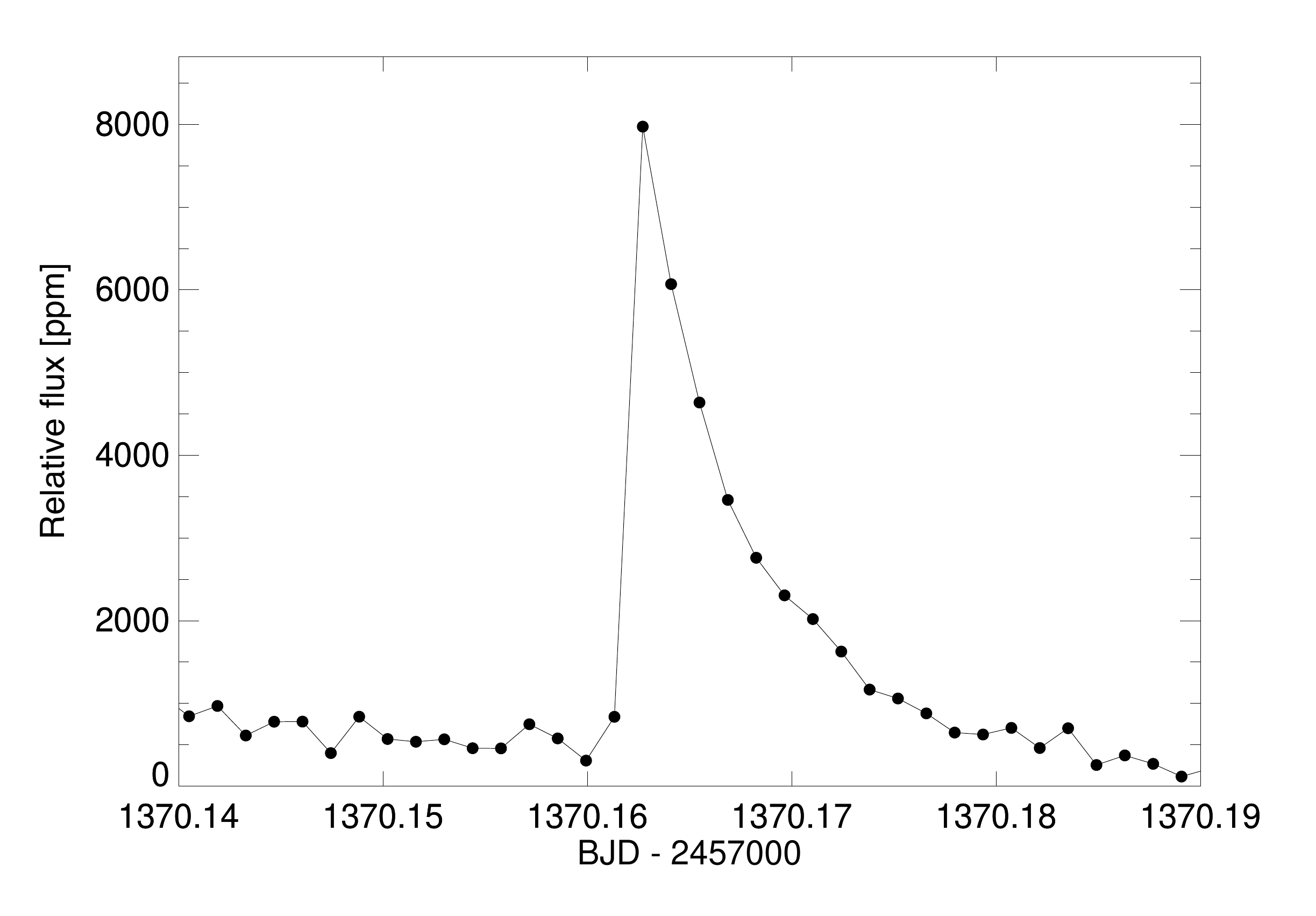}}
                \caption{Flare observed in the TESS data of $\upi$~PsA.  }
                \label{fig:flare}
\end{figure}

Based on TESS observations we find six statistically significant frequencies, which are 0.6721(2), 1.0242(2), 1.2688(3), 1.3814(3), 1.6385(5), 1.8376(3) d$^{-1}$, with additional peaks that are unfortunately unresolved, as this star was only observed for 27 days (Fig. \ref{TIC154842794}). Based on the extracted peaks no clear period spacing can be detected. This shows that $\gamma$~Dor stars require longer observations.

\begin{figure} 
\centering
\includegraphics[width=\columnwidth]{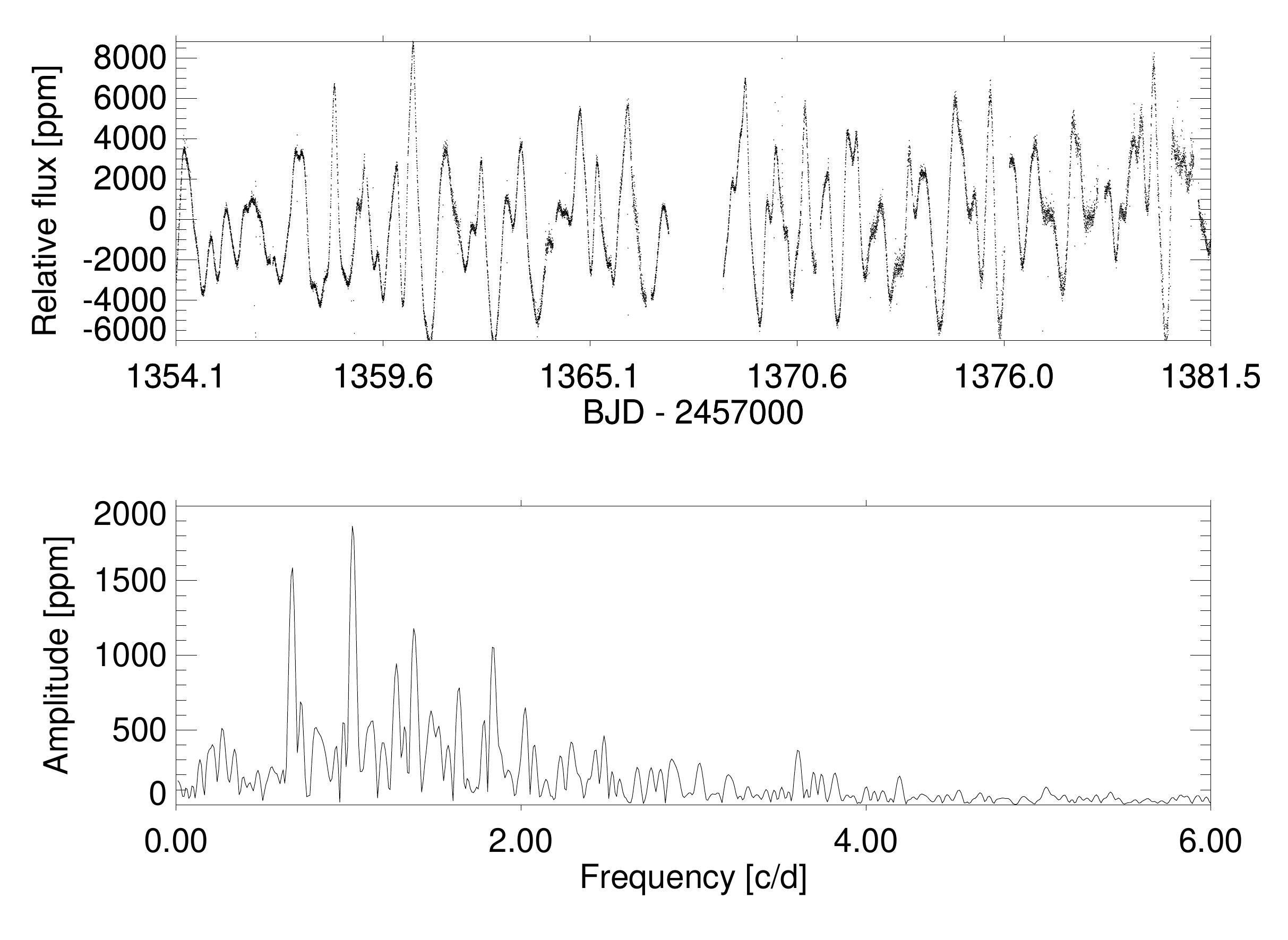} 
\caption{$\upi$~PsA. Upper panel: here we show the light curve that clearly resembles the variability of a typical $\gamma$ Dor star. Lower panel: Fourier spectrum of the TESS light curve. Here we find 6 statistically significant frequencies. See text for details. }
\label{TIC154842794} 
\end{figure}

\section{Conclusions}
\label{section: conclusions}

We highlighted the different aspects of the pulsational variability of intermediate-mass A and F stars observed by the TESS mission in its first two sectors. The 2-min cadence of the TESS mission data is particularly useful for studying high-frequency pulsations in $\delta$~Sct stars; many modes are found which are often higher than the Nyquist frequency of 24.5~d$^{-1}$ provided by 29.4-min cadence data of the {\it Kepler} mission (e.g., \citealt{2018MNRAS.476.3169B, 2017EPJWC.16003001M}). The homogenous, all-sky sample of $\delta$~Sct stars observed by TESS is a significant advantage over previous ensembles of stars observed by, for example, the {\it Kepler} mission, and offers opportunities to test the physics of pulsation driving for a range in stellar mass, metallicity and rotation. In addition we used Gaia DR2 data and SED $T_{\rm eff}$ values to calculate stellar luminosities and place our sample in the HR~diagram (Fig.~\ref{fig:stars}). 

In this study, we have also demonstrated that using models with time-dependent non-local convection treatment and mimicking He depletion in the outer envelopes as expected for Am stars, we can explain the driving of pulsations in these stars. The distribution of 31 known Am stars observed by TESS in the HR~diagram is consistent with our theoretical predictions, which show that turbulent pressure plays a very important role. Driving pulsations in Am stars has been a long-lasting mystery as the $\kappa$ mechanism alone could not explain the observations in the presence of He depletion through gravitational settling. Furthermore, the parameter space for which pulsation driving by turbulent pressure is predicted to be at its strongest is shown to be in the centre of the classical instability strip, corresponding to $T_{\rm eff} \simeq 7500$~K. This is compatible with the parameter space for which the pulsator fraction also reaches its maximum of $\sim 70$~per~cent within the classical instability strip \citep{2019MNRAS.tmp..585M}.

We highlighted the unprecedented asteroseismic potential that the TESS mission data provides for A and F stars, especially for pulsators that have previously not been observed with high-precision space telescopes such as the brightest known $\delta$ Sct and $\gamma$ Dor stars (e.g., $\alpha$ Pic, $\beta$ Pic, $\gamma$ Dor, etc). In addition, the near full-sky TESS observations will allow us to perform unbiased variability studies of A and F stars in general, but also in particular of, e.g., pre main-sequence and chemically peculiar stars, such as Am and Ap stars. We provided case studies of different variability types, including pulsators in eclipsing binaries, high-amplitude pulsators, chemically peculiar stars, and potential pre-main-sequence  stars. We also show that the rather short observing periods of each sector (27 days) impose in some cases severe constraints on our ability to perform asteroseismic studies of, e.g., $\gamma$ Dor, $\delta$ Sct or hybrid stars due to their unresolved pulsation spectra. A limited frequency resolution will prevent us from detecting consecutive radial orders of gravity and Rossby modes as well as  rotational splitting in pressure modes. Future studies using longer TESS light curves (obtained during the nominal and a possible extended mission), which are necessary for detailed asteroseismic modelling, will naturally build on this work and improve stellar structure and evolution theory for intermediate-mass stars.


\section*{Acknowledgements}

We thank the referee for useful comments and discussions. This paper includes data collected by the TESS mission. Funding for the TESS mission is provided by the NASA Explorer Program. Funding for the TESS Asteroseismic Science Operations Centre is provided by the Danish National Re- search Foundation (Grant agreement no.: DNRF106), ESA PRODEX (PEA 4000119301) and Stellar Astrophysics Centre (SAC) at Aarhus University. We thank the TESS and TASC/TASOC teams for their support of the present work. This research has made use of the SIMBAD database, operated at CDS, Strasbourg, France. Some of the data presented in this paper were obtained from the Mikulski Archive for Space Telescopes (MAST). STScI is operated by the Association of Universities for Research in Astronomy, Inc., under NASA contract NAS5-2655.

Funding for the Stellar Astrophysics Centre is provided by The Danish National Research Foundation (Grant agreement no.: DNRF106). MC was supported by FCT - Funda\c c\~ao para a Ci\^encia e a Tecnologia  through national funds and by FEDER through COMPETE2020 - Programa Operacional Competitividade e Internacionaliza\c c\~ao by these grants: UID/FIS/04434/2019, PTDC/FIS-AST/30389/2017 \& POCI-01-0145-FEDER-030389. MC is supported in the form of work contract funded by national funds through FCT (CEECIND/02619/2017).  JDD acknowledges support from the Polish National Science Center (NCN), grant no. 2018/29/B/ST9/02803. AGH acknowledges funding support from Spanish public funds for research under projects ESP2017-87676-2-2 and ESP2015-65712-C5-5-R of the Spanish Ministry of Science and Education. FKA gratefully acknowledge funding through grant 2015/18/A/ST9/00578 of the Polish National Science Centre (NCN). JPe acknowledges funding support from the NSF REU program under grant number PHY-1359195. APi and KK acknowledge support provided by the Polish National Science Center (NCN) grant No. 2016/21/B/ST9/01126. This project has been supported by the Lend\"ulet Program of the Hungarian Academy of Sciences, project No. LP2018-7/2018, and by the support provided from the National Research, Development and Innovation Fund of Hungary, financed under the K16 funding scheme, project No. 115709. JCS acknowledges funding support from Spanish public funds for research under projects ESP2017-87676-2-2 and ESP2015- 65712-C5-5-R, and from project RYC-2012-09913 under the `Ram\'on y Cajal' program of the Spanish Ministry of Science and Education. The research leading to these results has (partially) received funding from the European Research Council (ERC) under the European Union's Horizon 2020 research and innovation programme (grant agreement N$^\circ$670519: MAMSIE), from the KU\,Leuven Research Council (grant C16/18/005: PARADISE), from the Research Foundation Flanders (FWO) under grant agreement G0H5416N (ERC Runner Up Project), as well as from the BELgian federal Science Policy Office (BELSPO) through PRODEX grant PLATO. acknowledges support by the Spanish State Research Agency (AEI) through project No. ESP2017-87676-C5-1-R and No MDM-2017-0737 Unidad de Excelencia Mar\'ia de Maeztut-Centro de Astrobiolog\'ia (CSIC-INTA).ZsB acknowledges the support provided from the National Research, Development and Innovation Fund of Hungary, financed under the PD1717 funding scheme, project no. PD-123910. DLB acknowledges support from the Whitaker Foundation. SC gratefully acknowledge funding through grant 2015/18/A/ST9/00578 of the Polish National Science Centre (NCN). CCL gratefully acknowledges support from the Natural Sciences and Engineering Research Council of Canada. GMM acknowledges fundingby the STFC consolidated grant ST/R000603/1. RMO, SC and DR were supported in this work by the "Programme National de Physique Stellaire" (PNPS) of CNRS/INSU co-funded by CEA and CNES. IS acknowledges the partial support of projects DN 08-1/2016 and DN 18/13-12.12.2017. PS acknowledges financial support by the Polish NCN grant 2015/18/A/ST9/00578. MS acknowledges the Postdoc@MUNI project CZ.02.2.69/0.0/0.0/16-027/0008360. JPG, JRR and MLM acknowledges funding support from Spanish public funds for research under project ESP2017-87676-C5-5-R and from the State Agency for Research of the Spanish MCIU through the "Center of Excellence Severo Ochoa" award for the Instituto de Astrof\'{\i}sica de Andaluc\'{\i}a (SEV-2017-0709). JAE acknowledges STFC for funding support (reference ST/N504348/1). LFM acknowledges the financial support from the DGAPA-UNAM under grant PAPIIT IN100918. DM acknowledges his work as part of the research activity of the National Astronomical Research Institute of Thailand (NARIT), which is supported by the Ministry of Science and Technology of Thailand. MR acknowledges the support of the French Agence Nationale de la Recherche (ANR), under grant ESRR (ANR-16-CE31-0007-01). We acknowledge the International Space Science Institute (ISSI) for supporting the SoFAR international team \url{http://www.issi.unibe.ch/teams/sofar/}. SBF acknowledges support by the Spanish State Research Agency (AEI) through project No. `ESP2017-87676-C5-1-R' and No MDM-2017-0737 Unidad de Excelencia `Mar\'ia de Maeztu'-Centro de Astrobiolog\'ia (CSIC-INTA).

This work has made use of data from the European Space Agency (ESA) mission Gaia (https://www.cosmos.esa.int/gaia), processed by the Gaia Data Processing and Analysis Consortium (DPAC, https://www.cosmos.esa.int/web/gaia/dpac/consortium). Funding for the DPAC has been provided by national institutions, in particular the institutions participating in the Gaia Multilateral Agreement. The research leading to these results has received funding from the European Research Council (ERC) under the European Union’s Horizon 2020 research and innovation programme (grant agreement No.~670519: MAMSIE) and from the Fonds Wetenschappelijk Onderzoek - Vlaanderen (FWO) under the grant agreement G0H5416N (ERC Opvangproject). This research has made use of the VizieR catalogue access tool, CDS, Strasbourg, France (DOI: 10.26093/cds/vizier). The original description of the VizieR service was published in A\&AS 143, 23.

This publication makes use of data products from the Two Micron All
Sky Survey, which is a joint project of the University of Massachusetts
and the Infrared Processing and Analysis Center/California Institute of
Technology, funded by the National Aeronautics and Space Administration
and the National Science Foundation.




\bibliographystyle{mnras}
\bibliography{antoci} 

\begin{thebibliography}{}
\makeatletter
\relax
\def\mn@urlcharsother{\let\do\@makeother \do\$\do\&\do\#\do\^\do\_\do\%\do\~}
\def\mn@doi{\begingroup\mn@urlcharsother \@ifnextchar [ {\mn@doi@}
  {\mn@doi@[]}}
\def\mn@doi@[#1]#2{\def\@tempa{#1}\ifx\@tempa\@empty \href
  {http://dx.doi.org/#2} {doi:#2}\else \href {http://dx.doi.org/#2} {#1}\fi
  \endgroup}
\def\mn@eprint#1#2{\mn@eprint@#1:#2::\@nil}
\def\mn@eprint@arXiv#1{\href {http://arxiv.org/abs/#1} {{\tt arXiv:#1}}}
\def\mn@eprint@dblp#1{\href {http://dblp.uni-trier.de/rec/bibtex/#1.xml}
  {dblp:#1}}
\def\mn@eprint@#1:#2:#3:#4\@nil{\def\@tempa {#1}\def\@tempb {#2}\def\@tempc
  {#3}\ifx \@tempc \@empty \let \@tempc \@tempb \let \@tempb \@tempa \fi \ifx
  \@tempb \@empty \def\@tempb {arXiv}\fi \@ifundefined
  {mn@eprint@\@tempb}{\@tempb:\@tempc}{\expandafter \expandafter \csname
  mn@eprint@\@tempb\endcsname \expandafter{\@tempc}}}

\bibitem[\protect\citeauthoryear{{Abt}}{{Abt}}{2000}]{2000ApJ...544..933A}
{Abt} H.~A.,  2000, \mn@doi [\apj] {10.1086/317257}, \href
  {http://adsabs.harvard.edu/abs/2000ApJ...544..933A} {544, 933}

\bibitem[\protect\citeauthoryear{{Abt}}{{Abt}}{2005}]{2005ApJ...629..507A}
{Abt} H.~A.,  2005, \mn@doi [\apj] {10.1086/431207}, \href
  {http://adsabs.harvard.edu/abs/2005ApJ...629..507A} {629, 507}

\bibitem[\protect\citeauthoryear{{Aerts}, {Christensen-Dalsgaard}  \&
  {Kurtz}}{{Aerts} et~al.}{2010}]{2010aste.book.....A}
{Aerts} C.,  {Christensen-Dalsgaard} J.,   {Kurtz} D.~W.,  2010,
  {Asteroseismology}.
Springer

\bibitem[\protect\citeauthoryear{{Aerts}, {Van Reeth}  \& {Tkachenko}}{{Aerts}
  et~al.}{2017}]{2017ApJ...847L...7A}
{Aerts} C.,  {Van Reeth} T.,   {Tkachenko} A.,  2017, \mn@doi [\apjl]
  {10.3847/2041-8213/aa8a62}, \href
  {http://adsabs.harvard.edu/abs/2017ApJ...847L...7A} {847, L7}

\bibitem[\protect\citeauthoryear{{Aerts}, {Mathis}  \& {Rogers}}{{Aerts}
  et~al.}{2018a}]{2018arXiv180907779A}
{Aerts} C.,  {Mathis} S.,   {Rogers} T.,  2018a, \araa, in press, \href
  {http://adsabs.harvard.edu/abs/2018arXiv180907779A} {}

\bibitem[\protect\citeauthoryear{{Aerts} et~al.,}{{Aerts}
  et~al.}{2018b}]{2018ApJS..237...15A}
{Aerts} C.,  et~al., 2018b, \mn@doi [\apjs] {10.3847/1538-4365/aaccfb}, \href
  {http://adsabs.harvard.edu/abs/2018ApJS..237...15A} {237, 15}

\bibitem[\protect\citeauthoryear{{Allen}, {Burgasser}, {Faherty}  \&
  {Kirkpatrick}}{{Allen} et~al.}{2012}]{2012AJ....144...62A}
{Allen} P.~R.,  {Burgasser} A.~J.,  {Faherty} J.~K.,   {Kirkpatrick} J.~D.,
  2012, \mn@doi [\aj] {10.1088/0004-6256/144/2/62}, \href
  {http://adsabs.harvard.edu/abs/2012AJ....144...62A} {144, 62}

\bibitem[\protect\citeauthoryear{{Ammler-von Eiff} \& {Reiners}}{{Ammler-von
  Eiff} \& {Reiners}}{2012}]{2012A&A...542A.116A}
{Ammler-von Eiff} M.,  {Reiners} A.,  2012, \mn@doi [\aap]
  {10.1051/0004-6361/201118724}, \href
  {http://adsabs.harvard.edu/abs/2012A%26A...542A.116A} {542, A116}

\bibitem[\protect\citeauthoryear{{Ammons}, {Robinson}, {Strader}, {Laughlin},
  {Fischer}  \& {Wolf}}{{Ammons} et~al.}{2006}]{Ammons2006}
{Ammons} S.~M.,  {Robinson} S.~E.,  {Strader} J.,  {Laughlin} G.,  {Fischer}
  D.,   {Wolf} A.,  2006, \mn@doi [\apj] {10.1086/498490}, \href
  {https://ui.adsabs.harvard.edu/\#abs/2006ApJ...638.1004A} {638, 1004}

\bibitem[\protect\citeauthoryear{{Andrievsky} et~al.,}{{Andrievsky}
  et~al.}{2002}]{2002A&A...396..641A}
{Andrievsky} S.~M.,  et~al., 2002, \mn@doi [\aap] {10.1051/0004-6361:20021423},
  \href {http://adsabs.harvard.edu/abs/2002A%26A...396..641A} {396, 641}

\bibitem[\protect\citeauthoryear{{Antoci} et~al.,}{{Antoci}
  et~al.}{2011}]{2011Natur.477..570A}
{Antoci} V.,  et~al., 2011, \mn@doi [\nat] {10.1038/nature10389}, \href
  {https://ui.adsabs.harvard.edu/#abs/2011Natur.477..570A} {477, 570}

\bibitem[\protect\citeauthoryear{{Antoci} et~al.,}{{Antoci}
  et~al.}{2014}]{2014ApJ...796..118A}
{Antoci} V.,  et~al., 2014, \mn@doi [\apj] {10.1088/0004-637X/796/2/118}, \href
  {https://ui.adsabs.harvard.edu/#abs/2014ApJ...796..118A} {796, 118}

\bibitem[\protect\citeauthoryear{{Arenou}, {Grenon}  \& {Gomez}}{{Arenou}
  et~al.}{1992}]{arenou92}
{Arenou} F.,  {Grenon} M.,   {Gomez} A.,  1992, \aap, \href
  {http://esoads.eso.org/abs/1992A%26A...258..104A} {258, 104}

\bibitem[\protect\citeauthoryear{{Arenou} et~al.,}{{Arenou}
  et~al.}{2018}]{2018A&A...616A..17A}
{Arenou} F.,  et~al., 2018, \mn@doi [\aap] {10.1051/0004-6361/201833234}, \href
  {https://ui.adsabs.harvard.edu/abs/2018A&A...616A..17A} {616, A17}

\bibitem[\protect\citeauthoryear{{Asplund}, {Grevesse}, {Sauval}  \&
  {Scott}}{{Asplund} et~al.}{2009}]{2009ARA&A..47..481A}
{Asplund} M.,  {Grevesse} N.,  {Sauval} A.~J.,   {Scott} P.,  2009, \mn@doi
  [\araa] {10.1146/annurev.astro.46.060407.145222}, \href
  {http://adsabs.harvard.edu/abs/2009ARA%26A..47..481A} {47, 481}

\bibitem[\protect\citeauthoryear{{Auri{\`e}re} et~al.,}{{Auri{\`e}re}
  et~al.}{2010}]{2010A&A...523A..40A}
{Auri{\`e}re} M.,  et~al., 2010, \mn@doi [\aap] {10.1051/0004-6361/201014848},
  \href {http://adsabs.harvard.edu/abs/2010A%26A...523A..40A} {523, A40}

\bibitem[\protect\citeauthoryear{{Auvergne} et~al.,}{{Auvergne}
  et~al.}{2009}]{2009A&A...506..411A}
{Auvergne} M.,  et~al., 2009, \mn@doi [\aap] {10.1051/0004-6361/200810860},
  \href {http://adsabs.harvard.edu/abs/2009A%26A...506..411A} {506, 411}

\bibitem[\protect\citeauthoryear{{Baglin}, {Breger}, {Chevalier}, {Hauck}, {Le
  Contel}, {Sareyan}  \& {Valtier}}{{Baglin}
  et~al.}{1973}]{1973A&A....23..221B}
{Baglin} A.,  {Breger} M.,  {Chevalier} C.,  {Hauck} B.,  {Le Contel} J.~M.,
  {Sareyan} J.~P.,   {Valtier} J.~C.,  1973, \aap, \href
  {http://adsabs.harvard.edu/abs/1973A%26A....23..221B} {23, 221}

\bibitem[\protect\citeauthoryear{Bailer-Jones}{Bailer-Jones}{2011}]{bailer2011}
Bailer-Jones C.~A.,  2011, Monthly Notices of the Royal Astronomical Society,
  411, 435

\bibitem[\protect\citeauthoryear{{Balmforth}, {Cunha}, {Dolez}, {Gough}  \&
  {Vauclair}}{{Balmforth} et~al.}{2001}]{2001MNRAS.323..362B}
{Balmforth} N.~J.,  {Cunha} M.~S.,  {Dolez} N.,  {Gough} D.~O.,   {Vauclair}
  S.,  2001, \mn@doi [\mnras] {10.1046/j.1365-8711.2001.04182.x}, \href
  {http://adsabs.harvard.edu/abs/2001MNRAS.323..362B} {323, 362}

\bibitem[\protect\citeauthoryear{{Balona}}{{Balona}}{2014}]{2014MNRAS.439.3453B}
{Balona} L.~A.,  2014, \mn@doi [\mnras] {10.1093/mnras/stu193}, \href
  {http://adsabs.harvard.edu/abs/2014MNRAS.439.3453B} {439, 3453}

\bibitem[\protect\citeauthoryear{{Balona}}{{Balona}}{2016}]{2016MNRAS.459.1097B}
{Balona} L.~A.,  2016, \mn@doi [\mnras] {10.1093/mnras/stw671}, \href
  {http://cdsads.u-strasbg.fr/abs/2016MNRAS.459.1097B} {459, 1097}

\bibitem[\protect\citeauthoryear{{Balona} \& {Dziembowski}}{{Balona} \&
  {Dziembowski}}{2011}]{2011MNRAS.417..591B}
{Balona} L.~A.,  {Dziembowski} W.~A.,  2011, \mn@doi [\mnras]
  {10.1111/j.1365-2966.2011.19301.x}, \href
  {http://adsabs.harvard.edu/abs/2011MNRAS.417..591B} {417, 591}

\bibitem[\protect\citeauthoryear{{Balona}, {Hearnshaw}, {Koen}, {Collier},
  {Machi}, {Mkhosi}  \& {Steenberg}}{{Balona}
  et~al.}{1994a}]{1994MNRAS.267..103B}
{Balona} L.~A.,  {Hearnshaw} J.~B.,  {Koen} C.,  {Collier} A.,  {Machi} I.,
  {Mkhosi} M.,   {Steenberg} C.,  1994a, \mn@doi [\mnras]
  {10.1093/mnras/267.1.103}, \href
  {http://adsabs.harvard.edu/abs/1994MNRAS.267..103B} {267, 103}

\bibitem[\protect\citeauthoryear{{Balona}, {Krisciunas}  \& {Cousins}}{{Balona}
  et~al.}{1994b}]{1994MNRAS.270..905B}
{Balona} L.~A.,  {Krisciunas} K.,   {Cousins} A.~W.~J.,  1994b, \mn@doi
  [\mnras] {10.1093/mnras/270.4.905}, \href
  {http://adsabs.harvard.edu/abs/1994MNRAS.270..905B} {270, 905}

\bibitem[\protect\citeauthoryear{{Balona} et~al.,}{{Balona}
  et~al.}{1996}]{1996MNRAS.281.1315B}
{Balona} L.~A.,  et~al., 1996, \mn@doi [\mnras] {10.1093/mnras/281.4.1315},
  \href {http://adsabs.harvard.edu/abs/1996MNRAS.281.1315B} {281, 1315}

\bibitem[\protect\citeauthoryear{{Balona} et~al.,}{{Balona}
  et~al.}{2012}]{2012MNRAS.419.3028B}
{Balona} L.~A.,  et~al., 2012, \mn@doi [\mnras]
  {10.1111/j.1365-2966.2011.19939.x}, \href
  {http://cdsads.u-strasbg.fr/abs/2012MNRAS.419.3028B} {419, 3028}

\bibitem[\protect\citeauthoryear{{Balona}, {Daszy{\'n}ska-Daszkiewicz}  \&
  {Pamyatnykh}}{{Balona} et~al.}{2015}]{2015MNRAS.452.3073B}
{Balona} L.~A.,  {Daszy{\'n}ska-Daszkiewicz} J.,   {Pamyatnykh} A.~A.,  2015,
  \mn@doi [\mnras] {10.1093/mnras/stv1513}, \href
  {http://adsabs.harvard.edu/abs/2015MNRAS.452.3073B} {452, 3073}

\bibitem[\protect\citeauthoryear{{Barcel{\'o} Forteza}, {Michel}, {Roca
  Cort{\'e}s}  \& {Garc{\'{\i}}a}}{{Barcel{\'o} Forteza}
  et~al.}{2015}]{2015A&A...579A.133B}
{Barcel{\'o} Forteza} S.,  {Michel} E.,  {Roca Cort{\'e}s} T.,
  {Garc{\'{\i}}a} R.~A.,  2015, \mn@doi [\aap] {10.1051/0004-6361/201425507},
  \href {https://ui.adsabs.harvard.edu/abs/2015A%26A...579A.133B} {579, A133}

\bibitem[\protect\citeauthoryear{{Barcel{\'o} Forteza}, {Roca Cort{\'e}s},
  {Garc{\'{\i}}a Hern{\'a}ndez}  \& {Garc{\'{\i}}a}}{{Barcel{\'o} Forteza}
  et~al.}{2017}]{2017A&A...601A..57B}
{Barcel{\'o} Forteza} S.,  {Roca Cort{\'e}s} T.,  {Garc{\'{\i}}a Hern{\'a}ndez}
  A.,   {Garc{\'{\i}}a} R.~A.,  2017, \mn@doi [\aap]
  {10.1051/0004-6361/201628675}, \href
  {http://adsabs.harvard.edu/abs/2017A%26A...601A..57B} {601, A57}

\bibitem[\protect\citeauthoryear{{Barcel{\'o} Forteza}, {Roca Cort{\'e}s}  \&
  {Garc{\'{\i}}a}}{{Barcel{\'o} Forteza} et~al.}{2018}]{Barcelo2018a}
{Barcel{\'o} Forteza} S.,  {Roca Cort{\'e}s} T.,   {Garc{\'{\i}}a} R.~A.,
  2018, \mn@doi [\aap] {10.1051/0004-6361/201731803}, \href
  {http://adsabs.harvard.edu/abs/2018A%26A...614A..46B} {614, A46}

\bibitem[\protect\citeauthoryear{Barentsen et~al.,}{Barentsen
  et~al.}{2019}]{geert_barentsen_2019_2565212}
Barentsen G.,  et~al., 2019, KeplerGO/lightkurve: Lightkurve v1.0b29,
  \mn@doi{10.5281/zenodo.2565212}, \url
  {https://doi.org/10.5281/zenodo.2565212}

\bibitem[\protect\citeauthoryear{{Bigot} \& {Dziembowski}}{{Bigot} \&
  {Dziembowski}}{2002}]{2002AA...391..235B}
{Bigot} L.,  {Dziembowski} W.~A.,  2002, \mn@doi [\aap]
  {10.1051/0004-6361:20020824}, \href
  {http://adsabs.harvard.edu/abs/2002A%26A...391..235B} {391, 235}

\bibitem[\protect\citeauthoryear{{Bohlender}, {Gonzalez}  \&
  {Matthews}}{{Bohlender} et~al.}{1999}]{1999A&A...350..553B}
{Bohlender} D.~A.,  {Gonzalez} J.-F.,   {Matthews} J.~M.,  1999, \aap, \href
  {http://adsabs.harvard.edu/abs/1999A%26A...350..553B} {350, 553}

\bibitem[\protect\citeauthoryear{{Bowman}}{{Bowman}}{2017}]{2017ampm.book.....B}
{Bowman} D.~M.,  2017, {Amplitude Modulation of Pulsation Modes in Delta Scuti
  Stars}.
Springer International Publishing, 2017, \mn@doi{10.1007/978-3-319-66649-5}

\bibitem[\protect\citeauthoryear{{Bowman} \& {Kurtz}}{{Bowman} \&
  {Kurtz}}{2018}]{2018MNRAS.476.3169B}
{Bowman} D.~M.,  {Kurtz} D.~W.,  2018, \mn@doi [\mnras] {10.1093/mnras/sty449},
  \href {http://adsabs.harvard.edu/abs/2018MNRAS.476.3169B} {476, 3169}

\bibitem[\protect\citeauthoryear{{Bowman}, {Holdsworth}  \& {Kurtz}}{{Bowman}
  et~al.}{2015}]{2015MNRAS.449.1004B}
{Bowman} D.~M.,  {Holdsworth} D.~L.,   {Kurtz} D.~W.,  2015, \mn@doi [\mnras]
  {10.1093/mnras/stv364}, \href
  {http://adsabs.harvard.edu/abs/2015MNRAS.449.1004B} {449, 1004}

\bibitem[\protect\citeauthoryear{{Bowman}, {Kurtz}, {Breger}, {Murphy}  \&
  {Holdsworth}}{{Bowman} et~al.}{2016}]{2016MNRAS.460.1970B}
{Bowman} D.~M.,  {Kurtz} D.~W.,  {Breger} M.,  {Murphy} S.~J.,   {Holdsworth}
  D.~L.,  2016, \mn@doi [\mnras] {10.1093/mnras/stw1153}, \href
  {http://adsabs.harvard.edu/abs/2016MNRAS.460.1970B} {460, 1970}

\bibitem[\protect\citeauthoryear{{Bowman}, {Buysschaert}, {Neiner},
  {P{\'a}pics}, {Oksala}  \& {Aerts}}{{Bowman}
  et~al.}{2018}]{2018A&A...616A..77B}
{Bowman} D.~M.,  {Buysschaert} B.,  {Neiner} C.,  {P{\'a}pics} P.~I.,  {Oksala}
  M.~E.,   {Aerts} C.,  2018, \mn@doi [\aap] {10.1051/0004-6361/201833037},
  \href {http://adsabs.harvard.edu/abs/2018A%26A...616A..77B} {616, A77}

\bibitem[\protect\citeauthoryear{{Bowman} et~al.,}{{Bowman}
  et~al.}{2019}]{2019A&A...621A.135B}
{Bowman} D.~M.,  et~al., 2019, \mn@doi [\aap] {10.1051/0004-6361/201833662},
  \href {http://adsabs.harvard.edu/abs/2019A%26A...621A.135B} {621, A135}

\bibitem[\protect\citeauthoryear{{Breger}}{{Breger}}{1970}]{1970ApJ...162..597B}
{Breger} M.,  1970, \mn@doi [\apj] {10.1086/150691}, \href
  {http://adsabs.harvard.edu/abs/1970ApJ...162..597B} {162, 597}

\bibitem[\protect\citeauthoryear{{Breger}}{{Breger}}{2000}]{2000ASPC..210....3B}
{Breger} M.,  2000, in {Breger} M.,  {Montgomery} M.,  eds,  Astronomical
  Society of the Pacific Conference Series Vol. 210, Delta Scuti and Related
  Stars. p.~3

\bibitem[\protect\citeauthoryear{{Breger} et~al.,}{{Breger}
  et~al.}{1993}]{1993A&A...271..482B}
{Breger} M.,  et~al., 1993, \aap, \href
  {http://adsabs.harvard.edu/abs/1993A%26A...271..482B} {271, 482}

\bibitem[\protect\citeauthoryear{{Breger}, {Beck}, {Lenz}, {Schmitzberger},
  {Guggenberger}  \& {Shobbrook}}{{Breger} et~al.}{2006}]{2006A&A...455..673B}
{Breger} M.,  {Beck} P.,  {Lenz} P.,  {Schmitzberger} L.,  {Guggenberger} E.,
  {Shobbrook} R.~R.,  2006, \mn@doi [\aap] {10.1051/0004-6361:20065457}, \href
  {http://adsabs.harvard.edu/abs/2006A%26A...455..673B} {455, 673}

\bibitem[\protect\citeauthoryear{{Breger} et~al.,}{{Breger}
  et~al.}{2011}]{2011MNRAS.414.1721B}
{Breger} M.,  et~al., 2011, \mn@doi [\mnras]
  {10.1111/j.1365-2966.2011.18508.x}, \href
  {http://cdsads.u-strasbg.fr/abs/2011MNRAS.414.1721B} {414, 1721}

\bibitem[\protect\citeauthoryear{{Brunsden}, {Pollard}, {Wright}, {De Cat}  \&
  {Cottrell}}{{Brunsden} et~al.}{2018}]{2018MNRAS.475.3813B}
{Brunsden} E.,  {Pollard} K.~R.,  {Wright} D.~J.,  {De Cat} P.,   {Cottrell}
  P.~L.,  2018, \mn@doi [\mnras] {10.1093/mnras/sty034}, \href
  {http://adsabs.harvard.edu/abs/2018MNRAS.475.3813B} {475, 3813}

\bibitem[\protect\citeauthoryear{{Buscombe} \& {Morris}}{{Buscombe} \&
  {Morris}}{1961}]{1961MNRAS.123..183B}
{Buscombe} W.,  {Morris} P.~M.,  1961, \mn@doi [\mnras]
  {10.1093/mnras/123.2.183}, \href
  {http://adsabs.harvard.edu/abs/1961MNRAS.123..183B} {123, 183}

\bibitem[\protect\citeauthoryear{{Buzasi} et~al.,}{{Buzasi}
  et~al.}{2005}]{2005ApJ...619.1072B}
{Buzasi} D.~L.,  et~al., 2005, \mn@doi [\apj] {10.1086/426704}, \href
  {https://ui.adsabs.harvard.edu/abs/2005ApJ...619.1072B} {619, 1072}

\bibitem[\protect\citeauthoryear{{Casagrande}, {Sch{\"o}nrich}, {Asplund},
  {Cassisi}, {Ram{\'{\i}}rez}, {Mel{\'e}ndez}, {Bensby}  \&
  {Feltzing}}{{Casagrande} et~al.}{2011}]{Casagrande2011}
{Casagrande} L.,  {Sch{\"o}nrich} R.,  {Asplund} M.,  {Cassisi} S.,
  {Ram{\'{\i}}rez} I.,  {Mel{\'e}ndez} J.,  {Bensby} T.,   {Feltzing} S.,
  2011, \mn@doi [\aap] {10.1051/0004-6361/201016276}, \href
  {http://adsabs.harvard.edu/abs/2011A\%26A...530A.138C} {530, A138}

\bibitem[\protect\citeauthoryear{{Chaplin} \& {Miglio}}{{Chaplin} \&
  {Miglio}}{2013}]{2013ARA&A..51..353C}
{Chaplin} W.~J.,  {Miglio} A.,  2013, \mn@doi [\araa]
  {10.1146/annurev-astro-082812-140938}, \href
  {http://adsabs.harvard.edu/abs/2013ARA%26A..51..353C} {51, 353}

\bibitem[\protect\citeauthoryear{{Chen}, {Vergely}, {Valette}  \&
  {Carraro}}{{Chen} et~al.}{1998}]{chen98}
{Chen} B.,  {Vergely} J.~L.,  {Valette} B.,   {Carraro} G.,  1998, \aap, \href
  {http://esoads.eso.org/abs/1998A%26A...336..137C} {336, 137}

\bibitem[\protect\citeauthoryear{{Chen}, {Mittal}, {Kuchner}, {Forrest},
  {Lisse}, {Manoj}, {Sargent}  \& {Watson}}{{Chen}
  et~al.}{2014}]{2014ApJS..211...25C}
{Chen} C.~H.,  {Mittal} T.,  {Kuchner} M.,  {Forrest} W.~J.,  {Lisse} C.~M.,
  {Manoj} P.,  {Sargent} B.~A.,   {Watson} D.~M.,  2014, \mn@doi [\apjs]
  {10.1088/0067-0049/211/2/25}, \href
  {http://adsabs.harvard.edu/abs/2014ApJS..211...25C} {211, 25}

\bibitem[\protect\citeauthoryear{{Cheng}, {Neff}, {Johnson}, {Tarbell}, {Romo},
  {Gray}  \& {Corbally}}{{Cheng} et~al.}{2017}]{cheng2017}
{Cheng} K.-P.,  {Neff} J.~E.,  {Johnson} D.~M.,  {Tarbell} E.~S.,  {Romo}
  C.~A.,  {Gray} R.~O.,   {Corbally} C.~J.,  2017, \mn@doi [\aj]
  {10.3847/1538-3881/153/1/39}, \href
  {https://ui.adsabs.harvard.edu/\#abs/2017AJ....153...39C} {153, 39}

\bibitem[\protect\citeauthoryear{{Christophe}, {Ballot}, {Ouazzani}, {Antoci}
  \& {Salmon}}{{Christophe} et~al.}{2018}]{2018A&A...618A..47C}
{Christophe} S.,  {Ballot} J.,  {Ouazzani} R.-M.,  {Antoci} V.,   {Salmon}
  S.~J.~A.~J.,  2018, \mn@doi [\aap] {10.1051/0004-6361/201832782}, \href
  {http://adsabs.harvard.edu/abs/2018A%26A...618A..47C} {618, A47}

\bibitem[\protect\citeauthoryear{{Cohen} \& {Sarajedini}}{{Cohen} \&
  {Sarajedini}}{2012}]{2012MNRAS.419..342C}
{Cohen} R.~E.,  {Sarajedini} A.,  2012, \mn@doi [\mnras]
  {10.1111/j.1365-2966.2011.19697.x}, \href
  {http://cdsads.u-strasbg.fr/abs/2012MNRAS.419..342C} {419, 342}

\bibitem[\protect\citeauthoryear{{Conti}}{{Conti}}{1970}]{1970PASP...82..781C}
{Conti} P.~S.,  1970, \mn@doi [\pasp] {10.1086/128965}, \href
  {http://adsabs.harvard.edu/abs/1970PASP...82..781C} {82, 781}

\bibitem[\protect\citeauthoryear{Corbally}{Corbally}{1984}]{corbally1984}
Corbally C.,  1984, The Astronomical Journal, 89, 1887

\bibitem[\protect\citeauthoryear{{Cousins}}{{Cousins}}{1992}]{1992Obs...112...53C}
{Cousins} A.~W.~J.,  1992, The Observatory, \href
  {http://adsabs.harvard.edu/abs/1992Obs...112...53C} {112, 53}

\bibitem[\protect\citeauthoryear{{Cox}}{{Cox}}{1963}]{1963ApJ...138..487C}
{Cox} J.~P.,  1963, \mn@doi [\apj] {10.1086/147661}, \href
  {http://adsabs.harvard.edu/abs/1963ApJ...138..487C} {138, 487}

\bibitem[\protect\citeauthoryear{{Cunha}}{{Cunha}}{2001}]{2001MNRAS.325..373C}
{Cunha} M.~S.,  2001, \mn@doi [\mnras] {10.1046/j.1365-8711.2001.04434.x},
  \href {http://adsabs.harvard.edu/abs/2001MNRAS.325..373C} {325, 373}

\bibitem[\protect\citeauthoryear{{Cunha}}{{Cunha}}{2006}]{2006MNRAS.365..153C}
{Cunha} M.~S.,  2006, \mn@doi [\mnras] {10.1111/j.1365-2966.2005.09674.x},
  \href {http://adsabs.harvard.edu/abs/2006MNRAS.365..153C} {365, 153}

\bibitem[\protect\citeauthoryear{{Cunha} \& {Gough}}{{Cunha} \&
  {Gough}}{2000}]{2000MNRAS.319.1020C}
{Cunha} M.~S.,  {Gough} D.,  2000, \mn@doi [\mnras]
  {10.1046/j.1365-8711.2000.03896.x}, \href
  {http://adsabs.harvard.edu/abs/2000MNRAS.319.1020C} {319, 1020}

\bibitem[\protect\citeauthoryear{{Cunha} et~al.,}{{Cunha}
  et~al.}{2019}]{2019MNRAS.487.3523C}
{Cunha} M.~S.,  et~al., 2019, \mn@doi [\mnras] {10.1093/mnras/stz1332}, \href
  {https://ui.adsabs.harvard.edu/abs/2019MNRAS.487.3523C} {487, 3523}

\bibitem[\protect\citeauthoryear{{David} \& {Hillenbrand}}{{David} \&
  {Hillenbrand}}{2015a}]{2015ApJ...804..146D}
{David} T.~J.,  {Hillenbrand} L.~A.,  2015a, \mn@doi [\apj]
  {10.1088/0004-637X/804/2/146}, \href
  {http://adsabs.harvard.edu/abs/2015ApJ...804..146D} {804, 146}

\bibitem[\protect\citeauthoryear{{David} \& {Hillenbrand}}{{David} \&
  {Hillenbrand}}{2015b}]{David2015}
{David} T.~J.,  {Hillenbrand} L.~A.,  2015b, \mn@doi [\apj]
  {10.1088/0004-637X/804/2/146}, \href
  {https://ui.adsabs.harvard.edu/\#abs/2015ApJ...804..146D} {804, 146}

\bibitem[\protect\citeauthoryear{{Deeming}}{{Deeming}}{1975}]{1975Ap&SS..36..137D}
{Deeming} T.~J.,  1975, \mn@doi [\apss] {10.1007/BF00681947}, \href
  {http://adsabs.harvard.edu/abs/1975Ap%26SS..36..137D} {36, 137}

\bibitem[\protect\citeauthoryear{{Degroote} et~al.,}{{Degroote}
  et~al.}{2009}]{2009A&A...506..111D}
{Degroote} P.,  et~al., 2009, \mn@doi [\aap] {10.1051/0004-6361/200911782},
  \href {http://adsabs.harvard.edu/abs/2009A%26A...506..111D} {506, 111}

\bibitem[\protect\citeauthoryear{{D{\'\i}az}, {Gonz{\'a}lez}, {Levato}  \&
  {Grosso}}{{D{\'\i}az} et~al.}{2011}]{Diaz2011}
{D{\'\i}az} C.~G.,  {Gonz{\'a}lez} J.~F.,  {Levato} H.,   {Grosso} M.,  2011,
  \mn@doi [\aap] {10.1051/0004-6361/201016386}, \href
  {https://ui.adsabs.harvard.edu/\#abs/2011A&A...531A.143D} {531, A143}

\bibitem[\protect\citeauthoryear{{Drimmel}, {Cabrera-Lavers}  \&
  {L{\'o}pez-Corredoira}}{{Drimmel} et~al.}{2003}]{drimmel03}
{Drimmel} R.,  {Cabrera-Lavers} A.,   {L{\'o}pez-Corredoira} M.,  2003, \mn@doi
  [\aap] {10.1051/0004-6361:20031070}, \href
  {http://esoads.eso.org/abs/2003A%26A...409..205D} {409, 205}

\bibitem[\protect\citeauthoryear{{Dupret}, {Grigahc{\`e}ne}, {Garrido},
  {Gabriel}  \& {Scuflaire}}{{Dupret} et~al.}{2005}]{2005A&A...435..927D}
{Dupret} M.-A.,  {Grigahc{\`e}ne} A.,  {Garrido} R.,  {Gabriel} M.,
  {Scuflaire} R.,  2005, \mn@doi [\aap] {10.1051/0004-6361:20041817}, \href
  {http://adsabs.harvard.edu/abs/2005A%26A...435..927D} {435, 927}

\bibitem[\protect\citeauthoryear{{Dziembowski}}{{Dziembowski}}{1977}]{1977AcA....27...95D}
{Dziembowski} W.,  1977, \actaa, \href
  {http://adsabs.harvard.edu/abs/1977AcA....27...95D} {27, 95}

\bibitem[\protect\citeauthoryear{{Eggleton} \& {Tokovinin}}{{Eggleton} \&
  {Tokovinin}}{2008}]{Eggleton2008}
{Eggleton} P.~P.,  {Tokovinin} A.~A.,  2008, \mn@doi [\mnras]
  {10.1111/j.1365-2966.2008.13596.x}, \href
  {http://adsabs.harvard.edu/abs/2008MNRAS.389..869E} {389, 869}

\bibitem[\protect\citeauthoryear{Erspamer \& North}{Erspamer \&
  North}{2003}]{erspamer2003}
Erspamer D.,  North P.,  2003, Astronomy \& Astrophysics, 398, 1121

\bibitem[\protect\citeauthoryear{{Evans}, {Irwin}  \& {Helmer}}{{Evans}
  et~al.}{2002}]{2002A&A...395..347E}
{Evans} D.~W.,  {Irwin} M.~J.,   {Helmer} L.,  2002, \mn@doi [\aap]
  {10.1051/0004-6361:20021285}, \href
  {https://ui.adsabs.harvard.edu/abs/2002A%26A...395..347E} {395, 347}

\bibitem[\protect\citeauthoryear{Faraggiana, Bonifacio, Caffau, Gerbaldi  \&
  Nonino}{Faraggiana et~al.}{2004}]{faraggiana2004}
Faraggiana R.,  Bonifacio P.,  Caffau E.,  Gerbaldi M.,   Nonino M.,  2004,
  Astronomy \& Astrophysics, 425, 615

\bibitem[\protect\citeauthoryear{{Flower}}{{Flower}}{1996}]{Flower96}
{Flower} P.~J.,  1996, \mn@doi [\apj] {10.1086/177785}, \href
  {http://adsabs.harvard.edu/abs/1996ApJ...469..355F} {469, 355}

\bibitem[\protect\citeauthoryear{{Frandsen}, {Jones}, {Kjeldsen}, {Viskum},
  {Hjorth}, {Andersen}  \& {Thomsen}}{{Frandsen}
  et~al.}{1995}]{1995A&A...301..123F}
{Frandsen} S.,  {Jones} A.,  {Kjeldsen} H.,  {Viskum} M.,  {Hjorth} J.,
  {Andersen} N.~H.,   {Thomsen} B.,  1995, \aap, \href
  {http://adsabs.harvard.edu/abs/1995A%26A...301..123F} {301, 123}

\bibitem[\protect\citeauthoryear{{Gaia Collaboration}}{{Gaia
  Collaboration}}{2018}]{Gaia2018DR2}
{Gaia Collaboration} 2018, VizieR Online Data Catalog, \href
  {http://cdsads.u-strasbg.fr/abs/2018yCat.1345....0G} {1345}

\bibitem[\protect\citeauthoryear{{Gaia Collaboration} et~al.,}{{Gaia
  Collaboration} et~al.}{2018}]{2018A&A...616A...1G}
{Gaia Collaboration} et~al., 2018, \mn@doi [\aap]
  {10.1051/0004-6361/201833051}, \href
  {http://adsabs.harvard.edu/abs/2018A%26A...616A...1G} {616, A1}

\bibitem[\protect\citeauthoryear{{Garc{\'{\i}}a Hern{\'a}ndez},
  {Mart{\'{\i}}n-Ruiz}, {Monteiro}, {Su{\'a}rez}, {Reese}, {Pascual-Granado}
  \& {Garrido}}{{Garc{\'{\i}}a Hern{\'a}ndez} et~al.}{2015}]{GH2015}
{Garc{\'{\i}}a Hern{\'a}ndez} A.,  {Mart{\'{\i}}n-Ruiz} S.,  {Monteiro}
  M.~J.~P.~F.~G.,  {Su{\'a}rez} J.~C.,  {Reese} D.~R.,  {Pascual-Granado} J.,
  {Garrido} R.,  2015, \mn@doi [\apjl] {10.1088/2041-8205/811/2/L29}, \href
  {http://adsabs.harvard.edu/abs/2015ApJ...811L..29G} {811, L29}

\bibitem[\protect\citeauthoryear{{Garc{\'{\i}}a Hern{\'a}ndez}
  et~al.,}{{Garc{\'{\i}}a Hern{\'a}ndez} et~al.}{2017}]{GH2017a}
{Garc{\'{\i}}a Hern{\'a}ndez} A.,  et~al., 2017, \mn@doi [\mnras]
  {10.1093/mnrasl/slx117}, \href
  {http://adsabs.harvard.edu/abs/2017MNRAS.471L.140G} {471, L140}

\bibitem[\protect\citeauthoryear{{G{\'a}sp{\'a}r}, {Rieke}  \&
  {Ballering}}{{G{\'a}sp{\'a}r} et~al.}{2016}]{2016ApJ...826..171G}
{G{\'a}sp{\'a}r} A.,  {Rieke} G.~H.,   {Ballering} N.,  2016, \mn@doi [\apj]
  {10.3847/0004-637X/826/2/171}, \href
  {http://adsabs.harvard.edu/abs/2016ApJ...826..171G} {826, 171}

\bibitem[\protect\citeauthoryear{{Gonz{\'a}lez}, {Hubrig}  \&
  {Savanov}}{{Gonz{\'a}lez} et~al.}{2008}]{gonzalez2008}
{Gonz{\'a}lez} J.~F.,  {Hubrig} S.,   {Savanov} I.,  2008, Contributions of the
  Astronomical Observatory Skalnate Pleso, \href
  {http://ads.nao.ac.jp/abs/2008CoSka..38..411G} {38, 411}

\bibitem[\protect\citeauthoryear{{Gough}}{{Gough}}{1977a}]{1977LNP....71..349G}
{Gough} D.~O.,  1977a, in {Spiegel} E.~A.,  {Zahn} J.-P.,  eds,  Lecture Notes
  in Physics, Berlin Springer Verlag Vol. 71, Problems of Stellar Convection.
  pp 349--363, \mn@doi{10.1007/3-540-08532-7_56}

\bibitem[\protect\citeauthoryear{{Gough}}{{Gough}}{1977b}]{1977ApJ...214..196G}
{Gough} D.~O.,  1977b, \mn@doi [\apj] {10.1086/155244}, \href
  {http://adsabs.harvard.edu/abs/1977ApJ...214..196G} {214, 196}

\bibitem[\protect\citeauthoryear{{Gray}}{{Gray}}{2008}]{2008oasp.book.....G}
{Gray} D.~F.,  2008, {The Observation and Analysis of Stellar Photospheres}.
{Cambridge University Press}

\bibitem[\protect\citeauthoryear{{Gray} \& {Corbally}}{{Gray} \&
  {Corbally}}{1998}]{1998AJ....116.2530G}
{Gray} R.~O.,  {Corbally} C.~J.,  1998, \mn@doi [\aj] {10.1086/300613}, \href
  {http://adsabs.harvard.edu/abs/1998AJ....116.2530G} {116, 2530}

\bibitem[\protect\citeauthoryear{{Gray}, {Corbally}, {Garrison}, {McFadden},
  {Bubar}, {McGahee}, {O'Donoghue}  \& {Knox}}{{Gray}
  et~al.}{2006}]{2006AJ....132..161G}
{Gray} R.~O.,  {Corbally} C.~J.,  {Garrison} R.~F.,  {McFadden} M.~T.,  {Bubar}
  E.~J.,  {McGahee} C.~E.,  {O'Donoghue} A.~A.,   {Knox} E.~R.,  2006, \mn@doi
  [\aj] {10.1086/504637}, \href
  {https://ui.adsabs.harvard.edu/\#abs/2006AJ....132..161G} {132, 161}

\bibitem[\protect\citeauthoryear{{Green} et~al.,}{{Green}
  et~al.}{2015}]{green15}
{Green} G.~M.,  et~al., 2015, \mn@doi [\apj] {10.1088/0004-637X/810/1/25},
  \href {http://esoads.eso.org/abs/2015ApJ...810...25G} {810, 25}

\bibitem[\protect\citeauthoryear{{Green} et~al.,}{{Green}
  et~al.}{2018}]{green18}
{Green} G.~M.,  et~al., 2018, \mn@doi [\mnras] {10.1093/mnras/sty1008}, \href
  {http://adsabs.harvard.edu/abs/2018MNRAS.478..651G} {478, 651}

\bibitem[\protect\citeauthoryear{{Grigahc{\`e}ne} et~al.,}{{Grigahc{\`e}ne}
  et~al.}{2010}]{2010ApJ...713L.192G}
{Grigahc{\`e}ne} A.,  et~al., 2010, \mn@doi [\apjl]
  {10.1088/2041-8205/713/2/L192}, \href
  {http://adsabs.harvard.edu/abs/2010ApJ...713L.192G} {713, L192}

\bibitem[\protect\citeauthoryear{{Gruber} et~al.,}{{Gruber}
  et~al.}{2012}]{2012MNRAS.420..291G}
{Gruber} D.,  et~al., 2012, \mn@doi [\mnras]
  {10.1111/j.1365-2966.2011.20033.x}, \href
  {http://adsabs.harvard.edu/abs/2012MNRAS.420..291G} {420, 291}

\bibitem[\protect\citeauthoryear{{Guenther}, {Kallinger}, {Zwintz}, {Weiss}  \&
  {Tanner}}{{Guenther} et~al.}{2007}]{2007ApJ...671..581G}
{Guenther} D.~B.,  {Kallinger} T.,  {Zwintz} K.,  {Weiss} W.~W.,   {Tanner} J.,
   2007, \mn@doi [\apj] {10.1086/522880}, \href
  {http://adsabs.harvard.edu/abs/2007ApJ...671..581G} {671, 581}

\bibitem[\protect\citeauthoryear{{Guo}, {Gies}, {Matson}  \& {Garc{\'{\i}}a
  Hern{\'a}ndez}}{{Guo} et~al.}{2016}]{2016ApJ...826...69G}
{Guo} Z.,  {Gies} D.~R.,  {Matson} R.~A.,   {Garc{\'{\i}}a Hern{\'a}ndez} A.,
  2016, \mn@doi [\apj] {10.3847/0004-637X/826/1/69}, \href
  {http://adsabs.harvard.edu/abs/2016ApJ...826...69G} {826, 69}

\bibitem[\protect\citeauthoryear{{Guo}, {Gies}, {Matson}, {Garc{\'\i}a
  Hern{\'a}ndez}, {Han}  \& {Chen}}{{Guo} et~al.}{2017}]{2017ApJ...837..114G}
{Guo} Z.,  {Gies} D.~R.,  {Matson} R.~A.,  {Garc{\'\i}a Hern{\'a}ndez} A.,
  {Han} Z.,   {Chen} X.,  2017, \mn@doi [\apj] {10.3847/1538-4357/aa61a4},
  \href {https://ui.adsabs.harvard.edu/abs/2017ApJ...837..114G} {837, 114}

\bibitem[\protect\citeauthoryear{{Guzik}, {Kaye}, {Bradley}, {Cox}  \&
  {Neuforge}}{{Guzik} et~al.}{2000}]{2000ApJ...542L..57G}
{Guzik} J.~A.,  {Kaye} A.~B.,  {Bradley} P.~A.,  {Cox} A.~N.,   {Neuforge} C.,
  2000, \mn@doi [\apjl] {10.1086/312908}, \href
  {http://adsabs.harvard.edu/abs/2000ApJ...542L..57G} {542, L57}

\bibitem[\protect\citeauthoryear{Handberg}{Handberg}{2013}]{handbergphd}
Handberg R.,  2013, PhD thesis, Aarhus University

\bibitem[\protect\citeauthoryear{{Hatta}, {Sekii}, {Takata}  \&
  {Kurtz}}{{Hatta} et~al.}{2019}]{2019ApJ...871..135H}
{Hatta} Y.,  {Sekii} T.,  {Takata} M.,   {Kurtz} D.~W.,  2019, \mn@doi [\apj]
  {10.3847/1538-4357/aaf881}, \href
  {https://ui.adsabs.harvard.edu/abs/2019ApJ...871..135H} {871, 135}

\bibitem[\protect\citeauthoryear{{Heiter}}{{Heiter}}{2002}]{2002A&A...381..959H}
{Heiter} U.,  2002, \mn@doi [\aap] {10.1051/0004-6361:20011593}, \href
  {http://adsabs.harvard.edu/abs/2002A%26A...381..959H} {381, 959}

\bibitem[\protect\citeauthoryear{{Heiter} et~al.,}{{Heiter}
  et~al.}{2002}]{2002A&A...392..619H}
{Heiter} U.,  et~al., 2002, \mn@doi [\aap] {10.1051/0004-6361:20020788}, \href
  {http://adsabs.harvard.edu/abs/2002A%26A...392..619H} {392, 619}

\bibitem[\protect\citeauthoryear{{Helt}}{{Helt}}{1984}]{Helt1984}
{Helt} B.~E.,  1984, \aaps, \href
  {http://adsabs.harvard.edu/abs/1984A%26AS...56..457H} {56, 457}

\bibitem[\protect\citeauthoryear{{Henden}, {Levine}, {Terrell}  \&
  {Welch}}{{Henden} et~al.}{2015}]{2015AAS...22533616H}
{Henden} A.~A.,  {Levine} S.,  {Terrell} D.,   {Welch} D.~L.,  2015, in
  American Astronomical Society Meeting Abstracts \#225. p. 336.16

\bibitem[\protect\citeauthoryear{{Henden}, {Templeton}, {Terrell}, {Smith},
  {Levine}  \& {Welch}}{{Henden} et~al.}{2016}]{Henden16}
{Henden} A.~A.,  {Templeton} M.,  {Terrell} D.,  {Smith} T.~C.,  {Levine} S.,
  {Welch} D.,  2016, VizieR Online Data Catalog, \href
  {http://adsabs.harvard.edu/abs/2016yCat.2336....0H} {2336}

\bibitem[\protect\citeauthoryear{{Hoeg} et~al.,}{{Hoeg}
  et~al.}{1997}]{1997A&A...323L..57H}
{Hoeg} E.,  et~al., 1997, \aap, \href
  {https://ui.adsabs.harvard.edu/abs/1997A%26A...323L..57H} {323, L57}

\bibitem[\protect\citeauthoryear{{Houdek}}{{Houdek}}{2000}]{2000ASPC..210..454H}
{Houdek} G.,  2000, in {Breger} M.,  {Montgomery} M.,  eds,  Astronomical
  Society of the Pacific Conference Series Vol. 210, Delta Scuti and Related
  Stars. p.~454

\bibitem[\protect\citeauthoryear{{Houdek}, {Balmforth}, {Christensen-Dalsgaard}
   \& {Gough}}{{Houdek} et~al.}{1999}]{Houdek1999}
{Houdek} G.,  {Balmforth} N.~J.,  {Christensen-Dalsgaard} J.,   {Gough} D.~O.,
  1999, \aap, \href {https://ui.adsabs.harvard.edu/abs/1999A&A...351..582H}
  {351, 582}

\bibitem[\protect\citeauthoryear{{Howarth}}{{Howarth}}{1983}]{1983MNRAS.203..301H}
{Howarth} I.~D.,  1983, \mn@doi [\mnras] {10.1093/mnras/203.2.301}, \href
  {https://ui.adsabs.harvard.edu/abs/1983MNRAS.203..301H} {203, 301}

\bibitem[\protect\citeauthoryear{{Howell} et~al.,}{{Howell}
  et~al.}{2014}]{2014PASP..126..398H}
{Howell} S.~B.,  et~al., 2014, \mn@doi [\pasp] {10.1086/676406}, \href
  {http://adsabs.harvard.edu/abs/2014PASP..126..398H} {126, 398}

\bibitem[\protect\citeauthoryear{Huber et~al.,}{Huber et~al.}{2016}]{huber2016}
Huber D.,  et~al., 2016, The Astrophysical Journal Supplement Series, 224, 2

\bibitem[\protect\citeauthoryear{{Jancart}, {Jorissen}, {Babusiaux}  \&
  {Pourbaix}}{{Jancart} et~al.}{2005}]{2005A&A...442..365J}
{Jancart} S.,  {Jorissen} A.,  {Babusiaux} C.,   {Pourbaix} D.,  2005, \mn@doi
  [\aap] {10.1051/0004-6361:20053003}, \href
  {http://adsabs.harvard.edu/abs/2005A%26A...442..365J} {442, 365}

\bibitem[\protect\citeauthoryear{{Kallinger} et~al.,}{{Kallinger}
  et~al.}{2017}]{2017A&A...603A..13K}
{Kallinger} T.,  et~al., 2017, \mn@doi [\aap] {10.1051/0004-6361/201730625},
  \href {http://adsabs.harvard.edu/abs/2017A%26A...603A..13K} {603, A13}

\bibitem[\protect\citeauthoryear{{Kama}, {Folsom}  \& {Pinilla}}{{Kama}
  et~al.}{2015}]{2015A&A...582L..10K}
{Kama} M.,  {Folsom} C.~P.,   {Pinilla} P.,  2015, \mn@doi [\aap]
  {10.1051/0004-6361/201527094}, \href
  {http://adsabs.harvard.edu/abs/2015A%26A...582L..10K} {582, L10}

\bibitem[\protect\citeauthoryear{{Kamath}, {Wood}  \& {Van Winckel}}{{Kamath}
  et~al.}{2014}]{2014MNRAS.439.2211K}
{Kamath} D.,  {Wood} P.~R.,   {Van Winckel} H.,  2014, \mn@doi [\mnras]
  {10.1093/mnras/stt2033}, \href
  {http://adsabs.harvard.edu/abs/2014MNRAS.439.2211K} {439, 2211}

\bibitem[\protect\citeauthoryear{{Kharchenko}}{{Kharchenko}}{2001}]{kharchenko01}
{Kharchenko} N.~V.,  2001, Kinematika i Fizika Nebesnykh Tel, \href
  {http://adsabs.harvard.edu/abs/2001KFNT...17..409K} {17, 409}

\bibitem[\protect\citeauthoryear{{Kim}, {McNamara}  \& {Christensen}}{{Kim}
  et~al.}{1993}]{1993AJ....106.2493K}
{Kim} C.,  {McNamara} D.~H.,   {Christensen} C.~G.,  1993, \mn@doi [\aj]
  {10.1086/116817}, \href {http://adsabs.harvard.edu/abs/1993AJ....106.2493K}
  {106, 2493}

\bibitem[\protect\citeauthoryear{{King}}{{King}}{1994}]{1994MNRAS.269..209K}
{King} J.~R.,  1994, \mn@doi [\mnras] {10.1093/mnras/269.1.209}, \href
  {http://adsabs.harvard.edu/abs/1994MNRAS.269..209K} {269, 209}

\bibitem[\protect\citeauthoryear{{Koch} et~al.,}{{Koch}
  et~al.}{2010}]{2010ApJ...713L..79K}
{Koch} D.~G.,  et~al., 2010, \mn@doi [\apjl] {10.1088/2041-8205/713/2/L79},
  \href {http://adsabs.harvard.edu/abs/2010ApJ...713L..79K} {713, L79}

\bibitem[\protect\citeauthoryear{{Koen} \& {Eyer}}{{Koen} \&
  {Eyer}}{2002}]{2002MNRAS.331...45K}
{Koen} C.,  {Eyer} L.,  2002, \mn@doi [\mnras]
  {10.1046/j.1365-8711.2002.05150.x}, \href
  {http://adsabs.harvard.edu/abs/2002MNRAS.331...45K} {331, 45}

\bibitem[\protect\citeauthoryear{{Kopacki}}{{Kopacki}}{2015}]{2015AcA....65...81K}
{Kopacki} G.,  2015, \actaa, \href
  {http://cdsads.u-strasbg.fr/abs/2015AcA....65...81K} {65, 81}

\bibitem[\protect\citeauthoryear{{Kordopatis} et~al.,}{{Kordopatis}
  et~al.}{2013}]{Kordopatis2013}
{Kordopatis} G.,  et~al., 2013, \mn@doi [\aj] {10.1088/0004-6256/146/5/134},
  \href {http://cdsads.u-strasbg.fr/abs/2013AJ....146..134K} {146, 134}

\bibitem[\protect\citeauthoryear{{Kunder} et~al.,}{{Kunder}
  et~al.}{2017}]{Kunder2017}
{Kunder} A.,  et~al., 2017, VizieR Online Data Catalog, \href
  {http://cdsads.u-strasbg.fr/abs/2017yCat.3279....0K} {3279}

\bibitem[\protect\citeauthoryear{{Kurtz}}{{Kurtz}}{1978}]{1978ApJ...221..869K}
{Kurtz} D.~W.,  1978, \mn@doi [\apj] {10.1086/156090}, \href
  {http://adsabs.harvard.edu/abs/1978ApJ...221..869K} {221, 869}

\bibitem[\protect\citeauthoryear{{Kurtz}}{{Kurtz}}{1982}]{1982MNRAS.200..807K}
{Kurtz} D.~W.,  1982, \mn@doi [\mnras] {10.1093/mnras/200.3.807}, \href
  {http://adsabs.harvard.edu/abs/1982MNRAS.200..807K} {200, 807}

\bibitem[\protect\citeauthoryear{{Kurtz}}{{Kurtz}}{1985}]{1985MNRAS.213..773K}
{Kurtz} D.~W.,  1985, \mn@doi [\mnras] {10.1093/mnras/213.4.773}, \href
  {http://adsabs.harvard.edu/abs/1985MNRAS.213..773K} {213, 773}

\bibitem[\protect\citeauthoryear{{Kurtz}, {Saio}, {Takata}, {Shibahashi},
  {Murphy}  \& {Sekii}}{{Kurtz} et~al.}{2014}]{2014MNRAS.444..102K}
{Kurtz} D.~W.,  {Saio} H.,  {Takata} M.,  {Shibahashi} H.,  {Murphy} S.~J.,
  {Sekii} T.,  2014, \mn@doi [\mnras] {10.1093/mnras/stu1329}, \href
  {http://adsabs.harvard.edu/abs/2014MNRAS.444..102K} {444, 102}

\bibitem[\protect\citeauthoryear{{Kurtz}, {Shibahashi}, {Murphy}, {Bedding}  \&
  {Bowman}}{{Kurtz} et~al.}{2015}]{2015MNRAS.450.3015K}
{Kurtz} D.~W.,  {Shibahashi} H.,  {Murphy} S.~J.,  {Bedding} T.~R.,   {Bowman}
  D.~M.,  2015, \mn@doi [\mnras] {10.1093/mnras/stv868}, \href
  {http://adsabs.harvard.edu/abs/2015MNRAS.450.3015K} {450, 3015}

\bibitem[\protect\citeauthoryear{{Kurucz}}{{Kurucz}}{1993}]{1993KurCD..13.....K}
{Kurucz} R.,  1993, ATLAS9 Stellar Atmosphere Programs and 2 km/s grid.~Kurucz
  CD-ROM No.~13.~ Cambridge, Mass.: Smithsonian Astrophysical Observatory,
  1993., \href {http://esoads.eso.org/abs/1993KurCD..13.....K} {13}

\bibitem[\protect\citeauthoryear{{Kurucz} \& {Avrett}}{{Kurucz} \&
  {Avrett}}{1981}]{1981SAOSR.391.....K}
{Kurucz} R.~L.,  {Avrett} E.~H.,  1981, SAO Special Report, \href
  {http://adsabs.harvard.edu/abs/1981SAOSR.391.....K} {391}

\bibitem[\protect\citeauthoryear{{Lee}, {Hong}  \& {Kristiansen}}{{Lee}
  et~al.}{2019}]{2019AJ....157...17L}
{Lee} J.~W.,  {Hong} K.,   {Kristiansen} M.~H.,  2019, \mn@doi [\aj]
  {10.3847/1538-3881/aaf0fb}, \href
  {http://adsabs.harvard.edu/abs/2019AJ....157...17L} {157, 17}

\bibitem[\protect\citeauthoryear{{Lenz} \& {Breger}}{{Lenz} \&
  {Breger}}{2004}]{2004IAUS..224..786L}
{Lenz} P.,  {Breger} M.,  2004, in {Zverko} J.,  {Ziznovsky} J.,  {Adelman}
  S.~J.,   {Weiss} W.~W.,  eds,  IAU Symposium Vol. 224, The A-Star Puzzle. pp
  786--790, \mn@doi{10.1017/S1743921305009750}

\bibitem[\protect\citeauthoryear{{Lenz}, {Pamyatnykh}, {Breger}  \&
  {Antoci}}{{Lenz} et~al.}{2008}]{2008A&A...478..855L}
{Lenz} P.,  {Pamyatnykh} A.~A.,  {Breger} M.,   {Antoci} V.,  2008, \mn@doi
  [\aap] {10.1051/0004-6361:20078376}, \href
  {https://ui.adsabs.harvard.edu/#abs/2008A&A...478..855L} {478, 855}

\bibitem[\protect\citeauthoryear{{Li}, {Bedding}, {Murphy}, {Van Reeth},
  {Antoci}  \& {Ouazzani}}{{Li} et~al.}{2019a}]{2019MNRAS.482.1757L}
{Li} G.,  {Bedding} T.~R.,  {Murphy} S.~J.,  {Van Reeth} T.,  {Antoci} V.,
  {Ouazzani} R.-M.,  2019a, \mn@doi [\mnras] {10.1093/mnras/sty2743}, \href
  {http://adsabs.harvard.edu/abs/2019MNRAS.482.1757L} {482, 1757}

\bibitem[\protect\citeauthoryear{{Li}, {Van Reeth}, {Bedding}, {Murphy}  \&
  {Antoci}}{{Li} et~al.}{2019b}]{2019MNRAS.487..782L}
{Li} G.,  {Van Reeth} T.,  {Bedding} T.~R.,  {Murphy} S.~J.,   {Antoci} V.,
  2019b, \mn@doi [\mnras] {10.1093/mnras/stz1171}, \href
  {https://ui.adsabs.harvard.edu/abs/2019MNRAS.487..782L} {487, 782}

\bibitem[\protect\citeauthoryear{Liakos \& Niarchos}{Liakos \&
  Niarchos}{2016}]{liakos2016}
Liakos A.,  Niarchos P.,  2016, Monthly Notices of the Royal Astronomical
  Society, p. stw2756

\bibitem[\protect\citeauthoryear{{Liakos} \& {Niarchos}}{{Liakos} \&
  {Niarchos}}{2017}]{2017MNRAS.465.1181L}
{Liakos} A.,  {Niarchos} P.,  2017, \mn@doi [\mnras] {10.1093/mnras/stw2756},
  \href {http://adsabs.harvard.edu/abs/2017MNRAS.465.1181L} {465, 1181}

\bibitem[\protect\citeauthoryear{{Lindegren} et~al.,}{{Lindegren}
  et~al.}{2016}]{Gaia2}
{Lindegren} L.,  et~al., 2016, \mn@doi [\aap] {10.1051/0004-6361/201628714},
  \href {http://adsabs.harvard.edu/abs/2016A%26A...595A...4L} {595, A4}

\bibitem[\protect\citeauthoryear{{Lindegren} et~al.,}{{Lindegren}
  et~al.}{2018}]{2018A&A...616A...2L}
{Lindegren} L.,  et~al., 2018, \mn@doi [\aap] {10.1051/0004-6361/201832727},
  \href {https://ui.adsabs.harvard.edu/abs/2018A&A...616A...2L} {616, A2}

\bibitem[\protect\citeauthoryear{{Maeder}}{{Maeder}}{2009}]{2009pfer.book.....M}
{Maeder} A.,  2009, {Physics, Formation and Evolution of Rotating Stars}.
Springer, \mn@doi{10.1007/978-3-540-76949-1}

\bibitem[\protect\citeauthoryear{{Malkov}, {Tamazian}, {Docobo}  \&
  {Chulkov}}{{Malkov} et~al.}{2012}]{malkov2012}
{Malkov} O.~Y.,  {Tamazian} V.~S.,  {Docobo} J.~A.,   {Chulkov} D.~A.,  2012,
  \mn@doi [\aap] {10.1051/0004-6361/201219774}, \href
  {https://ui.adsabs.harvard.edu/\#abs/2012A&A...546A..69M} {546, A69}

\bibitem[\protect\citeauthoryear{{Marconi} \& {Palla}}{{Marconi} \&
  {Palla}}{1998}]{1998ApJ...507L.141M}
{Marconi} M.,  {Palla} F.,  1998, \mn@doi [\apjl] {10.1086/311704}, \href
  {http://adsabs.harvard.edu/abs/1998ApJ...507L.141M} {507, L141}

\bibitem[\protect\citeauthoryear{{Martell} et~al.,}{{Martell}
  et~al.}{2017}]{Martell2017}
{Martell} S.~L.,  et~al., 2017, \mn@doi [\mnras] {10.1093/mnras/stw2835}, \href
  {https://ui.adsabs.harvard.edu/\#abs/2017MNRAS.465.3203M} {465, 3203}

\bibitem[\protect\citeauthoryear{{Masana}, {Jordi}  \& {Ribas}}{{Masana}
  et~al.}{2006}]{Masana2006}
{Masana} E.,  {Jordi} C.,   {Ribas} I.,  2006, \mn@doi [\aap]
  {10.1051/0004-6361:20054021}, \href
  {http://adsabs.harvard.edu/abs/2006A%26A...450..735M} {450, 735}

\bibitem[\protect\citeauthoryear{{Matthews}}{{Matthews}}{2007}]{2007CoAst.150..333M}
{Matthews} J.~M.,  2007, \mn@doi [Communications in Asteroseismology]
  {10.1553/cia150s333}, \href
  {http://adsabs.harvard.edu/abs/2007CoAst.150..333M} {150, 333}

\bibitem[\protect\citeauthoryear{{McDonald}, {Zijlstra}  \& {Boyer}}{{McDonald}
  et~al.}{2012}]{McDonald2012}
{McDonald} I.,  {Zijlstra} A.~A.,   {Boyer} M.~L.,  2012, \mn@doi [\mnras]
  {10.1111/j.1365-2966.2012.21873.x}, \href
  {http://adsabs.harvard.edu/abs/2012MNRAS.427..343M} {427, 343}

\bibitem[\protect\citeauthoryear{{McDonald}, {Zijlstra}  \&
  {Watson}}{{McDonald} et~al.}{2017}]{McDonald2017}
{McDonald} I.,  {Zijlstra} A.~A.,   {Watson} R.~A.,  2017, \mn@doi [\mnras]
  {10.1093/mnras/stx1433}, \href
  {https://ui.adsabs.harvard.edu/\#abs/2017MNRAS.471..770M} {471, 770}

\bibitem[\protect\citeauthoryear{{McNamara}}{{McNamara}}{2000}]{2000ASPC..210..373M}
{McNamara} D.~H.,  2000, in {Breger} M.,  {Montgomery} M.,  eds,  Astronomical
  Society of the Pacific Conference Series Vol. 210, Delta Scuti and Related
  Stars. p.~373

\bibitem[\protect\citeauthoryear{{McNamara}}{{McNamara}}{2011}]{2011AJ....142..110M}
{McNamara} D.~H.,  2011, \mn@doi [\aj] {10.1088/0004-6256/142/4/110}, \href
  {http://adsabs.harvard.edu/abs/2011AJ....142..110M} {142, 110}

\bibitem[\protect\citeauthoryear{{McNamara}, {Clementini}  \&
  {Marconi}}{{McNamara} et~al.}{2007}]{2007AJ....133.2752M}
{McNamara} D.~H.,  {Clementini} G.,   {Marconi} M.,  2007, \mn@doi [\aj]
  {10.1086/513717}, \href {http://cdsads.u-strasbg.fr/abs/2007AJ....133.2752M}
  {133, 2752}

\bibitem[\protect\citeauthoryear{{Meynet}, {Ekstrom}, {Maeder}, {Eggenberger},
  {Saio}, {Chomienne}  \& {Haemmerl{\'e}}}{{Meynet}
  et~al.}{2013}]{2013LNP...865....3M}
{Meynet} G.,  {Ekstrom} S.,  {Maeder} A.,  {Eggenberger} P.,  {Saio} H.,
  {Chomienne} V.,   {Haemmerl{\'e}} L.,  2013, in {Goupil} M.,  {Belkacem} K.,
  {Neiner} C.,  {Ligni{\`e}res} F.,   {Green} J.~J.,  eds,  Lecture Notes in
  Physics, Berlin Springer Verlag Vol. 865, Lecture Notes in Physics, Berlin
  Springer Verlag. p.~3 (\mn@eprint {arXiv} {1301.2487}),
  \mn@doi{10.1007/978-3-642-33380-4_1}

\bibitem[\protect\citeauthoryear{{Michaud}}{{Michaud}}{1970}]{1970ApJ...160..641M}
{Michaud} G.,  1970, \mn@doi [\apj] {10.1086/150459}, \href
  {http://adsabs.harvard.edu/abs/1970ApJ...160..641M} {160, 641}

\bibitem[\protect\citeauthoryear{{Michel} et~al.,}{{Michel}
  et~al.}{2017}]{2017EPJWC.16003001M}
{Michel} E.,  et~al., 2017, in European Physical Journal Web of Conferences. p.
  03001 (\mn@eprint {arXiv} {1705.03721}),
  \mn@doi{10.1051/epjconf/201716003001}

\bibitem[\protect\citeauthoryear{{Mombarg}, {Van Reeth}, {Pedersen},
  {Molenberghs}, {Bowman}, {Johnston}, {Tkachenko}  \& {Aerts}}{{Mombarg}
  et~al.}{2019}]{2019MNRAS.tmp..495M}
{Mombarg} J.~S.~G.,  {Van Reeth} T.,  {Pedersen} M.~G.,  {Molenberghs} G.,
  {Bowman} D.~M.,  {Johnston} C.,  {Tkachenko} A.,   {Aerts} C.,  2019, \mn@doi
  [\mnras] {10.1093/mnras/stz501}, \href
  {http://adsabs.harvard.edu/abs/2019MNRAS.tmp..495M} {}

\bibitem[\protect\citeauthoryear{{Monet} et~al.,}{{Monet}
  et~al.}{2003}]{2003AJ....125..984M}
{Monet} D.~G.,  et~al., 2003, \mn@doi [\aj] {10.1086/345888}, \href
  {http://adsabs.harvard.edu/abs/2003AJ....125..984M} {125, 984}

\bibitem[\protect\citeauthoryear{{Montgomery} \& {O'Donoghue}}{{Montgomery} \&
  {O'Donoghue}}{1999}]{1999DSSN...13...28M}
{Montgomery} M.~H.,  {O'Donoghue} D.,  1999, Delta Scuti Star Newsletter, \href
  {http://adsabs.harvard.edu/abs/1999DSSN...13...28M} {13, 28}

\bibitem[\protect\citeauthoryear{{Munari} et~al.,}{{Munari}
  et~al.}{2014}]{2014AJ....148...81M}
{Munari} U.,  et~al., 2014, \mn@doi [\aj] {10.1088/0004-6256/148/5/81}, \href
  {http://cdsads.u-strasbg.fr/abs/2014AJ....148...81M} {148, 81}

\bibitem[\protect\citeauthoryear{{Murphy}}{{Murphy}}{2014}]{2014PhDT.......131M}
{Murphy} S.~J.,  2014, PhD thesis, Jeremiah Horrocks Institute, University of
  Central Lancashire, Preston, UK <EMAIL>murphy@physics.usyd.edu.au</EMAIL>

\bibitem[\protect\citeauthoryear{{Murphy}}{{Murphy}}{2015}]{2015MNRAS.453.2569M}
{Murphy} S.~J.,  2015, \mn@doi [\mnras] {10.1093/mnras/stv1842}, \href
  {http://adsabs.harvard.edu/abs/2015MNRAS.453.2569M} {453, 2569}

\bibitem[\protect\citeauthoryear{{Murphy} et~al.,}{{Murphy}
  et~al.}{2013}]{2013MNRAS.432.2284M}
{Murphy} S.~J.,  et~al., 2013, \mn@doi [\mnras] {10.1093/mnras/stt587}, \href
  {http://adsabs.harvard.edu/abs/2013MNRAS.432.2284M} {432, 2284}

\bibitem[\protect\citeauthoryear{{Murphy} et~al.,}{{Murphy}
  et~al.}{2015}]{2015PASA...32...36M}
{Murphy} S.~J.,  et~al., 2015, \mn@doi [\pasa] {10.1017/pasa.2015.34}, \href
  {http://adsabs.harvard.edu/abs/2015PASA...32...36M} {32, e036}

\bibitem[\protect\citeauthoryear{{Murphy}, {Hey}, {Van Reeth}  \&
  {Bedding}}{{Murphy} et~al.}{2019}]{2019MNRAS.tmp..585M}
{Murphy} S.~J.,  {Hey} D.,  {Van Reeth} T.,   {Bedding} T.~R.,  2019, \mn@doi
  [\mnras] {10.1093/mnras/stz590}, \href
  {http://adsabs.harvard.edu/abs/2019MNRAS.tmp..585M} {}

\bibitem[\protect\citeauthoryear{{Nemec} \& {Mateo}}{{Nemec} \&
  {Mateo}}{1990}]{1990ASPC...11...64N}
{Nemec} J.,  {Mateo} M.,  1990, in {Cacciari} C.,  {Clementini} G.,  eds,
  Astronomical Society of the Pacific Conference Series Vol. 11, Confrontation
  Between Stellar Pulsation and Evolution. pp 64--84

\bibitem[\protect\citeauthoryear{{Nemec}, {Nemec}  \& {Lutz}}{{Nemec}
  et~al.}{1994}]{1994AJ....108..222N}
{Nemec} J.~M.,  {Nemec} A. F.~L.,   {Lutz} T.~E.,  1994, \mn@doi [\aj]
  {10.1086/117062}, \href
  {https://ui.adsabs.harvard.edu/abs/1994AJ....108..222N} {108, 222}

\bibitem[\protect\citeauthoryear{{Nemec}, {Balona}, {Murphy}, {Kinemuchi}  \&
  {Jeon}}{{Nemec} et~al.}{2017}]{2017MNRAS.466.1290N}
{Nemec} J.~M.,  {Balona} L.~A.,  {Murphy} S.~J.,  {Kinemuchi} K.,   {Jeon}
  Y.-B.,  2017, \mn@doi [\mnras] {10.1093/mnras/stw3072}, \href
  {http://adsabs.harvard.edu/abs/2017MNRAS.466.1290N} {466, 1290}

\bibitem[\protect\citeauthoryear{{Niemczura}, {Scholz}, {Hubrig},
  {J{\"a}rvinen}, {Sch{\"o}ller}, {Ilyin}  \& {Kahraman
  Ali{\c{c}}avuș}}{{Niemczura} et~al.}{2017}]{Niemczura2017}
{Niemczura} E.,  {Scholz} R.~D.,  {Hubrig} S.,  {J{\"a}rvinen} S.~P.,
  {Sch{\"o}ller} M.,  {Ilyin} I.,   {Kahraman Ali{\c{c}}avuș} F.,  2017,
  \mn@doi [\mnras] {10.1093/mnras/stx1377}, \href
  {https://ui.adsabs.harvard.edu/\#abs/2017MNRAS.470.3806N} {470, 3806}

\bibitem[\protect\citeauthoryear{{Oelkers} et~al.,}{{Oelkers}
  et~al.}{2018}]{2018AJ....155...39O}
{Oelkers} R.~J.,  et~al., 2018, \mn@doi [\aj] {10.3847/1538-3881/aa9bf4}, \href
  {http://adsabs.harvard.edu/abs/2018AJ....155...39O} {155, 39}

\bibitem[\protect\citeauthoryear{{Ouazzani}, {Dupret}  \& {Reese}}{{Ouazzani}
  et~al.}{2012}]{2012A&A...547A..75O}
{Ouazzani} R.~M.,  {Dupret} M.~A.,   {Reese} D.~R.,  2012, \mn@doi [\aap]
  {10.1051/0004-6361/201219548}, \href
  {https://ui.adsabs.harvard.edu/abs/2012A&A...547A..75O} {547, A75}

\bibitem[\protect\citeauthoryear{{Ouazzani}, {Salmon}, {Antoci}, {Bedding},
  {Murphy}  \& {Roxburgh}}{{Ouazzani} et~al.}{2017}]{2017MNRAS.465.2294O}
{Ouazzani} R.-M.,  {Salmon} S.~J.~A.~J.,  {Antoci} V.,  {Bedding} T.~R.,
  {Murphy} S.~J.,   {Roxburgh} I.~W.,  2017, \mn@doi [\mnras]
  {10.1093/mnras/stw2717}, \href
  {http://adsabs.harvard.edu/abs/2017MNRAS.465.2294O} {465, 2294}

\bibitem[\protect\citeauthoryear{{Pamyatnykh}}{{Pamyatnykh}}{1999}]{1999AcA....49..119P}
{Pamyatnykh} A.~A.,  1999, \actaa, \href
  {http://adsabs.harvard.edu/abs/1999AcA....49..119P} {49, 119}

\bibitem[\protect\citeauthoryear{{Pamyatnykh}}{{Pamyatnykh}}{2000}]{2000ASPC..210..215P}
{Pamyatnykh} A.~A.,  2000, in {Breger} M.,  {Montgomery} M.,  eds,
  Astronomical Society of the Pacific Conference Series Vol. 210, Delta Scuti
  and Related Stars. p.~215 (\mn@eprint {} {astro-ph/0005276})

\bibitem[\protect\citeauthoryear{{Pamyatnykh}, {Dziembowski}, {Handler}  \&
  {Pikall}}{{Pamyatnykh} et~al.}{1998}]{1998A&A...333..141P}
{Pamyatnykh} A.~A.,  {Dziembowski} W.~A.,  {Handler} G.,   {Pikall} H.,  1998,
  \aap, \href {http://adsabs.harvard.edu/abs/1998A%26A...333..141P} {333, 141}

\bibitem[\protect\citeauthoryear{{Paparo}, {Sterken}, {Spoon}  \&
  {Birch}}{{Paparo} et~al.}{1996}]{paparo1996}
{Paparo} M.,  {Sterken} C.,  {Spoon} H.~W.~W.,   {Birch} P.~V.,  1996, \aap,
  \href {http://adsabs.harvard.edu/abs/1996A%26A...315..400P} {315, 400}

\bibitem[\protect\citeauthoryear{{Papar{\'o}}, {Benk{\H o}}, {Hareter}  \&
  {Guzik}}{{Papar{\'o}} et~al.}{2016a}]{Paparo2016b}
{Papar{\'o}} M.,  {Benk{\H o}} J.~M.,  {Hareter} M.,   {Guzik} J.~A.,  2016a,
  \mn@doi [\apjs] {10.3847/0067-0049/224/2/41}, \href
  {http://adsabs.harvard.edu/abs/2016ApJS..224...41P} {224, 41}

\bibitem[\protect\citeauthoryear{{Papar{\'o}}, {Benk{\H o}}, {Hareter}  \&
  {Guzik}}{{Papar{\'o}} et~al.}{2016b}]{Paparo2016a}
{Papar{\'o}} M.,  {Benk{\H o}} J.~M.,  {Hareter} M.,   {Guzik} J.~A.,  2016b,
  \mn@doi [\apj] {10.3847/0004-637X/822/2/100}, \href
  {http://adsabs.harvard.edu/abs/2016ApJ...822..100P} {822, 100}

\bibitem[\protect\citeauthoryear{{P{\'a}pics} et~al.,}{{P{\'a}pics}
  et~al.}{2012}]{2012A&A...542A..55P}
{P{\'a}pics} P.~I.,  et~al., 2012, \mn@doi [\aap]
  {10.1051/0004-6361/201218809}, \href
  {http://adsabs.harvard.edu/abs/2012A%26A...542A..55P} {542, A55}

\bibitem[\protect\citeauthoryear{{Pasinetti Fracassini}, {Pastori}, {Covino}
  \& {Pozzi}}{{Pasinetti Fracassini} et~al.}{2001}]{2001A&A...367..521P}
{Pasinetti Fracassini} L.~E.,  {Pastori} L.,  {Covino} S.,   {Pozzi} A.,  2001,
  \mn@doi [\aap] {10.1051/0004-6361:20000451}, \href
  {https://ui.adsabs.harvard.edu/\#abs/2001A&A...367..521P} {367, 521}

\bibitem[\protect\citeauthoryear{{Paunzen}, {Handler}, {Weiss}  \&
  {North}}{{Paunzen} et~al.}{1994}]{1994IBVS.4094....1P}
{Paunzen} E.,  {Handler} G.,  {Weiss} W.~W.,   {North} P.,  1994, Information
  Bulletin on Variable Stars, \href
  {http://adsabs.harvard.edu/abs/1994IBVS.4094....1P} {4094}

\bibitem[\protect\citeauthoryear{{Paunzen} et~al.,}{{Paunzen}
  et~al.}{2002}]{Paunzen2002}
{Paunzen} E.,  et~al., 2002, \mn@doi [\aap] {10.1051/0004-6361:20020854}, \href
  {http://adsabs.harvard.edu/abs/2002A%26A...392..515P} {392, 515}

\bibitem[\protect\citeauthoryear{{Paunzen}, {Heiter}, {Fraga}  \&
  {Pintado}}{{Paunzen} et~al.}{2012}]{2012MNRAS.419.3604P}
{Paunzen} E.,  {Heiter} U.,  {Fraga} L.,   {Pintado} O.,  2012, \mn@doi
  [\mnras] {10.1111/j.1365-2966.2011.20007.x}, \href
  {http://adsabs.harvard.edu/abs/2012MNRAS.419.3604P} {419, 3604}

\bibitem[\protect\citeauthoryear{{Pedersen}, {Antoci}, {Korhonen}, {White},
  {Jessen-Hansen}, {Lehtinen}, {Nikbakhsh}  \& {Viuho}}{{Pedersen}
  et~al.}{2017}]{2017MNRAS.466.3060P}
{Pedersen} M.~G.,  {Antoci} V.,  {Korhonen} H.,  {White} T.~R.,
  {Jessen-Hansen} J.,  {Lehtinen} J.,  {Nikbakhsh} S.,   {Viuho} J.,  2017,
  \mn@doi [\mnras] {10.1093/mnras/stw3226}, \href
  {https://ui.adsabs.harvard.edu/#abs/2017MNRAS.466.3060P} {466, 3060}

\bibitem[\protect\citeauthoryear{{Petersen} \&
  {Christensen-Dalsgaard}}{{Petersen} \&
  {Christensen-Dalsgaard}}{1996}]{1996A&A...312..463P}
{Petersen} J.~O.,  {Christensen-Dalsgaard} J.,  1996, \aap, \href
  {http://adsabs.harvard.edu/abs/1996A%26A...312..463P} {312, 463}

\bibitem[\protect\citeauthoryear{{Pigulski}}{{Pigulski}}{2014}]{2014IAUS..301...31P}
{Pigulski} A.,  2014, in {Guzik} J.~A.,  {Chaplin} W.~J.,  {Handler} G.,
  {Pigulski} A.,  eds,  IAU Symposium Vol. 301, Precision Asteroseismology. pp
  31--38 (\mn@eprint {arXiv} {1311.3954}), \mn@doi{10.1017/S1743921313014038}

\bibitem[\protect\citeauthoryear{{Pigulski} \& {Michalska}}{{Pigulski} \&
  {Michalska}}{2007}]{2007AcA....57...61P}
{Pigulski} A.,  {Michalska} G.,  2007, \actaa, \href
  {http://adsabs.harvard.edu/abs/2007AcA....57...61P} {57, 61}

\bibitem[\protect\citeauthoryear{{Pojma\'nski}}{{Pojma\'nski}}{1997}]{1997AcA....47..467P}
{Pojma\'nski} G.,  1997, \actaa, \href
  {http://cdsads.u-strasbg.fr/abs/1997AcA....47..467P} {47, 467}

\bibitem[\protect\citeauthoryear{{Pojma\'nski}}{{Pojma\'nski}}{2002}]{2002AcA....52..397P}
{Pojma\'nski} G.,  2002, \actaa, \href
  {http://cdsads.u-strasbg.fr/abs/2002AcA....52..397P} {52, 397}

\bibitem[\protect\citeauthoryear{{Poleski} et~al.,}{{Poleski}
  et~al.}{2010}]{2010AcA....60....1P}
{Poleski} R.,  et~al., 2010, \actaa, \href
  {http://cdsads.u-strasbg.fr/abs/2010AcA....60....1P} {60, 1}

\bibitem[\protect\citeauthoryear{{Poretti} et~al.,}{{Poretti}
  et~al.}{2005}]{2005A&A...440.1097P}
{Poretti} E.,  et~al., 2005, \mn@doi [\aap] {10.1051/0004-6361:20053463}, \href
  {http://cdsads.u-strasbg.fr/abs/2005A%26A...440.1097P} {440, 1097}

\bibitem[\protect\citeauthoryear{Poretti et~al.,}{Poretti
  et~al.}{2011}]{2011A&A...528A.147P}
Poretti E.,  et~al., 2011, \mn@doi [Astron. Astrophys.]
  {10.1051/0004-6361/201016045}, 528, A147

\bibitem[\protect\citeauthoryear{{Pourbaix} et~al.,}{{Pourbaix}
  et~al.}{2004a}]{2004A&A...424..727P}
{Pourbaix} D.,  et~al., 2004a, \mn@doi [\aap] {10.1051/0004-6361:20041213},
  \href {http://adsabs.harvard.edu/abs/2004A%26A...424..727P} {424, 727}

\bibitem[\protect\citeauthoryear{Pourbaix et~al.,}{Pourbaix
  et~al.}{2004b}]{pourbaix2004}
Pourbaix D.,  et~al., 2004b, Astronomy \& Astrophysics, 424, 727

\bibitem[\protect\citeauthoryear{{Press} \& {Rybicki}}{{Press} \&
  {Rybicki}}{1989}]{1989ApJ...338..277P}
{Press} W.~H.,  {Rybicki} G.~B.,  1989, \mn@doi [\apj] {10.1086/167197}, \href
  {http://adsabs.harvard.edu/abs/1989ApJ...338..277P} {338, 277}

\bibitem[\protect\citeauthoryear{{Preston}}{{Preston}}{1974}]{1974ARA&A..12..257P}
{Preston} G.~W.,  1974, \mn@doi [\araa] {10.1146/annurev.aa.12.090174.001353},
  \href {http://adsabs.harvard.edu/abs/1974ARA%26A..12..257P} {12, 257}

\bibitem[\protect\citeauthoryear{{Rainer} et~al.,}{{Rainer}
  et~al.}{2016}]{Rainer2016AJ}
{Rainer} M.,  et~al., 2016, \mn@doi [\aj] {10.3847/0004-6256/152/6/207}, \href
  {http://adsabs.harvard.edu/abs/2016AJ....152..207R} {152, 207}

\bibitem[\protect\citeauthoryear{{Rainer} et~al.,}{{Rainer}
  et~al.}{2017}]{Rainer2016}
{Rainer} M.,  et~al., 2017, VizieR Online Data Catalog, \href
  {http://adsabs.harvard.edu/abs/2017yCat..51520207R} {515}

\bibitem[\protect\citeauthoryear{{Reegen}}{{Reegen}}{2007}]{2007A&A...467.1353R}
{Reegen} P.,  2007, \mn@doi [\aap] {10.1051/0004-6361:20066597}, \href
  {http://adsabs.harvard.edu/abs/2007A%26A...467.1353R} {467, 1353}

\bibitem[\protect\citeauthoryear{Renson \& Manfroid}{Renson \&
  Manfroid}{2009}]{Renson2009}
Renson P.,  Manfroid J.,  2009, \mn@doi [Astron. Astrophys.]
  {10.1051/0004-6361/200810788}, 498, 961

\bibitem[\protect\citeauthoryear{Renson, Gerbaldi  \& Catalano}{Renson
  et~al.}{1991}]{renson1991}
Renson P.,  Gerbaldi M.,   Catalano F.,  1991, Astronomy and Astrophysics
  Supplement Series, 89, 429

\bibitem[\protect\citeauthoryear{{Ricker} et~al.,}{{Ricker}
  et~al.}{2014}]{2014SPIE.9143E..20R}
{Ricker} G.~R.,  et~al., 2014, in Space Telescopes and Instrumentation 2014:
  Optical, Infrared, and Millimeter Wave. p. 914320 (\mn@eprint {arXiv}
  {1406.0151}), \mn@doi{10.1117/12.2063489}

\bibitem[\protect\citeauthoryear{{Ricker} et~al.,}{{Ricker}
  et~al.}{2015}]{2015JATIS...1a4003R}
{Ricker} G.~R.,  et~al., 2015, \mn@doi [Journal of Astronomical Telescopes,
  Instruments, and Systems] {10.1117/1.JATIS.1.1.014003}, \href
  {http://adsabs.harvard.edu/abs/2015JATIS...1a4003R} {1, 014003}

\bibitem[\protect\citeauthoryear{Rodr{\'\i}guez \&
  L{\'o}pez-Gonz{\'a}lez}{Rodr{\'\i}guez \&
  L{\'o}pez-Gonz{\'a}lez}{2000}]{rodriguez2000}
Rodr{\'\i}guez E.,  L{\'o}pez-Gonz{\'a}lez M.,  2000, Astronomy and
  Astrophysics Supplement Series, 144, 469

\bibitem[\protect\citeauthoryear{{Royer}, {Grenier}, {Baylac}, {G{\'o}mez}  \&
  {Zorec}}{{Royer} et~al.}{2002}]{Royer2002}
{Royer} F.,  {Grenier} S.,  {Baylac} M.-O.,  {G{\'o}mez} A.~E.,   {Zorec} J.,
  2002, \mn@doi [\aap] {10.1051/0004-6361:20020943}, \href
  {http://adsabs.harvard.edu/abs/2002A%26A...393..897R} {393, 897}

\bibitem[\protect\citeauthoryear{Royer, Zorec  \& G{\'o}mez}{Royer
  et~al.}{2007}]{royer2007}
Royer F.,  Zorec J.,   G{\'o}mez A.,  2007, Astronomy \& Astrophysics, 463, 671

\bibitem[\protect\citeauthoryear{{Saio}}{{Saio}}{2005}]{2005MNRAS.360.1022S}
{Saio} H.,  2005, \mn@doi [\mnras] {10.1111/j.1365-2966.2005.09091.x}, \href
  {http://adsabs.harvard.edu/abs/2005MNRAS.360.1022S} {360, 1022}

\bibitem[\protect\citeauthoryear{{Saio} \& {Gautschy}}{{Saio} \&
  {Gautschy}}{2004}]{2004MNRAS.350..485S}
{Saio} H.,  {Gautschy} A.,  2004, \mn@doi [\mnras]
  {10.1111/j.1365-2966.2004.07659.x}, \href
  {http://adsabs.harvard.edu/abs/2004MNRAS.350..485S} {350, 485}

\bibitem[\protect\citeauthoryear{{Saio}, {Kurtz}, {Takata}, {Shibahashi},
  {Murphy}, {Sekii}  \& {Bedding}}{{Saio} et~al.}{2015}]{2015MNRAS.447.3264S}
{Saio} H.,  {Kurtz} D.~W.,  {Takata} M.,  {Shibahashi} H.,  {Murphy} S.~J.,
  {Sekii} T.,   {Bedding} T.~R.,  2015, \mn@doi [\mnras]
  {10.1093/mnras/stu2696}, \href
  {http://adsabs.harvard.edu/abs/2015MNRAS.447.3264S} {447, 3264}

\bibitem[\protect\citeauthoryear{{Saio}, {Kurtz}, {Murphy}, {Antoci}  \&
  {Lee}}{{Saio} et~al.}{2018a}]{2018MNRAS.474.2774S}
{Saio} H.,  {Kurtz} D.~W.,  {Murphy} S.~J.,  {Antoci} V.~L.,   {Lee} U.,
  2018a, \mn@doi [\mnras] {10.1093/mnras/stx2962}, \href
  {http://adsabs.harvard.edu/abs/2018MNRAS.474.2774S} {474, 2774}

\bibitem[\protect\citeauthoryear{{Saio}, {Bedding}, {Kurtz}, {Murphy},
  {Antoci}, {Shibahashi}, {Li}  \& {Takata}}{{Saio}
  et~al.}{2018b}]{2018MNRAS.477.2183S}
{Saio} H.,  {Bedding} T.~R.,  {Kurtz} D.~W.,  {Murphy} S.~J.,  {Antoci} V.,
  {Shibahashi} H.,  {Li} G.,   {Takata} M.,  2018b, \mn@doi [\mnras]
  {10.1093/mnras/sty784}, \href
  {http://adsabs.harvard.edu/abs/2018MNRAS.477.2183S} {477, 2183}

\bibitem[\protect\citeauthoryear{{Scargle}}{{Scargle}}{1982}]{1982ApJ...263..835S}
{Scargle} J.~D.,  1982, \mn@doi [\apj] {10.1086/160554}, \href
  {http://adsabs.harvard.edu/abs/1982ApJ...263..835S} {263, 835}

\bibitem[\protect\citeauthoryear{{Schlegel}, {Finkbeiner}  \&
  {Davis}}{{Schlegel} et~al.}{1998}]{schlegel98}
{Schlegel} D.~J.,  {Finkbeiner} D.~P.,   {Davis} M.,  1998, \mn@doi [\apj]
  {10.1086/305772}, \href {http://adsabs.harvard.edu/abs/1998ApJ...500..525S}
  {500, 525}

\bibitem[\protect\citeauthoryear{{Schmid} \& {Aerts}}{{Schmid} \&
  {Aerts}}{2016}]{2016A&A...592A.116S}
{Schmid} V.~S.,  {Aerts} C.,  2016, \mn@doi [\aap]
  {10.1051/0004-6361/201628617}, \href
  {http://adsabs.harvard.edu/abs/2016A%26A...592A.116S} {592, A116}

\bibitem[\protect\citeauthoryear{{Schwarzenberg-Czerny}}{{Schwarzenberg-Czerny}}{2003}]{2003ASPC..292..383S}
{Schwarzenberg-Czerny} A.,  2003, in {Sterken} C.,  ed.,  Astronomical Society
  of the Pacific Conference Series Vol. 292, Interplay of Periodic, Cyclic and
  Stochastic Variability in Selected Areas of the H-R Diagram. p.~383

\bibitem[\protect\citeauthoryear{{Shulyak}, {Tsymbal}, {Ryabchikova},
  {St{\"u}tz}  \& {Weiss}}{{Shulyak} et~al.}{2004}]{2004A&A...428..993S}
{Shulyak} D.,  {Tsymbal} V.,  {Ryabchikova} T.,  {St{\"u}tz} C.,   {Weiss}
  W.~W.,  2004, \mn@doi [\aap] {10.1051/0004-6361:20034169}, \href
  {https://ui.adsabs.harvard.edu/abs/2004A%26A...428..993S} {428, 993}

\bibitem[\protect\citeauthoryear{{Siebert} et~al.,}{{Siebert}
  et~al.}{2011}]{Siebert2011}
{Siebert} A.,  et~al., 2011, \mn@doi [\aj] {10.1088/0004-6256/141/6/187}, \href
  {http://adsabs.harvard.edu/abs/2011AJ....141..187S} {141, 187}

\bibitem[\protect\citeauthoryear{{Sikora} et~al.,}{{Sikora}
  et~al.}{2019}]{2019arXiv190508835S}
{Sikora} J.,  et~al., 2019, arXiv e-prints, \href
  {https://ui.adsabs.harvard.edu/abs/2019arXiv190508835S} {p. arXiv:1905.08835}

\bibitem[\protect\citeauthoryear{{Skrutskie} et~al.,}{{Skrutskie}
  et~al.}{2006}]{2006AJ....131.1163S}
{Skrutskie} M.~F.,  et~al., 2006, \mn@doi [\aj] {10.1086/498708}, \href
  {http://adsabs.harvard.edu/abs/2006AJ....131.1163S} {131, 1163}

\bibitem[\protect\citeauthoryear{{Sloan} et~al.,}{{Sloan}
  et~al.}{2004}]{2004ApJ...614L..77S}
{Sloan} G.~C.,  et~al., 2004, \mn@doi [\apjl] {10.1086/425324}, \href
  {http://adsabs.harvard.edu/abs/2004ApJ...614L..77S} {614, L77}

\bibitem[\protect\citeauthoryear{{Smalley} et~al.,}{{Smalley}
  et~al.}{2011}]{2011A&A...535A...3S}
{Smalley} B.,  et~al., 2011, \mn@doi [\aap] {10.1051/0004-6361/201117230},
  \href {http://adsabs.harvard.edu/abs/2011A%26A...535A...3S} {535, A3}

\bibitem[\protect\citeauthoryear{{Smalley} et~al.,}{{Smalley}
  et~al.}{2017}]{2017MNRAS.465.2662S}
{Smalley} B.,  et~al., 2017, \mn@doi [\mnras] {10.1093/mnras/stw2903}, \href
  {https://ui.adsabs.harvard.edu/#abs/2017MNRAS.465.2662S} {465, 2662}

\bibitem[\protect\citeauthoryear{{Soubiran}, {Le Campion}, {Cayrel de Strobel}
  \& {Caillo}}{{Soubiran} et~al.}{2010}]{Soubiran2010}
{Soubiran} C.,  {Le Campion} J.-F.,  {Cayrel de Strobel} G.,   {Caillo} A.,
  2010, \mn@doi [\aap] {10.1051/0004-6361/201014247}, \href
  {http://cdsads.u-strasbg.fr/abs/2010A%26A...515A.111S} {515, A111}

\bibitem[\protect\citeauthoryear{Soubiran, Le~Campion, Brouillet  \&
  Chemin}{Soubiran et~al.}{2016}]{soubiran2016}
Soubiran C.,  Le~Campion J.-F.,  Brouillet N.,   Chemin L.,  2016, Astronomy \&
  Astrophysics, 591, A118

\bibitem[\protect\citeauthoryear{{Stassun} et~al.,}{{Stassun}
  et~al.}{2018}]{TESSInput2018}
{Stassun} K.~G.,  et~al., 2018, \mn@doi [\aj] {10.3847/1538-3881/aad050}, \href
  {https://ui.adsabs.harvard.edu/\#abs/2018AJ....156..102S} {156, 102}

\bibitem[\protect\citeauthoryear{{St{\c e}pie{\'n}}}{{St{\c
  e}pie{\'n}}}{2000}]{2000A&A...353..227S}
{St{\c e}pie{\'n}} K.,  2000, \aap, \href
  {http://adsabs.harvard.edu/abs/2000A%26A...353..227S} {353, 227}

\bibitem[\protect\citeauthoryear{{Stevens}, {Stassun}  \& {Gaudi}}{{Stevens}
  et~al.}{2017}]{Stevens2017}
{Stevens} D.~J.,  {Stassun} K.~G.,   {Gaudi} B.~S.,  2017, \mn@doi [\aj]
  {10.3847/1538-3881/aa957b}, \href
  {https://ui.adsabs.harvard.edu/\#abs/2017AJ....154..259S} {154, 259}

\bibitem[\protect\citeauthoryear{{Strohmeier}}{{Strohmeier}}{1967}]{1967IBVS..195....1S}
{Strohmeier} W.,  1967, Information Bulletin on Variable Stars, \href
  {http://cdsads.u-strasbg.fr/abs/1967IBVS..195....1S} {195}

\bibitem[\protect\citeauthoryear{{Strohmeier}, {Knigge}  \& {Ott}}{{Strohmeier}
  et~al.}{1965}]{1965IBVS...86....1S}
{Strohmeier} W.,  {Knigge} R.,   {Ott} H.,  1965, Information Bulletin on
  Variable Stars, \href {http://adsabs.harvard.edu/abs/1965IBVS...86....1S}
  {86}

\bibitem[\protect\citeauthoryear{{Su{\'a}rez}, {Garc{\'\i}a Hern{\'a}ndez},
  {Moya}, {Rodrigo}, {Solano}, {Garrido}  \& {Rod{\'o}n}}{{Su{\'a}rez}
  et~al.}{2014}]{2014A&A...563A...7S}
{Su{\'a}rez} J.~C.,  {Garc{\'\i}a Hern{\'a}ndez} A.,  {Moya} A.,  {Rodrigo} C.,
   {Solano} E.,  {Garrido} R.,   {Rod{\'o}n} J.~R.,  2014, \mn@doi [\aap]
  {10.1051/0004-6361/201322270}, \href
  {https://ui.adsabs.harvard.edu/abs/2014A&A...563A...7S} {563, A7}

\bibitem[\protect\citeauthoryear{{Suran}, {Goupil}, {Baglin}, {Lebreton}  \&
  {Catala}}{{Suran} et~al.}{2001}]{2001A&A...372..233S}
{Suran} M.,  {Goupil} M.,  {Baglin} A.,  {Lebreton} Y.,   {Catala} C.,  2001,
  \mn@doi [\aap] {10.1051/0004-6361:20010485}, \href
  {http://adsabs.harvard.edu/abs/2001A%26A...372..233S} {372, 233}

\bibitem[\protect\citeauthoryear{{Tarrant}, {Chaplin}, {Elsworth}, {Spreckley}
  \& {Stevens}}{{Tarrant} et~al.}{2008}]{2008A&A...492..167T}
{Tarrant} N.~J.,  {Chaplin} W.~J.,  {Elsworth} Y.~P.,  {Spreckley} S.~A.,
  {Stevens} I.~R.,  2008, \mn@doi [\aap] {10.1051/0004-6361:200810952}, \href
  {http://adsabs.harvard.edu/abs/2008A%26A...492..167T} {492, 167}

\bibitem[\protect\citeauthoryear{{Templeton}, {Basu}  \&
  {Demarque}}{{Templeton} et~al.}{2002}]{2002ApJ...576..963T}
{Templeton} M.,  {Basu} S.,   {Demarque} P.,  2002, \mn@doi [\apj]
  {10.1086/341805}, \href {http://adsabs.harvard.edu/abs/2002ApJ...576..963T}
  {576, 963}

\bibitem[\protect\citeauthoryear{{Tkachenko}}{{Tkachenko}}{2015}]{2015A&A...581A.129T}
{Tkachenko} A.,  2015, \mn@doi [\aap] {10.1051/0004-6361/201526513}, \href
  {https://ui.adsabs.harvard.edu/abs/2015A%26A...581A.129T} {581, A129}

\bibitem[\protect\citeauthoryear{{Tokovinin} \& {Kiyaeva}}{{Tokovinin} \&
  {Kiyaeva}}{2016}]{Tokovinin2016}
{Tokovinin} A.,  {Kiyaeva} O.,  2016, \mn@doi [\mnras] {10.1093/mnras/stv2825},
  \href {http://adsabs.harvard.edu/abs/2016MNRAS.456.2070T} {456, 2070}

\bibitem[\protect\citeauthoryear{{Trilling} et~al.,}{{Trilling}
  et~al.}{2007}]{2007ApJ...658.1289T}
{Trilling} D.~E.,  et~al., 2007, \mn@doi [\apj] {10.1086/511668}, \href
  {http://adsabs.harvard.edu/abs/2007ApJ...658.1289T} {658, 1289}

\bibitem[\protect\citeauthoryear{{Tsymbal}}{{Tsymbal}}{1996}]{1996ASPC..108..198T}
{Tsymbal} V.,  1996, in {Adelman} S.~J.,  {Kupka} F.,   {Weiss} W.~W.,  eds,
  Astronomical Society of the Pacific Conference Series Vol. 108, M.A.S.S.,
  Model Atmospheres and Spectrum Synthesis. p.~198

\bibitem[\protect\citeauthoryear{{Turcotte}, {Richer}, {Michaud}  \&
  {Christensen-Dalsgaard}}{{Turcotte} et~al.}{2000}]{2000A&A...360..603T}
{Turcotte} S.,  {Richer} J.,  {Michaud} G.,   {Christensen-Dalsgaard} J.,
  2000, \aap, \href {http://adsabs.harvard.edu/abs/2000A%26A...360..603T} {360,
  603}

\bibitem[\protect\citeauthoryear{Uesugi \& Fukuda}{Uesugi \&
  Fukuda}{1970}]{uesugi1970}
Uesugi A.,  Fukuda I.,  1970, Contributions from the Institute of Astrophysics
  and Kwasan Observatory, University of Kyoto, Kyoto: University, Kwasan
  Observatory, Institute of Astrophysics, 1970

\bibitem[\protect\citeauthoryear{{Uytterhoeven} et~al.,}{{Uytterhoeven}
  et~al.}{2011}]{2011A&A...534A.125U}
{Uytterhoeven} K.,  et~al., 2011, \mn@doi [\aap] {10.1051/0004-6361/201117368},
  \href {http://cdsads.u-strasbg.fr/abs/2011A%26A...534A.125U} {534, A125}

\bibitem[\protect\citeauthoryear{{Van Reeth} et~al.,}{{Van Reeth}
  et~al.}{2015a}]{2015ApJS..218...27V}
{Van Reeth} T.,  et~al., 2015a, \mn@doi [\apjs] {10.1088/0067-0049/218/2/27},
  \href {http://adsabs.harvard.edu/abs/2015ApJS..218...27V} {218, 27}

\bibitem[\protect\citeauthoryear{{Van Reeth} et~al.,}{{Van Reeth}
  et~al.}{2015b}]{2015A&A...574A..17V}
{Van Reeth} T.,  et~al., 2015b, \mn@doi [\aap] {10.1051/0004-6361/201424585},
  \href {http://adsabs.harvard.edu/abs/2015A%26A...574A..17V} {574, A17}

\bibitem[\protect\citeauthoryear{{Van Reeth}, {Tkachenko}  \& {Aerts}}{{Van
  Reeth} et~al.}{2016}]{2016A&A...593A.120V}
{Van Reeth} T.,  {Tkachenko} A.,   {Aerts} C.,  2016, \mn@doi [\aap]
  {10.1051/0004-6361/201628616}, \href
  {http://adsabs.harvard.edu/abs/2016A%26A...593A.120V} {593, A120}

\bibitem[\protect\citeauthoryear{{Van Reeth} et~al.,}{{Van Reeth}
  et~al.}{2018}]{2018A&A...618A..24V}
{Van Reeth} T.,  et~al., 2018, \mn@doi [\aap] {10.1051/0004-6361/201832718},
  \href {http://adsabs.harvard.edu/abs/2018A%26A...618A..24V} {618, A24}

\bibitem[\protect\citeauthoryear{{Venn} \& {Lambert}}{{Venn} \&
  {Lambert}}{1990}]{1990ApJ...363..234V}
{Venn} K.~A.,  {Lambert} D.~L.,  1990, \mn@doi [\apj] {10.1086/169334}, \href
  {http://adsabs.harvard.edu/abs/1990ApJ...363..234V} {363, 234}

\bibitem[\protect\citeauthoryear{{Weiss} et~al.,}{{Weiss}
  et~al.}{2014}]{2014PASP..126..573W}
{Weiss} W.~W.,  et~al., 2014, \mn@doi [\pasp] {10.1086/677236}, \href
  {http://adsabs.harvard.edu/abs/2014PASP..126..573W} {126, 573}

\bibitem[\protect\citeauthoryear{{White} et~al.,}{{White}
  et~al.}{2017}]{2017MNRAS.471.2882W}
{White} T.~R.,  et~al., 2017, \mn@doi [\mnras] {10.1093/mnras/stx1050}, \href
  {http://adsabs.harvard.edu/abs/2017MNRAS.471.2882W} {471, 2882}

\bibitem[\protect\citeauthoryear{{Xiong}, {Deng}, {Zhang}  \& {Wang}}{{Xiong}
  et~al.}{2016}]{2016MNRAS.457.3163X}
{Xiong} D.~R.,  {Deng} L.,  {Zhang} C.,   {Wang} K.,  2016, \mn@doi [\mnras]
  {10.1093/mnras/stw047}, \href
  {http://adsabs.harvard.edu/abs/2016MNRAS.457.3163X} {457, 3163}

\bibitem[\protect\citeauthoryear{{Ziaali}, {Bedding}, {Murphy}, {Van Reeth}  \&
  {Hey}}{{Ziaali} et~al.}{2019}]{2019MNRAS.486.4348Z}
{Ziaali} E.,  {Bedding} T.~R.,  {Murphy} S.~J.,  {Van Reeth} T.,   {Hey} D.~R.,
   2019, \mn@doi [\mnras] {10.1093/mnras/stz1110}, \href
  {https://ui.adsabs.harvard.edu/abs/2019MNRAS.486.4348Z} {486, 4348}

\bibitem[\protect\citeauthoryear{{Zinn}, {Pinsonneault}, {Huber}  \&
  {Stello}}{{Zinn} et~al.}{2019}]{2019ApJ...878..136Z}
{Zinn} J.~C.,  {Pinsonneault} M.~H.,  {Huber} D.,   {Stello} D.,  2019, \mn@doi
  [\apj] {10.3847/1538-4357/ab1f66}, \href
  {https://ui.adsabs.harvard.edu/abs/2019ApJ...878..136Z} {878, 136}

\bibitem[\protect\citeauthoryear{{Zorec} \& {Royer}}{{Zorec} \&
  {Royer}}{2012a}]{2012A&A...537A.120Z}
{Zorec} J.,  {Royer} F.,  2012a, \mn@doi [\aap] {10.1051/0004-6361/201117691},
  \href {https://ui.adsabs.harvard.edu/\#abs/2012A&A...537A.120Z} {537, A120}

\bibitem[\protect\citeauthoryear{Zorec \& Royer}{Zorec \&
  Royer}{2012b}]{zorec2012}
Zorec J.,  Royer F.,  2012b, Astronomy \& Astrophysics, 537, A120

\bibitem[\protect\citeauthoryear{{Zwintz}, {Fossati}, {Ryabchikova}, {Kaiser},
  {Gruberbauer}, {Barnes}, {Baglin}  \& {Chaintreuil}}{{Zwintz}
  et~al.}{2013}]{2013A&A...550A.121Z}
{Zwintz} K.,  {Fossati} L.,  {Ryabchikova} T.,  {Kaiser} A.,  {Gruberbauer} M.,
   {Barnes} T.~G.,  {Baglin} A.,   {Chaintreuil} S.,  2013, \mn@doi [\aap]
  {10.1051/0004-6361/201220127}, \href
  {http://adsabs.harvard.edu/abs/2013A%26A...550A.121Z} {550, A121}

\bibitem[\protect\citeauthoryear{{Zwintz} et~al.,}{{Zwintz}
  et~al.}{2014}]{2014Sci...345..550Z}
{Zwintz} K.,  et~al., 2014, \mn@doi [Science] {10.1126/science.1253645}, \href
  {http://adsabs.harvard.edu/abs/2014Sci...345..550Z} {345, 550}

\bibitem[\protect\citeauthoryear{{Zwitter} et~al.,}{{Zwitter}
  et~al.}{2010}]{Zwitter2010}
{Zwitter} T.,  et~al., 2010, \mn@doi [\aap] {10.1051/0004-6361/201014922},
  \href {https://ui.adsabs.harvard.edu/\#abs/2010A&A...522A..54Z} {522, A54}

\bibitem[\protect\citeauthoryear{{van Leeuwen}}{{van
  Leeuwen}}{2007}]{Leeuwen07}
{van Leeuwen} F.,  2007, \mn@doi [\aap] {10.1051/0004-6361:20078357}, \href
  {http://adsabs.harvard.edu/abs/2007A%26A...474..653V} {474, 653}

\makeatother
\end{thebibliography}




\appendix

\section{Tables}

\onecolumn

\begin{landscape}

\setlength\LTcapwidth{\textwidth} 
\setlength\LTleft{0pt}            
\setlength\LTright{0pt}           

\begin{longtable}{|r|c|l|c|c|c|c|c|c|c|c|c|c|c|}

\caption{Stellar parameters for the TESS $\delta$~Sct and $\gamma$~Dor pulsators observed at 2-min cadence during the first two pointings (Sectors 1 and 2) in the ecliptic Southern hemisphere. The columns indicate: the TESS Input Catalogue (TIC) number, the TESS magnitude, an alternative identifier of the star, spectroscopic (Spec.) and photometric (Phot) effective temperatures and log g values and their associated uncertainties from literature respectively. The reference is indicated next to each value. In addition, we list the projected rotational velocity and binarity status as available in literature.  }
\label{table: star_parameters1}

\\ \hline

\multicolumn{1}{|c|}{TIC}  &  TESS  & \multicolumn{1}{|c|}{alternative}   &  $\teff [K]$  &  $\sigma_{\teff}$  &   $\teff [K]$  &  $\sigma_{\teff}$  &   $\logg$  &  $\sigma_{\logg}$  &   $\logg$  &  $\sigma_{\logg}$  &  $v \sin i$  &  $\sigma_{\text{vsini}}$ &   binarity \\

 & mag &\multicolumn{1}{|c|}{identifier} & Spec. & Spec. & Phot. & Phot. & Spec. & Spec. & Phot. & Phot. &[km~s$^{-1}$]&[km~s$^{-1}$]& \\\hline

\endhead
\hline
\endfoot

9632550 & 9.06 & BS Aqr & 7245\footnotemark[1] & 75\footnotemark[1] & - & - & 3.82\footnotemark[1] & 0.12\footnotemark[1] & - & - & 23\footnotemark[1] & - &  - \\
12470841 & 8.44 & HD 923 & - & - & 7832 \footnotemark[2] & - & - & - & - & - & - & - & - \\
12784216 & 8.92 & HD 213125 & - & - & 6519\footnotemark[3] & 178\footnotemark[3] & - & - & 4.03\footnotemark[3] & 0.26\footnotemark[3] & - & - & -\\
12974182 & 6.78 & HD 218003 & 7471\footnotemark[4] & 95\footnotemark[4] & 7584\footnotemark[5] & 208\footnotemark[5] & 3.86\footnotemark[4] & 0.18\footnotemark[4] & 4.05\footnotemark[5] & - & - & - & - \\
29281339 & 6.63 & HD 198501 & - & - & 7511\footnotemark[5] & 284\footnotemark[5] & - & - & 3.88\footnotemark[5] & - & - & - & - \\
30531417 & 11.03 & XX Dor & - & - & 6769\footnotemark[6] & 202\footnotemark[6] & - & - & - & 1.82\footnotemark[6] & - & - & - \\
32176627 & 6.90 & HD 209970 & - & - & 7156\footnotemark[5] & 347\footnotemark[5] & - & - & 4.10\footnotemark[5] & - & - & - & - \\
32197339 & 6.97 & HD 210139 & - & - & 7766\footnotemark[5] & 125\footnotemark[5] & - & - & 3.50\footnotemark[5] & - & - & - & \\
33911462 & 8.06 & HD 222828 & - & - & 6987\footnotemark[7] & 98\footnotemark[7] & - & - & 3.53\footnotemark[7] & - & - & - & - \\
33984043 & 6.37 & HD 223466 & - & - & 8433\footnotemark[8] & 58\footnotemark[8] & - & - & 4.23\footnotemark[9]{} & - & 69.5\footnotemark[10]{} & 1\footnotemark[10]{} & -\\
34038105 & 6.74 & HD 223991A & - & - & 8854\footnotemark[11] & & - & - & - & - & - & - & \\ 
38587180 & 7.69 & TX Ret & 8009\footnotemark[12] & - & 7258\footnotemark[2] & - & 3.62\footnotemark[12] & - & - & - & - & - & -\\
38602305 & 6.13 & $\theta$ Ret A & - & - & 12442\footnotemark[2]{} & - & - & - & - & - & - & - & binary\footnotemark[13]\\ 
38847248 & 5.86 & $\theta$ Tuc & - & - & 7638\footnotemark[14] & 90\footnotemark[14] & - & - & 3.59\footnotemark[14] & 0.08\footnotemark[14] & 80\footnotemark[15] & - & binary\footnotemark[16]\\ 
44627561 & 8.98 & HD 215559 & - & - & - & - & - & - & - & - & - & - & -\\ 
49677785 & 9.35 & HD 220687 & 7033\footnotemark[1] & 77\footnotemark[1] & - & - & 3.39\footnotemark[1] & 0.12\footnotemark[1] & - & - & 91\footnotemark[1] & - & - \\ 
51991595 & 12.07 & 2MASS J01011055-6843069 & - & - & 6794\footnotemark[6] & 201\footnotemark[6] & - & - & 4.19\footnotemark[6] & - & - & - & - \\
52244754 & 9.04 & HD 8438 & - & - & - & - & - & - & - & - & - & - & -\\ 
52258534 & 7.71 & BG Hyi & 7289\footnotemark[12] & - & 6928\footnotemark[2] & - & 2.91\footnotemark[12] & - & - & - & - & - & - \\ 
66434034 & 8.88 & HD 5497 & - & - & - & - & - & - & - & - & - & - & - \\ 
70744515 & 8.80 & HD 225077 & - & - & - & - & - & - & - & - & - & - & - \\ 
71334169 & 8.52 & HD 209430 & - & - & 6872\footnotemark[17] & 125\footnotemark[17] & - & - & - & - & - & - & - \\ 
79394646 & 3.58 & $\theta$ Ind A & 7870\footnotemark[18] & - & - & - & 4.14\footnotemark[18] & - & - & - & 135\footnotemark[10] & 2\footnotemark[10] & binary\footnotemark[19] \\
80474886 & 6.23 & AZ Phe & - & - & 7278\footnotemark[14] & 35\footnotemark[14] & - & - & 3.96\footnotemark[5] & 0.11\footnotemark[14] & & - & - \\ 
88277481 & 9.04 & HD 210740 & - & - & - & - & - & - & - & - & - & - & -\\ 
89464315 & 5.29 & WX PsA & 7430\footnotemark[39] & 34\footnotemark[39] & 7509\footnotemark[9] & 255\footnotemark[9]{} & - & - & 3.67\footnotemark[9] & 0.14\footnotemark[9] & 91\footnotemark[8] & - & - \\ 
89542582 & 9.71 & HD 223587 & - & - & 7295\footnotemark[5] & 208\footnotemark[5] & - & - & 3.97\footnotemark[5] & - & - & - & - \\
92734713 & 8.32 & HD 200835 & - & - & - & - & - & - & - & - & - & - & - \\
99839685 & 8.46 & HD 204972 & - & - & 7153 \footnotemark[2] & - & - & - & - & - & - & - & - \\
102090493 & 9.20 & HD 7454 & - & - & 7195\footnotemark[11] & - & - & - & - & - & - & - & -\\ 
116157537 & 5.88 & BB Phe & 6973\footnotemark[39] & 107\footnotemark[39]{} & 7000\footnotemark[2] & - & - & - & - & - & 83\footnotemark[15]{} & - & - \\ 
126659093 & 9.18 & ZZ Mic & 7217\footnotemark[1] & 54\footnotemark[1] & 7613\footnotemark[11] & - & - & - & - & - & 35\footnotemark[1] & - & -\\ 
137796620 & 9.87 & HD 216728 & - & - & 7559\footnotemark[6] & 213\footnotemark[6] & - & - & 4.13\footnotemark[6] & 0.48\footnotemark[6] & - & - & - \\
139131968 & 5.52 & Tau$^{3}$ Gru & - & - & 7820\footnotemark[21] & 100\footnotemark[21] & - & - & 3.723\footnotemark[5] & - & - & - & - \\ 
139232635 & 3.94 & $\theta$ Gru & - & - & - & - & - & - & - & - & 64\footnotemark[22] & - & binary\footnotemark[19]\\ 
139825582 & 7.40 & CF Ind & 6938\footnotemark[23] & - & 7075\footnotemark[2] & - & 3.3\footnotemark[23] & - & - & - & - & - & - \\ 
139845816 & 8.02 & RS Gru & 7853\footnotemark[12] & - & 6933\footnotemark[7] & 81\footnotemark[7] & - & - & 3.53\footnotemark[7] & - & 40\footnotemark[15] & - & binary\footnotemark[24]\\
141026903 & 6.37 & HD 35184 & - & - & 8017\footnotemark[8]{} & 206\footnotemark[8]{} & - & - & 3.7\footnotemark[5] & - & 94\footnotemark[25] & 6\footnotemark[25] & -\\
144309524 & 11.33 & TYC 8459-201-1 & - & - & 7818\footnotemark[6] & 217\footnotemark[6] & - & - & 4.11\footnotemark[6] & 1.76\footnotemark[6] & - & - & - \\
144387364 & 7.16 & BF Phe & - & - & 7241\footnotemark[5] & 185\footnotemark[5] & - & - & 4.02\footnotemark[5] & 0.17\footnotemark[14] & 80 & - & - \\ 
147085268 & 8.02 & HD 203880 & - & - & 7046\footnotemark[26] & - & - & - & - & - & - & - & - \\
147113185 & 5.30 & HD 204018 & - & - & 7118\footnotemark[2] & - & - & - & - & - & - & - & binary\footnotemark[27] \\ 
150101501 & 9.49 & HD 42780 & - & - & 8051\footnotemark[28] & 200\footnotemark[28] & - & - & - & - & - & - & - \\ 
150394126 & 6.54 & HD 46190 & 9790\footnotemark[29] & 97\footnotemark[29] & 8346\footnotemark[5] & 669\footnotemark[5] & - & - & 4.21\footnotemark[5] & - & - & - & - \\ 
152864226 & 7.84 & HD 217417 & - & - & - & - & - & - & - & - & - & - & - \\ 
154842794 & 5.15 & $\pi$ PsA & 7194\footnotemark[18] & - & - & & 4.24 & - & - & - & - & & binary\footnotemark[30]\\
161172103 & 6.34 & CC Gru & - & - & 7031\footnotemark[14] & 65\footnotemark[14] & - & - & 3.55\footnotemark[14] & 0.14\footnotemark[9]{} & 122 \footnotemark[18] & - & - \\ 
166808854 & 4.88 & $\eta$ Hor & - & - & 7203\footnotemark[31] & 90\footnotemark[31] & - & - & - & - & 7\footnotemark[31] & 2\footnotemark[31] & binary\footnotemark[32]\\
167602316 & 3.04 & $\alpha$ Pic & - & - & 7551\footnotemark[33] & 35\footnotemark[33] & - & - & - & - & 206\footnotemark[8] & - & - \\
183532876 & 11.24 & CD-34 16262 & - & - & 6844\footnotemark[6] & 203\footnotemark[6] & - & - & 4.18\footnotemark[6] & 1.49\footnotemark[6] & - & - & - \\ 
183595451 & 5.73 & AI Scl & 7147\footnotemark[39] & 145\footnotemark[39] & 7502\footnotemark[9] & 255\footnotemark[9] & 3.21\footnotemark[23] & - & & - & - & - & - \\
197686479 & 5.86 & BZ Gru & - & - & 6982\footnotemark[14] & 81\footnotemark[14] & - & - & 3.26\footnotemark[14] & - & - & - & - \\ 
197759259 & 8.68 & HD 209689 & - & - & - & - & - & - & - & - & - & - & - \\ 
198035211 & 9.45 & HD 26058 & 7335\footnotemark[1] & 442\footnotemark[1] & 7950\footnotemark[1] & 330\footnotemark[1] & - & - & - & - & - & - & - \\ 
201250317 & 9.23 & HD 225032 & 7474\footnotemark[1] & 83\footnotemark[1] & 8115\footnotemark[11] & - & 4.25\footnotemark[1] & 0.16\footnotemark[1] & - & - & - & - & -\\ 
206660904 & 7.65 & HD 216290 & - & - & - & - & - & - & - & - & - & - & - \\ 
211379298 & 9.46 & HD 203513 & - & - & 8019\footnotemark[28] & 200\footnotemark[28] & - & - & - & - & - & - & - \\ 
219332123 & 7.16 & DR Gru & - & - & 7447\footnotemark[14] & 137\footnotemark[14] & - & - & 3.82\footnotemark[14] & 0.1\footnotemark[34] & - & - & - \\ 
224280541 & 7.21 & HD 222701 & - & - & 7420\footnotemark[2]{} & - & - & - & - & - & - & - & - \\ 
224285142 & 8.68 & HD 222987 & - & - & - & - & - & - & - & - & - & - & - \\ 
224285325 & 7.01 & SX Phe & 9138\footnotemark[23] & - & - & - & 5.02\footnotemark[23] & - & - & - & 18\footnotemark[15]{} & - & - \\ 
229059574 & 6.15 & HD 210111 & 7400\footnotemark[14] & 100\footnotemark[14] & 7550\footnotemark[14] & 123\footnotemark[14] & 3.8\footnotemark[14] & 0.1\footnotemark[14] & 3.84\footnotemark[14] & - & 30\footnotemark[14] & - & binary\footnotemark[35]\\ 
229150702 & 5.76 & BD Phe & 7950\footnotemark[39] & - & 8050\footnotemark[9] & 274\footnotemark[9] & 3.83 & - & 3.99\footnotemark[9]{} & 0.09\footnotemark[14] & 118\footnotemark[31] & 15\footnotemark[31]{} & -\\
229154157 & 9.53 & HD 11667 & 7061\footnotemark[1] & 81\footnotemark[1] & 7179\footnotemark[4] & 146\footnotemark[4] & 3.48\footnotemark[1] & 0.12\footnotemark[1] & - & - & 31\footnotemark[1] & - & - \\ 
229810275 & 9.03 & HD 19098 & - & - & - & - & - & - & - & - & - & - & -\\ 
231014033 & 9.23 & HD 10961 & 7482\footnotemark[1] & 83\footnotemark[1] & 8069\footnotemark[11] & - & 4.34\footnotemark[1] & 0.16\footnotemark[1] & - & - & 39\footnotemark[1] & - & -\\ 
231020078 & 9.01 & HD 11333 & - & - & - & - & - & - & - & - & - & - & -\\ 
231048083 & 6.55 & FG Eri & 8986\footnotemark[23] & 400\footnotemark[23] & 7998\footnotemark[8] & 93\footnotemark[8] & 3.93\footnotemark[23] & 0.5\footnotemark[23] & 3.15\footnotemark[5] & - & 107\footnotemark[31] & 4\footnotemark[31] & -\\
231632224 & 12.72 & 2MASS J21101928-5750453 & - & - & 7027\footnotemark[6] & 202\footnotemark[6] & - & - & 4.17\footnotemark[6] & - & - & - & - \\
234498473 & 5.21 & HD 4125 & 7943\footnotemark[39] & 92\footnotemark[39] & - & - & - & - & - & - & 15\footnotemark[31] & 5\footnotemark[31] & binary\footnotemark[32]{} \\
234516307 & 8.78 & HD 5648 & - & - & - & - & - & - & - & - & - & - & - \\ 
234528371 & 8.97 & HD 6622 & - & - & 7230\footnotemark[36] & 27\footnotemark[36] & - & - & - & - & - & - & - \\
234548714 & 7.21 & BS Tuc & 7337\footnotemark[37] & - & 7328\footnotemark[14] & 102\footnotemark[14] & - & - &3.84\footnotemark[14] & 0.17\footnotemark[38] &130\footnotemark[15] & - & - \\ 
237318602 & 9.18 & HD 216941 & - & - & 7175\footnotemark[4] & 496\footnotemark[4] & - & - & - & - & - & - & -\\ 
237881239 & 9.38 & HD 8052 & 7477\footnotemark[1] & 83\footnotemark[1] & 7238\footnotemark[4] & 469\footnotemark[4] & 4.15\footnotemark[1] & 0.16\footnotemark[1] & - & - & 198\footnotemark[1] & - & - \\ 
238185398 & 9.14 & HD 23537 & - & - & 8173\footnotemark[11] & - & - & - & - & - & - & - & - \\
246845553 & 8.35 & HD 1619 & - & - & - & - & - & - & - & - & - & - & -\\ 
246853154 & 8.03 & HD 2026 & - & - & 6577\footnotemark[2]{} & - & - & - & - & - & - & - & -\\
253917376 & 8.07 & UV PsA & 7918\footnotemark[12] & - & 6889\footnotemark[7] & 154\footnotemark[7] & 4.00\footnotemark[12] & - & - & - & - & - & - \\
260353074 & 8.96 & HD 44596 & - & - & 7031\footnotemark[26] & - & - & - & - & - & - & - & - \\ 
260416268 & 5.47 & $\nu$ Pic & 7537\footnotemark[20] & 76\footnotemark[20] & - & - & - & - & - & - & - & - & - \\ 
260654645 & 12.44 & 2MASS J06324522-5748198 & - & - & 6826\footnotemark[6] & 201\footnotemark[6] & - & - & 4.18\footnotemark[6] & - & - & - & - \\ 
261089835 & 11.45 & TYC 9492-1017-1 & - & - & 6978\footnotemark[6] & 205\footnotemark[6] & - & - & 4.17\footnotemark[6] & - & - & - & -\\ 
265566844 & 7.52 & BX Ind & 6968\footnotemark[12] & - & 6762\footnotemark[7] & 88\footnotemark[7] & 3.62\footnotemark[7] & - & 0.19\footnotemark[7] & - & - & - & - \\
267094416 & 8.51 & HD 18006 & - & - & 7172\footnotemark[2] & - & - & - & - & - & - & - & - \\ 
269994543 & 8.63 & HD 197648 & - & - & - & - & - & - & - & - & - & - & - \\ 
270067755 & 7.09 & 2MASS J20472005-2624547 & - & - & 7098\footnotemark[5] & 163\footnotemark[5] & - & - & 3.83\footnotemark[5] & - & - & - & \\ 
277682809 & 8.32 & HD 208154 & - & - & - & - & - & - & - & - & - & - & - \\
278611926 & 7.51 & HD 210767 & - & - & - & - & - & - & - & - & - & - & binary\footnotemark[19]\\ 
279361762 & 8.04 & V383 Car & 6838\footnotemark[7] & 95\footnotemark[7] & - & - & 3.68\footnotemark[7] & - & - & - & - & - & - \\ 
279613634 & 8.77 & HD 54480 & - & - & 7147\footnotemark[28] & 150\footnotemark[28] & - & - & - & - & - & - & - \\ 
303584611 & 6.74 & BS Scl & 8216\footnotemark[23] & 400\footnotemark[23] & 7764\footnotemark[5] & 175\footnotemark[5] & 3.76\footnotemark[23] & 0.5\footnotemark[23] & 3.83\footnotemark[5] & - & - & - & - \\ 
308396022 & 10.99 & TYC 8928-1300-1 & - & - & 6730\footnotemark[5] & 248\footnotemark[5] & - & - & 3.81\footnotemark[5] & - & - & - & - \\
340705975 & 6.51 & AW Scl & - & - & 8103\footnotemark[5] & 143\footnotemark[5] & - & - & 3.60\footnotemark[5] & - & - & - & binary\footnotemark[30]\\ 
348762920 & 5.42 & HD 20313 & - & - & 7000\footnotemark[2]{} & - & - & - & 3.98 & - & 75\footnotemark[15] & - & binary\footnotemark[16]\\ 
348772511 & 7.25 & CP Oct & 7028\footnotemark[39] & 120\footnotemark[39] & 6900\footnotemark[38] & 100\footnotemark[38] & 3.44\footnotemark[23] & - & 3.63\footnotemark[38] & - & 72\footnotemark[40] & - & binary\footnotemark[41]\\ 
350431472 & 9.57 & HD 38081 & - & - & 7624\footnotemark[28] & 75\footnotemark[28] & - & - & - & - & - & - & - \\ 
350563225 & 9.07 & HD 39131 & 7632\footnotemark[42] & 95\footnotemark[42] &7827\footnotemark[11] & - & 4.1\footnotemark[42] & 0.18\footnotemark[42] &-& - &157\footnotemark[42] & - & - \\ 
355547586 & 12.56 & 2MASS J21242025-5751539 & - & - & 7073\footnotemark[6] & 206\footnotemark[6] & - & - & 4.16\footnotemark[6] & - & - & - & - \\
355687188 & 9.72 & HD 224852 & 6834\footnotemark[43] & 300\footnotemark[43] & 6865\footnotemark[5] & 360\footnotemark[5] & 3.50\footnotemark[43] & 0.3\footnotemark[43] & 3.55\footnotemark[5] & - & - & - & - \\
358070081 & 9.47 & HD 16938 & 7090\footnotemark[44] & - & 7032\footnotemark[4] & 208\footnotemark[4] & 3.95\footnotemark[44] & - & - & - & 109\footnotemark[1] & - & -\\
358502706 & 11.60 & 2MASS J04032009-8341586 & - & - & 7369\footnotemark[6] & 210\footnotemark[6] & - & - & - & - & - & - & - \\
364399376 & 7.16 & V393 Car & - & - & 7400\footnotemark[34] & 100\footnotemark[34] & - & - & 3.7\footnotemark[34] & - & - & - & - \\ 
381204458 & 8.76 & HD 19532 & - & - & - & - & - & - & - & - & - & - & - \\ 
381857833 & 8.15 & HD 204352 & - & - & 7308\footnotemark[45] & 81\footnotemark[45] & - & - & - & - & - & - & -\\ 
382551468 & 10.44 & CD-53 251 & 6761\footnotemark[46] & 93\footnotemark[46] & 5836\footnotemark[6] & 190\footnotemark[6] & 4.03\footnotemark[46] & 0.10\footnotemark[46] & 3.64\footnotemark[6] & - & 17\footnotemark[46] & 1\footnotemark[46] & - \\
394015973 & 7.96 & BE Ind & 7830\footnotemark[23] & - & - & - & 3.56\footnotemark[23] & - & - & - & 39\footnotemark[23] & 1\footnotemark[23] & - \\
396720223 & 9.57 & HD 28001 & 8720\footnotemark[52] & - & 7233\footnotemark[4] & 186\footnotemark[4] & - & - & - & - & - & - & - \\ 
399572664 & 9.78 & HD 215353 & - & - & 7826\footnotemark[5] & 247\footnotemark[5] & - & - & 4.10\footnotemark[5] & 1.56 \footnotemark[6] & - & - & - \\
402047030 & 8.05 & HD 221098 & - & - & - & - & - & - & - & - & - & - & - \\ 
402318229 & 8.96 & HD 221576 & - & - & - & - & - & - & - & - & - & - & - \\ 
426012953 & 9.62 & HD 8096 & 8000\footnotemark[1] & 83\footnotemark[1] & 8919\footnotemark[11] & - & 4.50\footnotemark[1] & 0.16\footnotemark[1] & - & - & - & - & - \\ 
431589510 & 11.81 & TYC 9158-919-1 & 7023\footnotemark[47] & 41\footnotemark[47] & - & - & 4.04\footnotemark[47] & 0.17\footnotemark[47] & - & - & - & - & - \\
439399707 & 8.82 & HD 225186 & - & - & 7518\footnotemark[2] & 377\footnotemark[2] & - & - & - & - & - & - & - \\ 
441110063 & 5.54 & 8 PsA & - & - & - & - & - & - & - & - & 23\footnotemark[25] & - & - \\
469844770 & 9.62 & HD 199247 & - & - & 7950\footnotemark[5] & 208\footnotemark[5] & - & - & 4.30\footnotemark[5] & - & - & - & - \\
469933721 & 5.88 & DQ Gru & - & - & - & - & - & - & - & - & 179\footnotemark[25] & - & -\\

\end{longtable}
\end{landscape}

\footnotetext[1]{\cite{Kunder2017}} 
\footnotetext[2]{\cite{McDonald2012}}
\footnotetext[3]{\cite{huber2016}} 
\footnotetext[4]{\cite{Stevens2017}} 
\footnotetext[5]{\cite{McDonald2017}} 
\footnotetext[6]{\cite{TESSInput2018}} 
\footnotetext[7]{\cite{Casagrande2011}} 
\footnotetext[8]{\cite{zorec2012}} 
\footnotetext[9]{\cite{David2015}}
\footnotetext[10]{\cite{Diaz2011}}
\footnotetext[11]{\cite{Ammons2006}} 
\footnotetext[12]{\cite{Rainer2016}} 
\footnotetext[13]{\cite{renson1991}} 
\footnotetext[14]{\cite{Paunzen2002}} 
\footnotetext[15]{\cite{rodriguez2000}}
\footnotetext[16]{\cite{2017MNRAS.465.1181L}} 
\footnotetext[17]{\cite{bailer2011}} 
\footnotetext[18]{\cite{erspamer2003}}
\footnotetext[19]{\cite{Tokovinin2016}}
\footnotetext[20]{\cite{Soubiran2010}} 
\footnotetext[21]{\cite{Eggleton2008}} 
\footnotetext[22]{\cite{uesugi1970}}
\footnotetext[23]{\cite{Rainer2016AJ}} 
\footnotetext[24]{\cite{liakos2016}} 
\footnotetext[25]{\cite{Royer2002}} 
\footnotetext[26]{\cite{2011A&A...535A...3S}} 
\footnotetext[27]{\cite{corbally1984}}
\footnotetext[28]{\cite{Gaia2018DR2}} 
\footnotetext[29]{\cite{2004ApJ...614L..77S}} 
\footnotetext[30]{\cite{pourbaix2004}}
\footnotetext[31]{\cite{2012A&A...542A.116A}}
\footnotetext[32]{\cite{malkov2012}}
\footnotetext[33]{\cite{royer2007}} 
\footnotetext[34]{\cite{Helt1984}} 
\footnotetext[35]{\cite{faraggiana2004}} 
\footnotetext[36]{\cite{2014AJ....148...81M}} 
\footnotetext[37]{\cite{cheng2017}} 
\footnotetext[38]{\cite{Niemczura2017}} 
\footnotetext[39]{\cite{soubiran2016}}
\footnotetext[40]{\cite{gonzalez2008}} 
\footnotetext[41]{\cite{paparo1996}} 
\footnotetext[42]{\cite{Kordopatis2013}} 
\footnotetext[43]{\cite{Zwitter2010}} 
\footnotetext[44]{\cite{Siebert2011}} 
\footnotetext[45]{\cite{Masana2006}} 
\footnotetext[46]{\cite{2014Sci...345..550Z}} 
\footnotetext[47]{\cite{Martell2017}} 


\clearpage
\twocolumn

\clearpage
\onecolumn

\begin{landscape}
\setlength\LTcapwidth{\textwidth} 
\setlength\LTleft{0pt}            
\setlength\LTright{0pt}           

\begin{longtable}{|r|l|c|c|c|c|c|c|l|c|} 

\caption{Stellar parameters and variability type of the 2-min cadence TESS $\delta$ Scuti and $\gamma$~Dor pulsators determined in this work. In this table we list the TESS Input Catalogue (TIC) name, an alternative name for each star, the effective temperatures derived using the SED method (Section \ref{SED}), the luminosities derived based on the Gaia and when available Hipparcos parallaxes and the respective uncertainties (Section \ref{Gaia}). In column 9 we describe the variability type determined from the TESS data. The question mark means that the variability type is uncertain (e.g. due to unresolved peaks).  HAGD denotes a High Amplitude $\gamma$ Dor star. The last column describes the chemical peculiarity as described in \citet{Renson2009}.}


\label{tab:stellar_parameters_new1}
\\ \hline

\multicolumn{1}{|c|}{TIC} 	&	\multicolumn{1}{|c|}{alternative}   &	$T_{\rm eff}$ [K] &	 $\sigma T_{\rm eff}$ & $\log (\rm L/{\rm L}_{\odot}$) & $\sigma \log (\rm L/{\rm L}_{\odot}$) & $\log (\rm L/{\rm L}_{\odot}$) & $\sigma \log (\rm L/{\rm L}_{\odot}$) 	&	\multicolumn{1}{|c|}{variability}  	& chem.	\\
	
			&\multicolumn{1}{|c|}{identifier} &	SED	 	& 		&	Gaia		&			&	HIPP.	&		&	\multicolumn{1}{|c|}{type} 	&	pec.	\\ \hline			
\endhead
\hline
\endfoot			

9632550 &  BS Aqr  & 6760 & 140 & 1.62 & 0.03 & 1.96 & 0.94 &  HADS &\\
12470841 &  HD 923   & 8090 & 230 & 1.64 & 0.02 & 2.01 & 0.43 &  one large peak only &\\
12784216 &  HD 213125  & 6390 & 120 & 1.14 & 0.01 &  &  &  $\delta$~Sct/hybrid?&\\
12974182 &  HD 218003  & 7670 & 170 & 1.103 & 0.008 & 1.08 & 0.04 &  rot?/binary?& \\
29281339 &  HD 198501  & 7430 & 160 & 1.205 & 0.007 & 1.17 & 0.04 &  almost const &\\
30531417 &  XX Dor  & 6800 & 150 & 1.64 & 0.03 &  &  &  HAGD &\\
32176627 &  HD 209970  & 7140 & 140 & 0.883 & 0.006 & 0.92 & 0.04 &  const &  \\ 
32197339 &  HD 210139  & 8120 & 180 & 1.9 & 0.1 & 1.69 & 0.13 &  $\gamma$~Dor (unresolved)&   \\
33911462 &  HD 222828  & 6700 & 140 & 1.28 & 0.02 & 1.52 & 0.33 &  $\delta$~Sct &  \\
33984043 &  HD 223466  & 8340 & 80 & 1.257 & 0.008 & 1.33 & 0.05 &  const  &  Am \\
34038105 &  HD 223991A  & 8640 & 90 & 1.131 & 0.007 & 1.02 & 0.06 &  const&   \\
38587180 &  TX Ret  & 7230 & 160 & 1.343 & 0.006 & 1.38 & 0.1 &  $\delta$~Sct  & Am \\
38602305 &  $\theta$ Ret A  & 12000 & 390 & 2.101 & 0.006 & 2.09 & 0.04 &  binary or multiple system&    \\
38847248 &  $\theta$ Tuc  & 7510 & 150 & 1.831 & 0.007 & 1.77 & 0.03 &  $\delta$~Sct/EB  & \\
44627561 &  HD 215559  & 6410 & 140 & 1.44 & 0.02 &  &  &  $\delta$~Sct/binary?/Ap?&   Am \\
49677785 &  HD 220687  & 7360 & 180 & 1.42 & 0.02 &  &  &  $\delta$~Sct /binary &  \\
51991595 &  2MASS J01011055-6843069  & 6890 & 150 & 1.25 & 0.03 &  &  &  HADS & Am \\
52244754 &  HD 8438  & 6650 & 140 & 1.240 & 0.009 &  &  &  $\delta$~Sct  & Am \\
52258534 &  BG Hyi  & 6990 & 140 & 1.233 & 0.006 & 1.33 & 0.1 &  $\delta$~Sct &\\
66434034 &  HD 5497  & 7830 & 200 & 0.94 & 0.01 &  &  &  $\delta$~Sct   &\\
70744515 &  HD 225077  & 9870 & 550 & 1.27 & 0.03 & 1.21 & 0.18 &  const &  \\
71334169 &  HD 209430  & 7150 & 160 & 1.21 & 0.01 & 0.62 & 0.21 &  $\delta$~Sct   & \\
79394646 &  $\theta$ Ind A  & 8090 & 60 & 1.074 & 0.008 & 1.1 & 0.01 &  $\delta$~Sct/binary/rot  & Am \\
80474886 &  AZ Phe  & 7030 & 120 & 1.308 & 0.006 & 1.39 & 0.09 &  $\delta$~Sct &   Am \\
88277481 &  HD 210740  & 7520 & 170 & 0.852 & 0.008 &  &  &  $\gamma$~Dor/hybrid/1 peak only in $\delta$~Sct regime &   \\
89464315 &  WX PsA  & 7250 & 150 & 1.68 & 0.01 & 1.7 & 0.02 &  $\delta$~Sct  & Am \\
89542582 &  HD 223587  & 7430 & 170 & 1.09 & 0.03 & 0.81 & 0.36 &  $\delta$~Sct &  \\
92734713 &  HD 200835  & 8420 & 270 & 1.12 & 0.01 &  &  &  $\delta$~Sct &  \\
99839685 &  HD 204972  & 7210 & 160 & 1.10 & 0.01 & 0.84 & 0.16 &  $\delta$~Sct/ hybrid? &  \\
102090493 &  HD 7454  & 5890 & 110 & 1.28 & 0.01 & 0.79 & 0.21 &  $\delta$~Sct /Ap/binary?  &  \\
116157537 &  BB Phe  & 7070 & 120 & 1.749 & 0.006 & 1.7 & 0.05 &  $\delta$~Sct  & \\
126659093 &  ZZ Mic  & 8110 & 210 & 1.2 & 0.02 & 1.03 & 0.33 &  HADS  & \\
137796620 &  HD 216728  & 7840 & 200 & 0.89 & 0.03 &  &  &  $\delta$~Sct  & \\
139131968 &  $\tau^{3}$ Gru  & 7550 & 160 & 1.457 & 0.009 & 1.38 & 0.02 &  const  & Am \\
139232635 &  $\theta$ Gru  & 6990 & 90 & 1.4 & 0.02 & 1.45 & 0.02 &  rot &  \\
139825582 &  CF Ind  & 7250 & 140 & 1.504 & 0.009 & 1.47 & 0.14 &  $\delta$~Sct/ hybrid? &   Am \\
139845816 &  RS Gru  & 7780 & 170 & 1.45 & 0.01 & 1.41 & 0.25 &  HADS &  \\
141026903 &  HD 35184   & 8120 & 160 & 1.455 & 0.005 & 1.43 & 0.03 &  const  & \\
144309524 &  TYC 8459-201-1   & 7220 & 240 & 1.32 & 0.04 &  &  &  HADS &   \\
144387364 &  BF Phe  & 7350 & 170 & 1.013 & 0.005 & 0.97 & 0.06 &  $\delta$~Sct/ hybrid? &  \\
147085268 &  HD 203880  & 7290 & 160 & 1.062 & 0.008 &  &  &  $\delta$~Sct/ hybrid? &  \\
147113185 &  HD 204018  & 6900 & 120 & 1.175 & 0.006 & 1.21 & 0.03 &  const  & Am \\
150101501 &  HD 42780   & 7920 & 240 & 1.09 & 0.01 &  &  &  $\delta$~Sct  & \\
150394126 &  HD 46190   & 8850 & 100 & 1.167 & 0.005 & 1.16 & 0.02 &  $\delta$~Sct  &  \\
152864226 &  HD 217417  & 7200 & 150 & 0.93 & 0.01 & 0.91 & 0.1 &  $\delta$~Sct/ hybrid  & \\
154842794 &  $\pi$PsA  & 7420 & 110 & 0.783 & 0.007 & 0.82 & 0.02 &  $\gamma$~Dor   &Am \\
161172103 &  CC Gru  & 7160 & 150 & 1.466 & 0.006 & 1.57 & 0.06 &  $\delta$~Sct  & \\
166808854 &  $\eta$Hor  & 7840 & 500 & 1.06 & 0.02 & 1.14 & 0.04 &  $\delta$~Sct   &Am \\
167602316 &  $\alpha$ Pic   & 7350 & 660 &   &   & 1.56 & 0.05 &  $\delta$~Sct/hybrid &    \\
183532876 &  CD-34 16262  & 7140 & 280 & 1.55 & 0.04 &  &  &  HADS  & Am \\
183595451 &  AI Scl  & 7360 & 140 & 1.234 & 0.009 & 1.23 & 0.02 &  $\delta$~Sct &  Am \\ 
197686479 &  BZ Gru  & 6950 & 130 & 1.69 & 0.01 & 1.77 & 0.05 &  $\delta$~Sct &  \\
197759259 &  HD 209689  & 7510 & 180 & 1.135 & 0.009 &  &  &  $\delta$~Sct/ hybrid &  \\
198035211 &  HD 26058   & 7530 & 170 & 0.95 & 0.01 &  &  &  $\delta$~Sct  & \\
201250317 &  HD 225032  & 7410 & 250 & 1.01 & 0.01 &  &  &  $\delta$~Sct  &  Am \\
206660904 &  HD 216290  & 7450 & 170 & 1.33 & 0.01 & 1.26 & 0.12 &  const  & \\
211379298 &  HD 203513  & 8110 & 210 & 1.20 & 0.03 &  &  &  $\delta$~Sct /binary   & Am \\
219332123 &  DR Gru  & 7830 & 160 & 1.062 & 0.007 & 1.02 & 0.07 &  $\delta$~Sct/ hybrid? &  \\
224280541 &  HD 222701  & 7420 & 140 & 1.35 & 0.01 & 1.46 & 0.08 &  const &  \\
224285142 &  HD 222987  & 7730 & 190 & 1.23 & 0.01 & 1.09 & 0.18 &  $\delta$~Sct&   \\
224285325 &  SX Phe  & 7320 & 170 & 0.844 & 0.009 & 0.79 & 0.04 &  HADS  & Am \\
229059574 &  HD 210111  & 7910 & 110 & 1.157 & 0.006 & 1.15 & 0.03 &  $\delta$~Sct &   \\
229150702 &  BD Phe  & 8300 & 80 & 1.343 & 0.005 & 1.33 & 0.02 &  $\delta$~Sct  & Am \\
229154157 &  HD 11667   & 6660 & 140 & 1.33 & 0.01 &  &  &  $\delta$~Sct &  \\
229810275 &  HD 19098   & 8250 & 220 & 1.17 & 0.008 &  &  &  const &  \\
231014033 &  HD 10961   & 7150 & 150 & 0.88 & 0.01 &  &  &  $\delta$~Sct  & \\
231020078 &  HD 11333   & 7450 & 180 & 0.94 & 0.02 &  &  &  $\delta$~Sct/ hybrid? &  \\
231048083 &  FG Eri  & 7940 & 140 & 1.984 & 0.007 & 2 & 0.07 &  developed $\delta$~Sct? &  \\
231632224 &  2MASS J21101928-5750453  & 7210 & 280 & 1.56 & 0.08 &  &  &  HADS  &  \\
234498473 &  HD 4125  & 7960 & 160 & 1.61 & 0.08 &  &  &  $\delta$~Sct  & Am \\
234516307 &  HD 5648  & 7520 & 180 & 1.097 & 0.008 &  &  &  $\delta$~Sct &  \\
234528371 &  HD 6622  & 7180 & 220 & 0.91 & 0.007 &  &  &  $\delta$~Sct &  Am \\
234548714 &  BS Tuc  & 7290 & 160 & 0.981 & 0.005 & 0.95 & 0.04 &  $\delta$~Sct/ hybrid? &  \\
237318602 &  HD 216941  &  &  &  &  &  &  &  $\delta$~Sct  & Am \\
237881239 &  HD 8052  & 7400 & 180 & 0.913 & 0.008 &  &  &  $\delta$~Sct/ hybrid?  & \\
238185398 &  HD 23537   & 8100 & 210 & 1.011 & 0.007 &  &  &   roAp candidate? &   \\
246845553 &  HD 1619  & 7200 & 160 & 1.00 & 0.01 &  &  &  spots &  Am \\
246853154 &  HD 2026  & 8340 & 160 & 1.39 & 0.01 & 1.27 & 0.12 &  rot?/binary?  & \\
253917376 &  UV PsA  & 6850 & 140 & 1.24 & 0.01 &  &  &  $\delta$~Sct  &  Am \\
260353074 &  HD 44596   & 7010 & 150 & 1.26 & 0.02 &  &  &  $\delta$~Sct/ hybrid?  & \\
260416268 &  $\nu$ Pic  & 7830 & 300 & 1.20 & 0.02 & 1.04 & 0.02 &  rot?/binary?  & Am \\
260654645 &  2MASS J06324522-5748198  & 7050 & 160 & 1.24 & 0.03 &  &  &  HADS  & \\
261089835 &  TYC 9492-1017-1  & 6310 & 160 & 1.32 & 0.03 &  &  &  HADS &  Am \\
265566844 &  BX Ind  & 6640 & 130 & 1.46 & 0.01 & 1.09 & 0.1 &  $\delta$~Sct?  1 peak &\\
267094416 &  HD 18006   & 7300 & 160 & 1.08 & 0.02 & 1 & 0.1 &  almost const   \\
269994543 &  HD 197648  & 7430 & 170 & 0.95 & 0.02 &  &  &  $\delta$~Sct/ hybrid? &  Am \\
270067755 &  2MASS J20472005-2624547  & 7790 & 200 & 1.077 & 0.008 & 1.13 & 0.1 &  almost  const &  Am \\
277682809 &  HD 208154  & 8320 & 240 & 0.936 & 0.007 & 0.98 & 0.14 &  $\delta$~Sct/ hybrid? &  Am \\
278611926 &  HD 210767  & 7370 & 140 & 4 & 3 &  &  &  rot?/binary? &   \\
279361762 &  V383 Car  & 6950 & 140 & 1.612 & 0.008 & 1.38 & 0.1 &  $\delta$~Sct &  \\
279613634 &  HD 54480   & 7190 & 160 & 1.00 & 0.01 &  &  &  $\delta$~Sct   & \\
303584611 &  BS Scl  & 8050 & 140 & 1.370 & 0.008 & 1.27 & 0.04 &  $\delta$~Sct  & \\
308396022 &  TYC 8928-1300-1  & 6860 & 150 & 0.94 & 0.01 &  &  &  $\delta$~Sct/HADS/hybrid? &  \\
340705975 &  AW Scl  & 8010 & 150 & 1.757 & 0.009 & 1.63 & 0.06 &  Ap?  & \\
348762920 &  HD 20313   & 7370 & 130 & 1.461 & 0.007 & 1.48 & 0.04 &  $\delta$~Sct/hybrid  &  \\
348772511 &  CP Oct  & 6810 & 140 & 1.47 & 0.01 & 1.63 & 0.12 &  $\delta$~Sct  &  \\
350431472 &  HD 38081   & 7020 & 170 & 0.998 & 0.007 &  &  &  $\delta$~Sct &  \\
350563225 &  HD 39131   & 7610 & 180 & 1.07 & 0.01 &  &  &  $\delta$~Sct &  \\
355547586 &  2MASS J21242025-5751539  & 7240 & 170 & 1.05 & 0.08 &  &  &  HADS   &\\
355687188 &  HD 224852  & 6510 & 140 & 1.45 & 0.02 &  &  &  HADS  & \\
358070081 &  HD 16938   & 6540 & 140 & 0.9 & 0.2 &  &  &  $\delta$~Sct &  \\
358502706 &  TYC 9492-02623-1  & 7220 & 190 & 1.12 & 0.02 &  &  &  HADS &  \\
364399376 &  V393 Car  & 7330 & 170 & 1.489 & 0.008 &  &  &  $\delta$~Sct  & \\
381204458 &  HD 19532   & 7730 & 180 & 1.48 & 0.01 &  &  &  $\delta$~Sct /binary  & Am \\
381857833 &  HD 204352  & 7460 & 170 & 1.102 & 0.008 & 0.98 & 0.11 &  $\delta$~Sct  &  \\
382551468 &  CD-53 251  & 6080 & 110 & 0.83 & 0.01 &  &  &  const  & \\
394015973 &  BE Ind  & 7520 & 180 & 1.33 & 0.01 & 1.42 & 0.23 &  $\delta$~Sct/ hybrid?  & \\
396720223 &  HD 28001   & 7200 & 160 & 1.13 & 0.02 &  &  &  $\delta$~Sct/ hybrid? &  \\
399572664 &  HD 215353  & 8050 & 210 & 1.26 & 0.02 &  &  &  $\delta$~Sct /binary?/rot? &  \\
402047030 &  HD 221098  & 7080 & 150 & 1.213 & 0.008 & 1.25 & 0.16 &  const &  \\
402318229 &  HD 221576  & 7650 & 180 & 1.07 & 0.02 &  &  &  $\delta$~Sct/ hybrid?  & Am \\
426012953 &  HD 8096  & 8360 & 290 & 1.38 & 0.02 &  &  &  const &   \\
431589510 &  TYC 9158-919-1   & 6930 & 190 & 1.2 & 0.3 &  &  &  HADS   \\
439399707 &  HD 225186  & 7250 & 170 & 0.99 & 0.02 & 1.19 & 0.25 &  $\delta$~Sct/binary/flares &  \\
441110063 &  8 PsA  & 7720 & 160 & 1.122 & 0.008 & 1.17 & 0.02 &  $\delta$~Sct  & \\
469844770 &  HD 199247  & 7980 & 210 & 0.92 & 0.01 &  &  &  $\delta$~Sct &  \\
469933721 &  V* DQ Gru  & 0 & 0 & 1.691 & 0.008 & 1.63 & 0.07 &  $\gamma$~Dor/  $\delta$~Sct/hybrid? &   \\
			
\hline

\end{longtable}

\end{landscape}
\clearpage
\twocolumn

\section{Author Affiliations}

$^{1}$Stellar Astrophysics Centre, Ny Munkegade 120, 8000, Aarhus, Denmark \\
$^{2}$Instituto de Astrof{\' i}sica e Ci{\^ e}ncias do Espa{\c c}o, Universidade do Porto, CAUP, Rua das Estrelas, PT4150-762 Porto, Portugal \\
$^{3}$Institute of Astronomy, KU Leuven, Celestijnenlaan 200D, 3001 Leuven, Belgium \\
$^{4}$Sydney Institute for Astronomy (SIfA), School of Physics, University of Sydney, NSW 2006, Australia \\
$^{5}$Jeremiah Horrocks Institute, University of Central Lancashire, Preston PR1~2HE, UK\\
$^{6}$LESIA, Observatoire de Paris, Universit\'e PSL, CNRS, Sorbonne Universit\'e, Universit\'e de Paris, 5 place Jules Janssen, 92195 Meudon, France \\
$^{7}$Instytut Astronomiczny, Uniwersytet Wroc{\l}awski, Kopernika 11, 51-622 Wroc{\l}aw, Poland\\
$^{8}$Instituto de Astronom\'{\i}a, Universidad Nacional Aut\'onoma de M\'exico, Ap. P. 877, Ensenada, BC 22860, Mexico\\
$^{9}$Dept. F\'{\i}sica Te\'orica y del Cosmos, Universidad de Granada, Campus de Fuentenueva s/n, E-18071 Granada, Spain \\
$^{10}$Instituto de Astrof\'isica de Andaluc\'ia, Glorieta de la Astronom\'ia s/n, Granada, Spain\\
$^{11}$Department of Physics, Institute for Advanced Studies in Basic Sciences (IASBS), Zanjan 45137-66731, Iran\\
$^{12}$Department of Physics, Faculty of Science, University of Zanjan P.O. Box 45195-313, Zanjan, Iran\\
$^{13}$Nicolaus Copernicus Astronomical Center, Bartycka 18, PL-00-716 Warsaw, Poland \\
$^{14}$ \c{C}anakkale Onsekiz Mart University, Faculty of Sciences and Arts, Physics Department, 17100 \c{C}anakkale, Turkey\\ 
$^{15}$Center for Astrophysics ${\rm \mid}$ Harvard {\rm \&} Smithsonian, 60 Garden Street, Cambridge, MA 02138, USA\\
$^{16}$Department of Physics and Astronomy, University of British Columbia, Vancouver, Canada\\
$^{17}$Department of Theoretical Physics and Astrophysics, Masaryk University, Kotlarska 2, 611 37 Brno, Czech Republic\\
$^{18}$ Instituto de Astrof\'isica La Plata, CONICET-UNLP, Paseo del Bosque s/n, 1900 La Plata, Argentina\\
$^{19}$Department of Physics, Lehigh University, 16 Memorial Drive East, Bethlehem, PA, 18015, USA\\
$^{20}$School of Earth and Space Exploration, Arizona State University, Tempe, AZ 85281, USA\\ 
$^{21}$Department of Physics and Kavli Institute for Astrophysics and Space Research, Massachusetts Institute of Technology, Cambridge, MA 02139, USA\\
$^{22}$Department of Earth, Atmospheric and Planetary Sciences, Massachusetts Institute of Technology, Cambridge, MA 02139, USA\\
$^{23}$Department of Aeronautics and Astronautics, MIT, 77 Massachusetts Avenue, Cambridge, MA 02139, USA\\
$^{24}$Astrophysics Group, Keele University, Staffordshire, ST5 5BG, United Kingdom\\
$^{25}$Department of Physics, University of York, Heslington, York, YO10 5DD, UK\\
$^{26}$Konkoly Observatory, MTA CSFK, H-1121, Konkoly Thege Mikl\'os \'ut 15-17, Budapest, Hungary \\
$^{27}$22 MTA CSFK Lend\"ulet Near-Field Cosmology Research Group \\
$^{28}$Dept. F\'{\i}sica Te\'orica y del Cosmos, Universidad de Granada, Campus de Fuentenueva s/n, E-18071, Granada, Spain\\
$^{29}$Instituto de Astrof\'isica de Andaluc\'ia, Glorieta de la Astronom\'ia s/n, Granada, Spain \\
$^{30}$Yunnan Observatories, Chinese Academy of Sciences, 396 Yangfangwang, Guandu District, Kunming, 650216, P. R. China\\
$^{31}$Key Laboratory for the Structure and Evolution of Celestial Objects, Chinese Academy of Sciences, 396 Yangfangwang, Guandu District, Kunming, 650216, P. R. China\\
$^{32}$Center for Astronomical Mega-Science, Chinese Academy of Sciences, 20A Datun Road, Chaoyang District, Beijing, 100012, P. R. China\\
$^{33}$Universit\"at Innsbruck, Institut f\"ur Astro- und Teilchenphysik, Technikerstrasse 25, A-6020 Innsbruck, Austria\\
$^{34}$Dpto. de Astrof\'isica, Centro de Astrobiolog\'ia (CSIC-INTA), ESAC, Camino Bajo del Castillo s/n, 28692, Spain\\
$^{35}$Department of Chemistry and Physics, Florida Gulf Coast University, 10501 FGCU Blvd., Fort Myers, FL 33965 USA\\
$^{36}$Royal Observatory of Belgium, Ringlaan 3, 1180, Brussel, Belgium\\
$^{37}$Center for Exoplanets and Habitable Worlds, 525 Davey Laboratory, The Pennsylvania State University, University Park, PA 16802, USA \\
$^{38}$Los Alamos National Laboratory, MS T-082, Los Alamos, NM 87545 USA\\
$^{39}$Physics Department, Bloomsburg University, 200 E 2nd St, Bloomsburg, PA 17815, USA\\
$^{40}$Physics Department, Mount Allison University, Sackville, NB, E4L 1E6, Canada\\
$^{41}$Astrophysics Research Group, Faculty of Engineering and Physical Sciences, University of Surrey, Guildford GU2 7XH, UK\\
$^{42}$National Astronomical Research Institue of Thailand, 260 Moo 4, T. Donkaew, A. Maerim, Chiangmai, 50180 Thailand\\
$^{43}$Departamento de F\'{\i}sica e Astronomia, Faculdade de Ci\^encias da Universidade do Porto, Portugal\\
$^{44}$Department of Physics \& Astronomy, Camosun College, Victoria, BC, Canada.\\
$^{45}$IRAP, Universit\'e de Toulouse, CNRS, UPS, CNES, 14, avenue Edouard Belin, F-31400 Toulouse, France\\
$^{46}$Department of Theoretical Physics and Astrophysics, Masaryk University, Kotl\'{a}\v{r}sk\'{a} 2, 61137 Brno, Czech Republic \\
$^{47}$Astronomical Institute, Czech Academy of Sciences, Fri\v{c}ova 298, 25165, Ond\v{r}ejov, Czech Republic \\
$^{48}$Institute of Astronomy and NAO, Bulgarian Academy of Sciences, blvd.Tsarigradsko chaussee 72, Sofia 1784, Bulgaria\\
$^{49}$Institute for Astrophysics, University of Vienna, T\"urkenschanzstrasse 17, 1180 Vienna, Austria \\


\bsp	
\label{lastpage}
\end{document}